\documentstyle[12pt,psfig]{article}
\def\Del{\Delta}
\def\dx{\Del\chi^2}
\def\halfsq{\frac{1}{\sqrt{2}}}
\def\diag{{\rm diag}}
\def\eff{{\rm eff}}
\def\susy{{\rm SUSY}}
\def\delg{\ov{\delta}_G^{}}
\def\ddelg{\Del\ov{\delta}_G^{}}
\def\dgfa{\Del g_\alpha^{f}}
\def\dgnll{\Del g_L^{\nu_l}}
\def\dgnl{\Del g_L^{\nu}}
\def\dgel{\Del g_L^{e}}
\def\dgll{\Del g_L^l}
\def\dglr{\Del g_R^l}
\def\dger{\Del g_R^{e}}
\def\dgul{\Del g_L^{u}}
\def\dgur{\Del g_R^{u}}
\def\dgdl{\Del g_L^{d}}
\def\dgdr{\Del g_R^{d}}
\def\dgntaul{\Del g_L^{\nu_\tau}}
\def\dgtaul{\Del g_L^{\tau}}
\def\dgtaur{\Del g_R^{\tau}}
\def\dgbl{\Del g_L^{b}}
\def\dgbr{\Del g_R^{b}}
\def\alps{\alpha_s}
\def\ztautau{Z\tau\tau}
\def\zbb{Zbb}
\def\zblbl{Z b_L^{} b_L^{}}
\def\zbrbr{Z b_R^{} b_R^{}}
\def\zqq{Zqq}
\def\zff{Zff}
\def\zll{Zll}
\def\simgt{\,{\rlap{\lower 3.5pt\hbox{$\mathchar\sim$}}\raise 1pt\hbox{$>$}}\,}
\def\simlt{\,{\rlap{\lower 3.5pt\hbox{$\mathchar\sim$}}\raise 1pt\hbox{$<$}}\,}

\def\mhlight{m_h}
\def\mhheavy{m_H}
\def\mpseudo{m_A}
\def\mcharged{m_{H^-}}
\def\chargino#1{\widetilde{\chi}^-_#1}
\def\neutralino#1{\widetilde{\chi}^0_#1}
\def\sfermi{\wt{f}}
\def\squark{\wt{q}}
\def\slepton{\wt{l}}
\def\sup{\wt{u}}
\def\sdown{\wt{d}}
\def\stop{\wt{t}}
\def\sbottom{\wt{b}}
\def\sneutrino{\wt{\nu}}
\def\selectron{\wt{e}}
\def\smuon{\wt{\mu}}
\def\stau{\wt{\tau}}
\def\gluino{\wt{g}}
\def\mn#1{m_{\widetilde{\chi}^0_#1}}
\def\mch#1{m_{\widetilde{\chi}^-_#1}}
\def\cunitary#1{U^C_#1}
\def\nunitary#1{U^N_#1}
\def\rttwo{\sqrt{2}}
\def\re{{\rm Re}}
\def\gf{G_F}
\def\bite{\begin{itemize}}
\def\eite{\end{itemize}}
\def\half{\frac{1}{2}}
\def\quarter{\frac{1}{4}}
\def\wt{\widetilde}
\def\dof{{\rm d.o.f.}}
\def\to{\rightarrow}
\def\vsk#1{\noalign{\vskip#1 cm}}
\def\vsp#1{\vspace{#1 cm}}
\def\hsp#1{\hspace{#1 cm}}
\def\ov{\overline}
\def\disp{\displaystyle}
\def\smr{{\rm SM}}
\def\hph{\hphantom{-}}
\def\hpz{\hphantom{0}}
\def\fb{{\rm FB}}
\def\gev{~{\rm GeV}}
\def\tev{~{\rm TeV}}
\def\mev{~{\rm MeV}}
\def\mb{m_b^{}}
\def\mt{m_t^{}}
\def\mtsq{m_t^2}
\def\mh{m_{H_{\smr}}}
\def\xa{x_\alpha^{}}
\def\xs{x_s^{}}
\def\xt{x_t^{}}
\def\xh{x_h^{}}
\def\mw{m_W^{}}
\def\mwsq{m_W^2}
\def\mz{m_Z^{}}
\def\mzsq{m_Z^2}
\def\msbar{\ov{{\rm MS}}}
\def\dsz{\Del S_Z}
\def\dtz{\Del T_Z}

\def\dr{\Del R}
\def\sz{S_Z}
\def\tz{T_Z}

\def\ds{\Del S}
\def\dt{\Del T}
\def\du{\Del U}
\def\dgbl{\Del g_L^b}
\def\dmw{\Del m_W^{}}
\def\ebarsq{\bar{e}^2}
\def\sbarsq{\bar{s}^2}
\def\cbarsq{\bar{c}^2}
\def\gzbarsq{\bar{g}_Z^2}
\def\gwbarsq{\bar{g}_W^2}
\def\abar{\bar{\alpha}}
\def\ehat{\hat{e}}
\def\shat{\hat{s}}
\def\chat{\hat{c}}
\def\gzhat{\hat{g}_Z}
\def\ghat{\hat{g}}
\def\ehatsq{\hat{e}^2}
\def\shatsq{\hat{s}^2}
\def\chatsq{\hat{c}^2}
\def\gzhatsq{\hat{g}_Z^2}
\def\ghatsq{\hat{g}^2}

\def\dgzbarsq{\Del \gzbarsq}
\def\dsbarsq{\Del \sbarsq}
\def\etal{{\it et al.~}}

\newcommand {\bdd} {{\stackrel{\leftrightarrow}{\partial}}} 
\newcommand{\beq}{\begin{equation}}
\newcommand{\eeq}{\end{equation}}
\newcommand{\bea}{\begin{eqnarray}}
\newcommand{\eea}{\end{eqnarray}}
\newcommand{\bsub}{\begin{subequations}}
\newcommand{\esub}{\end{subequations} \noindent}
\newcommand{\clean}{\setcounter{equation}{0}}
\renewcommand{\theequation}{\thesection.\arabic{equation}}

\def\MPLA#1#2#3{Mod. Phys. Lett. {\bf A#1} (19#2) #3}
\def\PRD#1#2#3{Phys. Rev. {\bf D#1} (19#2) #3}

\def\NPB#1#2#3{Nucl. Phys. {\bf B#1} (19#2) #3}
\def\PTP#1#2#3{Prog. Theor. Phys. {\bf #1} (19#2) #3}
\def\ZPC#1#2#3{Z. Phys. {\bf C#1} (19#2) #3}
\def\EPJC#1#2#3{Eur. Phys. J. {\bf C#1} (19#2) #3}
\def\PLB#1#2#3{Phys. Lett. {\bf B#1} (19#2) #3}
\def\PRL#1#2#3{Phys. Rev. Lett. {\bf #1} (19#2) #3}

\def\lep_table{
	\begin{table}[p]%[htbp]
	\begin{center}
	\begin{tabular}{|r|c|c|c|}
	\hline
	 & data  & SM$^*$ & pull$^*$\\ \hline
	\makebox[45mm][l]{{\bf LEP 1}~\cite{lepewwg98}} & & & \\
	\begin{tabular}{l}
	line-shape \& FB asym.: \end{tabular}
	& & & \\
	$m^{}_Z$ (GeV) & 91.1867 $\pm$  0.0021 &------ &----- \\ 
	$\Gamma_Z^{}$ (GeV)  & $2.4939\pm 0.0024$  & 2.4972 & $-1.4$   \\
	$\sigma^0_h$(nb) & $41.491\pm 0.058\hpz$ & 41.474 & $\hph 0.3$ \\ 
	$R_{\ell}$ & $20.765\pm 0.026\hpz$ & 20.747 & $\hph (0.7)$ \\
	$A^{0,\ell}_\fb$ & $0.01683\pm 0.00096$ & 0.01651 & $\hph (0.3)$ \\
	\begin{tabular}{l}
	for each lepton: \end{tabular} &&&\\
	$\left\{ \begin{array}{r}
        ~R_e^{}\; \\ ~R_\mu^{}\; \\ ~R_\tau^{}\; \end{array} \right.$ & 
        $\begin{array}{c} 
        20.783\pm 0.052\hpz \\ 20.789\pm 0.034\hpz \\ 20.764\pm 0.045\hpz 
        \end{array}$ & 
        $\begin{array}{c} 20.747 \\ 20.747 \\ 20.795 \end{array}$ & 
        $\begin{array}{c} \hph 0.7 \\ \hph 1.3 \\ -0.7 \end{array}$ \\ 
	$\left\{ \begin{array}{r}
        A^{0,e}_\fb \\ A^{0,\mu}_\fb \\ A^{0,\tau}_\fb \end{array} \right.$ & 
        $\begin{array}{c} 
        0.0153\pm 0.0025\\0.0164\pm 0.0013\\0.0183\pm 0.0017 \end{array}$ & 
        $\begin{array}{c} 0.0165 \\ 0.0165 \\ 0.0165 \end{array}$ & 
        $\begin{array}{c} -0.5 \\-0.1 \\ \hph 1.1 \end{array}$ \\ 
	\begin{tabular}{l}
	$\tau$ polarization: \end{tabular}
	& & & \\
	$A_{\tau}$& 0.1431 $\pm$ 0.0045 & 0.1484 & $-1.2$ \\
	$A_{e} $  & 0.1479 $\pm$ 0.0051 & 0.1484 & $-0.1$ \\
	\begin{tabular}{l}
	$b$ and $c$ quark results: \end{tabular}
	& & & \\
	$R_b$ & 0.21656 $\pm$ 0.00074 & 0.21566 & $\hph 1.2$ \\
	$R_c$ & 0.1735 $\pm$ 0.0044 & 0.1721 & $\hph 0.3$ \\
	$A^{0,b}_{FB}$ & 0.0990 $\pm$ 0.0021 & 0.1040 & $-2.4$ \\
	$A^{0,c}_{FB}$ & 0.0709 $\pm$ 0.0044 & 0.0744 & $-0.8$ \\
	\begin{tabular}{l}
	jet charge asymmetry: \end{tabular}
	& & & \\
	$\sin^2\theta_{\rm eff}^{\rm lept}$ 
	& 0.2321 $\pm$ 0.0010 & 0.2314 & $\hph 0.7$ \\
	\makebox[45mm][l]{{\bf SLC}~\cite{lepewwg98}} & & & \\
	$A^0_{LR}$ & 0.1510 $\pm$ 0.0025 & 0.1484 & $\hph 1.0$ \\
	$A_b$ & $0.867\pm 0.035$ & $0.935\hpz$ & $-1.9$ \\
	$A_c$ & $0.647\pm 0.040$ & $0.668\hpz$ & $-0.5$ \\
	\makebox[45mm][l]{{\bf Tevatron + LEP 2}~\cite{wboson_moriond}}
			& & & \\
	$m^{}_W$ (GeV) & 80.410 $\pm$ 0.044& 80.402 & $\hph 0.18$ \\ 
	\hline	
	$\chi^2_{\rm tot}$ (19 data points)& & & 19.8 \\ 
	\hline	\hline
	\makebox[45mm][l]{\bf Parameters } & {\bf Constraints}& & \\
	$m^{}_t$ (GeV)~\cite{PDG98} & 173.8 $\pm$ 5.2\hpz\hpz & 175.0 
			& --- \\ 
	$\alps(\mz)$~\cite{PDG98} & 0.119 $\pm$ 0.002 &0.118  & ---\\ 
	$1/\alpha(\mzsq)$~\cite{EJ} & 128.90 $\pm$ 0.09 \hpz&  
			128.90 & --- \\ 
	                \cite{DH} & 128.94 $\pm$ 0.04 \hpz&  
			--- & --- \\ \hline
	\end{tabular} 
	\end{center}
\caption{{\small 
Electroweak measurements at LEP, SLC and Tevatron. 
The average $W$-boson mass is found in ref.~\cite{wboson_moriond}.  
The reference SM predictions and the corresponding 
`pull' factors are given for 
$\mt=175\gev$, $\mh=100\gev$, $\alps(\mz)=0.118$ and 
$1/\alpha(\mzsq)=128.90$. 
Correlation matrix elements of the $Z$ line-shape parameters 
and those for the heavy-quark parameters 
are found in ref.~\cite{lepewwg98}. 
The data $R_\ell^{}$ and $A^{0,\ell}_\fb$ are obtained 
by assuming the $e$-$\mu$-$\tau$ universality, and are not used 
in our $\chi^2$ analysis. }}
\label{tab:ewdata98}
\end{table}
}
\def\cvca_tab{	
	\begin{table}[t]
	\begin{center}
	\begin{tabular}{|l|c|c|}
	\hline
	& ${\it C_{fV}}$ & ${\it C_{fA}}$  \\ \hline
	$u$     & 3.1166 + 0.0030$x_s$ & 3.1351 + 0.0040$x_s$ \\ \hline
	$d = s$ & 3.1166 + 0.0030$x_s$ & 3.0981 + 0.0021$x_s$ \\ \hline
	$c$     & 3.1167 + 0.0030$x_s$ & 3.1343 + 0.0041$x_s$ \\ \hline
	$b$     & 3.1185 + 0.0030$x_s$ & 3.0776 + 0.0030$x_s$ \\ \hline
        $\nu$   & 1 & 1 \\ \hline
	$e=\mu$ & 1 & 1 \\ \hline
	$\tau$  & 1 & 0.9977 \\ \hline
	\end{tabular}
	\end{center}
	\caption{{\small Numerical values of factors $C_{fV}, 
	C_{fA}$ for quarks and leptons used in 
	eq.~(\ref{eq:partial_width}).  }}
	\vsp{0.3}
	\label{tab:cvca}
	\end{table}
	}
\makeatletter
\newtoks\@stequation

\def\subequations{\refstepcounter{equation}%
  \edef\@savedequation{\the\c@equation}%
  \@stequation=\expandafter{\theequation}%   %only want \theequation
  \edef\@savedtheequation{\the\@stequation}% %expanded once
  \edef\oldtheequation{\theequation}%
  \setcounter{equation}{0}%
  \def\theequation{\oldtheequation\alph{equation}}}

\def\endsubequations{%
  \ifnum\c@equation < 2 \@warning{Only \the\c@equation\space subequation
    used in equation \@savedequation}\fi
  \setcounter{equation}{\@savedequation}%
  \@stequation=\expandafter{\@savedtheequation}%
  \edef\theequation{\the\@stequation}%
  \global\@ignoretrue}

\def\eqnarray{\stepcounter{equation}\let\@currentlabel\theequation
\global\@eqnswtrue\m@th
\global\@eqcnt\z@\tabskip\@centering\let\\\@eqncr
$$\halign to\displaywidth\bgroup\@eqnsel\hskip\@centering
     $\displaystyle\tabskip\z@{##}$&\global\@eqcnt\@ne
      \hfil$\;{##}\;$\hfil
     &\global\@eqcnt\tw@ $\displaystyle\tabskip\z@{##}$\hfil
   \tabskip\@centering&\llap{##}\tabskip\z@\cr}

\makeatother

\begin{document}
\thispagestyle{empty}
\vspace*{-15mm}
%----------
\baselineskip 10pt
\begin{flushright}
\begin{tabular}{l}
{\bf KEK-TH-648}\\
{\bf SNS-PH/99-16}\\
{\bf hep-ph/9912260}
\end{tabular}
\end{flushright}
\baselineskip 24pt 
\vglue 10mm 
%%%%%%%%%%%%%%%%%%%%%%%%%%%%%%%%%%%%%%%%%%%%%%
%                Title 
%%%%%%%%%%%%%%%%%%%%%%%%%%%%%%%%%%%%%%%%%%%%%%
\begin{center}
{\Large\bf
Supersymmetry versus precision experiments revisited
}
\vspace{3mm}

\baselineskip 18pt 
\def\thefootnote{\fnsymbol{footnote}}
\setcounter{footnote}{0}
{\bf
Gi-Chol Cho\footnote{Research Fellow of the Japan Society 
for the Promotion of Science}$^{,1),2)}$ and 
Kaoru Hagiwara$^{1)}$}
\vspace{2mm}

$^{1)}${\it Theory Group, KEK, Tsukuba, Ibaraki 305-0801, Japan}\\
$^{2)}${\it Scuola Normale Superiore, Piazza dei Cavalieri 7, 
I-56126 Pisa, Italy}\\
\vspace{4mm}
\end{center}
%%%%%%%%%%%%%%%%%%%%%%%%%%%%%%%%%%%%%%%%
%%%%%                              %%%%%
%%%%%          Abstract            %%%%%
%%%%%                              %%%%%
%%%%%%%%%%%%%%%%%%%%%%%%%%%%%%%%%%%%%%%%
\begin{center}
{\bf Abstract}\\[7mm]
\begin{minipage}{16cm}
\baselineskip 16pt
\noindent
%%%%%----------------------------------
We study constraints on the supersymmetric standard model from the 
updated electroweak precision measurements --- the $Z$-pole 
experiments and the $W$-boson mass measurements. 
The supersymmetric-particle contributions to the universal 
gauge-boson-propagator corrections are parametrized by the three 
oblique parameters $\sz$, $\tz$ and $\mw$.	
The oblique corrections, the $\zqq$ and $\zll$ vertex corrections, 
and the vertex and box corrections to the $\mu$-decay width are 
separately studied in detail.	
We first study individual contribution from the four sectors of the 
model, the squarks, the sleptons, the supersymmetric fermions 
(charginos and neutralinos), and the supersymmetric Higgs bosons, 
to the universal oblique parameters, where the sum of individual 
contributions gives the total correction.	
We find that the light squarks or sleptons, whose masses just above 
the present direct search limits, always make the fit worse than 
that of the Standard Model (SM), whereas the light charginos and 
neutralinos generally make the fit slightly better. 
The contribution from the supersymmetric Higgs sector is found small. 
We then study the vertex/box corrections carefully when both the 
supersymmetric fermions (-inos) and the supersymmetric scalars 
(squarks and sleptons) are light, and find that no significant 
improvement over the SM fit is achieved.  
The best overall fit to the precision measurements are found 
when charginos of mass $\sim 100\gev$ with a dominant wino-component 
are present and the doublet squarks and sleptons are all much heavier. 
The improvement over the SM is marginal, however, where the total 
$\chi^2$ of the fit to the 22 data points decreases by about one 
unit, due mainly to a slightly better fit to the $Z$-boson total width. 
%%%%%----------------------------------  
\end{minipage}
\end{center}
%%%%%------------------------------------------
%%%%% PACS number(s) & Key words
%%%%%------------------------------------------
\baselineskip 18pt 
{\small 
\begin{flushleft}
\end{flushleft}
}
%%%%%----------------------------------  
\newpage
\baselineskip 16pt 
\def\thefootnote{\fnsymbol{footnote}}
\setcounter{footnote}{0}
%%%%%----------------------------------------------------
%%%%%
%%%%%	Section: Introduction
%%%%%
%%%%%----------------------------------------------------
\section{Introduction}
%%%--------------------------------------------
The supersymmetric standard model has been the leading candidate for 
the theory beyond the Standard Model (SM) of elementary particles.  
The supersymmetry (SUSY), the symmetry between bosons and fermions, 
gives us an elegant solution~\cite{susygut81} to the hierarchy 
problem~\cite{hierarchy76,technicolor79}, 
the stability of the electroweak scale ($\sim 100\gev$) against 
the more fundamental scale of physics, the Grand Unified Theory (GUT) 
scale ($\sim 10^{16}\gev$) where the three gauge interactions may be 
unified, or the Planck scale ($\sim 10^{19}\gev$) where the gauge 
interactions may be unified with the gravity interactions.  
It also offers us an attractive scenario that the electroweak symmetry 
breaking may occur radiatively~\cite{susyrge82} if the top quark is 
sufficiently heavy but not too much ($100\gev \simlt \mt \simlt 
200\gev$)~\cite{radsb82}. 
The observation that the minimal supersymmetric standard model 
(MSSM) leads to the unification of the three gauge couplings at 
the `right' scale, $\sim 2\times 10^{16}\gev$, high enough for 
the proton longevity, if the supersymmetric particles exist with 
masses at or below the $1\tev$ scale~\cite{sgut91}, and the discovery 
of the top quark~\cite{top94} in the `right'  mass range, jointly have 
made us take the MSSM as a serious candidate for the theory just 
beyond the present new-particle search front.  
%%%----------------------------------
%%%	a new paragraph

%%%	a new paragraph
%%%----------------------------------
Despite the high expectations, however, various efforts in search of 
an evidence for supersymmetric particles have so far been fruitless.  
There had been occasions when a short-lived experimental `anomaly' 
such as the mismatch of $\alps$ determined from the low-energy 
experiments and that obtained from the $Z$-boson decays and/or the 
significantly large partial width $\Gamma(Z\to b\bar{b})$ as compared 
to the SM prediction, were considered as a possible evidence for 
relatively light supersymmetric particles~\cite{hm90,zbb_susy,yhm95}. 
All such anomalies were short-lived and the precision electroweak 
data~\cite{lepewwg98} from the completed experiments at LEP1 and 
the Tevatron run-I, and the on-going experiments at SLC and LEP2 
are all consistent with the SM predictions.  
Alongside, the direct search limits for the supersymmetric particle 
masses have steadily risen with time. 
The only additional encouragement from the precision electroweak 
experiments is that the preferred range of the SM Higgs boson mass 
is consistent with the stringent upper bound~\cite{susymh91,susymh93,
higgs_twoloop_ren,higgs_twoloop_eff,higgs_twoloop_dia} of the lightest 
Higgs boson mass in the supersymmetric theories.   
The good news is that the supersymmetric theories have not been 
ruled out yet, while the bad news is that studies of precision 
electroweak experiments give us only the lower bounds for the 
supersymmetric particle masses.  
%%%----------------------------------
%%%	a new paragraph

%%%	a new paragraph
%%%----------------------------------
There have recently been numerous attempts~\cite{ewmssm97,chks98} 
to identify constraints on the MSSM parameters from precision 
electroweak measurements. 
In this report, we present the first results of our comprehensive  
study of the constraints on the MSSM parameters from the electroweak 
precision measurements.  
%%%----------------------------------
%%%	a new paragraph

%%%	a new paragraph
%%%----------------------------------
We perform a systematic study of the MSSM parameter space by observing 
that the concept of the universal gauge-boson-propagator corrections, 
or the oblique corrections~\cite{stu90,ht90_gw91,mr90_kl90,ab91}, 
is useful in the MSSM in the sense that they dominate the radiative 
corrections if either the supersymmetric scalars (squarks and 
sleptons) or the supersymmetric fermions (charginos and neutralinos) 
are all sufficiently heavy.  
In this limit, all the precision measurements on the $Z$-boson 
properties can be parametrized by just two oblique parameters, 
$\sz$ and $\tz$, whereas the $W$-boson mass itself makes the 
third oblique parameter.  
Here we closely follow the formalism developed in 
refs.~\cite{hhkm94,chu98,arnps98}, slightly modified to suit our  
MSSM studies.  
In this limit of the dominant oblique corrections, we can study 
in detail the contributions of the four sectors of the MSSM 
separately~\cite{dh90}; 
the squarks, the sleptons, the supersymmetric fermions 
(charginos and neutralinos), and the supersymmetric Higgs sector.  
We find that although the contributions from relatively light 
squarks and/or sleptons generally make the fit to the electroweak 
data worse than the SM fit, those from relatively light charginos 
and neutralinos can slightly improve the fit.  
%%%----------------------------------
%%%	a new paragraph

%%%	a new paragraph
%%%----------------------------------
In the second stage, we examine the case where both the supersymmetric 
scalars and fermions are light, by studying their contributions to 
the muon lifetime, $\ddelg$, and to all the $Z$-boson decay 
amplitudes, $\dgnl$, $\dgel$, $\dger$, $\dgul$, $\dgur$, $\dgdl$, 
$\dgdr$, $\dgntaul$, $\dgtaul$, $\dgtaur$, $\dgbl$, and $\dgbr$.   
Here $\dgfa$ stands for the deviation of the non-oblique correction 
to the $Zf_\alpha f_\alpha$ vertex from the reference SM prediction.    
Within our approximation of the MSSM Lagrangian, the amplitudes for 
the first two generation quarks and leptons are identical, and only 
the above 12 distinctive corrections appear. 
Because we find an indication that the existence of relatively 
light ($\sim 100\gev$) charginos and neutralinos can somewhat 
improve the SM fit to the electroweak data, we study the consequences 
of having light sleptons and/or squarks in addition to the light 
charginos and neutralinos.  
We generally find that the goodness of the fit worsens in such cases. 
Studies of the consequences of more specific models of the 
supersymmetry breaking mechanism, such as the supergravity models and 
the gauge-mediated supersymmetry breaking models, will be presented 
elsewhere~\cite{chk99}. 
%%%----------------------------------
%%%	a new paragraph

%%%	a new paragraph
%%%----------------------------------
The paper is organized as follows. 
In section 2, we present the formalism which we calculate all 
the electroweak observables in the MSSM.  The dependences of those 
observables in terms of the SM parameters, $\mt$, $\mh$, 
$\alps=\alps(\mz)_{\msbar}$, and $\alpha(\mzsq)$ are given 
in a compactly parametrized form~\cite{chu98,arnps98}.  
The MSSM Lagrangian and constraints on its parameters are summarized 
in section 3, and all the MSSM contributions to the one-loop functions 
are presented in the appendices.
In section 4, we study oblique corrections from the squarks and 
sleptons (sec.~4.1), the supersymmetric Higgs bosons (sec.~4.2), 
and the supersymmetric fermions (charginos and neutralinos) (sec.~4.3), 
separately.  
In section 5, we study consequences of vertex and box corrections 
in addition to the gauge-boson-propagator corrections: 
the vertex corrections due to the MSSM Higgs bosons (sec.~5.1), 
those from squarks and gluinos (sec.~5.2), 
those from squarks and charginos/neutralinos (sec.~5.3), 
and the vertex/box corrections from sleptons and charginos/neutralinos 
(sec.~5.4).  
In section 6, we examine quantitative significance of the MSSM 
contributions to the electroweak radiative corrections.  
Section 7 gives conclusions.  
Appendix A summarizes our notation~\cite{mssml99} for the MSSM mixing 
matrices and the couplings among the mass eigenstates.  
Appendix B gives the MSSM contribution to the gauge-boson-propagator 
corrections and the oblique parameters $S,T,U$ and $R$. 
Appendix C gives the MSSM contribution to the $Z$-boson decay 
amplitudes. 
Appendix D gives the MSSM contribution to the muon-decay amplitude. 
%%%%%----------------------------------------------------
%%%%%
%%%%%	Section: Electroweak observables in the MSSM
%%%%%
%%%%%----------------------------------------------------
\clean
\section{Electroweak observables in the MSSM}
%-------------------
\lep_table
%------------------- 
In this section, we give all the theoretical predictions for 
the electroweak observables which are used in our analysis. 
The experimental data of the $Z$-pole experiments and the 
$W$-boson mass measurement are summarized in Table~\ref{tab:ewdata98}. 
Because the MSSM predicts slightly different $Z$-decay amplitudes into  
the $\tau$-leptons as compared to those into $e$ or $\mu$, we will 
use in the following analysis the data set which do not assume the 
$e$-$\mu$-$\tau$ universality.  By removing the two entries, 
$R_\ell$ and $A_\fb^{0,\ell}$ that assume the lepton universality, 
there are 19 data points in the Table.  
The total $\chi^2$ is obtained by using the correlation matrices 
of ref.~\cite{lepewwg98}. 
%%%----------------------------------
%%%	a new paragraph

%%%	a new paragraph
%%%----------------------------------
The reference SM predictions are given in the table for an arbitrarily 
chosen set of the four inputs, $\mt=175\gev$, $\mh=100\gev$, 
$1/\alpha(\mzsq)=128.90$\footnote{
$\alpha(\mzsq)$ is the running QED coupling constant at $q^2=\mzsq$ 
where only the quark and lepton contributions to the deviation 
from its $q^2=0$ value, $\alpha(0) = 1/137.036$, are included. 
Its magnitude is commonly referred to in the 
literatures~\cite{PDG98,EJ,DH}. 
It is denoted as $\alpha(\mzsq)_{\rm f}$ in refs.~\cite{hhkm94,arnps98}, 
and is related to the coupling $\abar(\mzsq)$ that contains the 
$W$-boson contribution by 
$1/\alpha(\mzsq) = 1/\abar(\mzsq) + 0.15$~\cite{hhkm94,arnps98}. 
} 
and $\alps=\alps(\mz)_{\msbar}=0.118$.  
In the following, we present simple parametrizations of the SM 
predictions for the parameter sets around the above reference point 
in terms of the normalized variables; 
%%%------------------------------ 
\bsub\label{xtxhxaxs}
\bea
\xt &=& \frac{\mt({\rm GeV})-175}{10}, \\
\xh &=& \ln{\frac{\mh({\rm GeV})}{100}}, \\
\xa &=& \frac{1/\alpha(\mzsq) -128.90}{0.09},  
%    =  \frac{1/\bar{\alpha}(\mzsq) -128.75}{0.09} 
\label{eq:xa}
\\
\xs &=& \frac{\alps(\mz)_{\msbar} -0.118}{0.003}. 
\label{eq:param_xs}
\eea
\label{eq:param_sm}
\esub
%%%------------------------------ 
The SM predictions for an arbitrary set of the four input parameters 
($\mt$, $\mh$, $\alpha(\mzsq)$, $\alps(\mz)$) are then obtained 
easily for all the electroweak observables. 
The predictions of the MSSM are then expressed as the sum of the 
SM predictions and the difference between the predictions of 
the SM and those of the MSSM. 
This separation is useful because the SM predictions include 
parts of the known two- and three-loop corrections whereas 
the MSSM contributions are evaluated strictly in the one-loop 
order in the present analysis. 
%%%%%%%%%%%%%%%%%%%%%%%%%%%%%%%%%%%%%%%%%%%%%
%%%%%
%%%%%
%%%%%	Subsection: Observables at $Z$-pole experiments
%%%%%
%%%%%
%%%%%%%%%%%%%%%%%%%%%%%%%%%%%%%%%%%%%%%%%%%%%
\subsection{Observables at $Z$-pole experiments}
%-------------------
The amplitude for the decay process $Z \to f_\alpha \ov{f_\alpha}$ 
is written as
%%%------------
\bea
T (Z \to f_\alpha \ov{f_\alpha}) 
	&=& M_\alpha^f \epsilon_Z^{} \cdot J_{f_\alpha}, 
\eea
%%%------------
where $\epsilon_Z^\mu$ is the polarization vector of the $Z$-boson 
and $J_{f_\alpha}^\mu = \ov{f_\alpha} \gamma^\mu f_\alpha
= \ov{f} \gamma^\mu P_\alpha f$ is the fermion current 
with definite chirality, $\alpha = L$ or $R$. 
The pseudo-observables of the $Z$-pole experiments are expressed 
in terms of the scalar amplitudes $M_\alpha^f$ with the 
following normalization~\cite{lepewwg98}: 
%%%------------
\bea
g_\alpha^f &=& \frac{M_\alpha^f}{\sqrt{4\rttwo \gf \mzsq}} 
	\approx \frac{M_\alpha^f}{0.74070}. 
\eea
%%%------------
A convenient parametrization of the effective couplings in generic 
electroweak theories has been given in refs.~\cite{chu98,arnps98}: 
%%%------------ 
\bsub
\bea
g^{\nu_l}_L  
	&=& \makebox[3.3mm]{ } 0.50214 + 0.453\, \dgzbarsq 
	\hphantom{+ 1.001\,{\Del \bar{s}^2}\,\,\,}
	+ \dgnll \label{eq:amp_nue},  \\
g^l_L &=& - 0.26941 - 0.244\, \dgzbarsq 
	+ 1.001\, \dsbarsq + \dgll,  \label{eq:amp_el}\\
g^l_R &=& \makebox[3.3mm]{ } 0.23201 + 0.208\,
	\dgzbarsq + 1.001\, \dsbarsq 
	+ \dglr, \label{eq:amp_er}\\
g^u_L &=&  \makebox[3.3mm]{ } 0.34694 + 0.314\,
	\dgzbarsq - 0.668\, \dsbarsq 
	+ \dgul, \label{eq:amp_ul}\\
g^u_R &=& - 0.15466 - 0.139\, \dgzbarsq 
	- 0.668 \, \dsbarsq + \dgur, \label{eq:amp_ur}\\
g^d_L &=& - 0.42451 - 0.383\, \dgzbarsq 
	+ 0.334\, \dsbarsq + \dgdl, \label{eq:amp_dl}\\
g^d_R &=& \makebox[3.3mm]{ }  0.07732
	+ 0.069\, \dgzbarsq + 0.334\, \dsbarsq 
	+ \dgdr, \label{eq:amp_dr}\\
g^b_L &=& - 0.42109 - 0.383\, \dgzbarsq 
	+ 0.334\, \dsbarsq + \dgbl. 
\label{eq:amp_bl}
\eea
\label{eq:amp_sm}
\esub
%%%----------------- 
Here the first numerical terms in the r.h.s.\  of the above 
equations are the SM predictions at the reference point, 
$(\mt({\rm GeV}),\mh({\rm GeV}),1/\alpha(\mzsq),\alps(\mz))$ 
= (175, 100, 128.90, 0.118), 
$\dgzbarsq$ and $\dsbarsq$ are the universal gauge-boson-propagator 
corrections~\cite{hhkm94}, and $\Del g_\alpha^f$ $(\alpha=L,R)$ 
are the process specific corrections. 
In the SM the following universality relations hold very accurately: 
%%%----------------- 
\bsub
\bea
(\Del g_L^{\nu_e})_{\smr} 
	&=& (\Del g_L^{\nu_\mu})_{\smr} 
	  = (\Del g_L^{\nu_\tau})_{\smr} = 0, 
\\
(\Del g_\alpha^{e})_{\smr} 
	&=& (\Del g_\alpha^{\mu})_{\smr} 
	  = (\Del g_\alpha^{\tau})_{\smr} = 0, 
\\
(\Del g_\alpha^{u})_{\smr} 
	&=& (\Del g_\alpha^{c})_{\smr} = 0, 
\\
(\Del g_R^{d})_{\smr} 
	&=& (\Del g_R^{s})_{\smr} 
	  = (\Del g_R^{b})_{\smr} = 0, 
\\
(\Del g_L^{d})_{\smr} 
	&=& (\Del g_L^{s})_{\smr} = 0 
	  \neq (\Del g_L^{b})_{\smr}. 
\eea
\label{eq:dgfa}
\esub
%%%----------------- 
Only the term $(\dgbl)_{\smr}$ has non-trivial $\mt$ and 
$\mh$ dependence. 
In the MSSM, we find that all the $(\dgfa)$ terms 
are non-vanishing, and the flavor-universality holds only 
among the first two generations. 
We study $\dgntaul, \dgtaul, \dgtaur, \dgbl$ and $\dgbr$ 
separately in the following sections. 
%%%------------

%%%------------
The universal part of the corrections $\dgzbarsq$ and $\dsbarsq$ 
are defined as the shift in the effective couplings 
$\gzbarsq(\mzsq)$ and $\sbarsq(\mzsq)$~\cite{hhkm94}  
from their SM reference values at 
$(\mt({\rm GeV}),\mh({\rm GeV}),1/\alpha(\mzsq))$ = (175, 100, 128.90): 
%%%------------
\bsub
\bea
\gzbarsq(\mzsq) &=& 0.55635 + \dgzbarsq,\\
\sbarsq(\mzsq) &=& 0.23035 + \dsbarsq.  
\eea
\esub
%---------------------- 
We find it convenient to express the above shifts in the two 
effective couplings in terms of the two parameters $\dsz$ and $\dtz$, 
%---------------------- 
\bsub
\bea
\dgzbarsq &=& 0.00412 \dtz, \\
\dsbarsq &=& 0.00360 \dsz - 0.00241 \dtz. 
\eea
\label{gzb_sb}
\esub
%---------------------- 
The parameters $S_Z$ and $T_Z$ are related to the $S$ and 
$T$ parameters~\cite{stu90}
%----------------------  
\bsub
\bea
S_Z &\equiv& S + R - 0.064 \xa, 
\\
T_Z &\equiv& T + 1.49 R - \frac{\ddelg}{\alpha}, 
\eea
\label{eq:sz_tz}
\esub
%%%------------
in the notation of refs.~\cite{hhkm94,chu98,arnps98}.  
A compact summary of the definitions of the effective charges of 
ref.~\cite{hhkm94} and the oblique parameters $S,T,U,R$ are given 
in Appendix B. The shift $R$  
%%%-------------------------
\bea
\frac{4\pi}{\gzbarsq(\mzsq)} - \frac{4\pi}{\gzbarsq(0)} 
	\equiv -\quarter R = -\quarter (1.1879+\dr), 
\label{eq:running_of_gzbar}
\eea
%%%-------------------------
accounts for the difference between the $T$ parameter which 
measures the neutral current strength at the zero-momentum 
transfer~\cite{stu90,veltman} and the $T_Z$ parameter which 
measures the quantity on the $Z$-pole. 
The $\xa$-dependent term in $S_Z$ reflects the fact that 
only a combination of $S$ and $1/\alpha(\mzsq)$ is constrained 
by the $Z$-pole asymmetry experiments~\cite{hhkm94,arnps98,swartz95}. 
The factor $\delg$ denotes the vertex and box corrections to the 
muon decay constant 
%%%------------------------- 
\bea
\gf = \frac{\gwbarsq(0) + \ghatsq \delg}{4\sqrt{2}\mwsq}, 
\label{eq:gf}
\eea
%%%------------------------- 
which takes $\delg = 0.0055$~\cite{hhkm94} in the SM. 
In the MSSM, vertex and box corrections affect its magnitude 
and we define the shift $\ddelg$ as 
%%%------------------------- 
\bea
\delg &=& 0.00550 + \ddelg. 
\eea
%%%------------------------- 
Although the supersymmetric contribution to $\ddelg$ is the 
correction specific to the muon decay rather than a part of 
the universal gauge-boson-propagator corrections, 
it appears in all the predictions of the electroweak 
observables because we use the muon decay constant $\gf$ 
as one of the basic inputs of our calculations. 
We therefore include the effect $\ddelg$ as a part of 
our ``universal'' parameter, $T_Z$. 
%%%----------------------------------
%%%	a new paragraph

%%%	a new paragraph
%%%----------------------------------
By inserting (\ref{eq:running_of_gzbar}) into (\ref{eq:sz_tz}) 
and by using the shifts $\ds$ and $\dt$~\cite{hhkm94} 
from the reference SM values of the $S$ and $T$ parameters, 
we can express the parameters $\dsz$ and $\dtz$ as 
%%%----------------- 
\bsub
\bea
\dsz &=& \sz - 0.955 = \ds + \dr - 0.064 \xa, 
\label{eq:dsz}
\\
\dtz &=& \tz - 2.65\hphantom{5} 
	= \dt + 1.49 \dr - \frac{\ddelg}{\alpha}. 
\label{eq:dtz}
\eea
\label{eq:sztzdr}
\esub
%%%----------------- 
The SM contribution to the shift $\ds,\dt$ and $\dr$ are 
parametrized as~\cite{hhkm94,chu98}
%%%----------------- 
\bsub
\bea
(\ds)_{\rm SM} &=& -0.007 \xt +0.091 \xh -0.010 x^2_h , \\
(\dt)_{\rm SM} &=& (0.130 - 0.003 \xh) \xt +0.003 x_t^2
	- 0.079 \xh - 0.028 x^2_h \nonumber \\ 
	& & +0.0026 x^3_h, \\
(\dr)_{\rm SM} &=& -0.124\biggl\{ 
	\log\biggl[ 1+\biggl( \frac{26}{\mh({\rm GeV})} \biggr)^2 \biggr] 
	- \log\biggl[ 1+\biggl( \frac{26}{100} \biggr)^2 \biggr] 
	\biggr\}, 
\eea
\esub
%--------------
and we set $(\ddelg)_{\smr}=0$ in our calculation. 
The explicit form of the MSSM contributions to the parameters 
$\ds,\dt,\dr$ are given in Appendix~\ref{section:two_point}. 
The MSSM contributions to the muon-decay parameter 
$\ddelg$ is given in Appendix~\ref{section:muon_decay}, 
and those to the effective $Z$-boson decay amplitudes, 
$\Del g^f_\alpha$, are given in Appendix~\ref{section:zdecay_amplitudes}. 
%%%------------------------

%%%------------------------
The pseudo-observables of the $Z$-pole experiments are then 
obtained by using the above 12 effective couplings $g_\alpha^f$; 
8 couplings of eq.~(\ref{eq:amp_sm}) that are distinct in the SM, 
and the 4 additional coupling, $g^{\nu_\tau}_L$, $g^{\tau}_L$, 
$g^{\tau}_R$ and $g^{b}_R$, which can be distinct in the MSSM. 
The partial widths of the $Z$-boson are 
%---------------  
\bea
\Gamma_f &=& \frac{\gf m_Z^{3}}{3\rttwo \pi} \left\{ 
	\left| g^f_L + g^f_R \right|^2\frac{C_{fV}}{2} 
	+ \left| g^f_L - g^f_R \right|^2 \frac{C_{fA}}{2}  \right\}
	\left( 1+\frac{3}{4}Q^2_f\frac{\alpha(\mzsq)}{\pi}\right)
        \makebox[10mm][l]{,}
\label{eq:partial_width}
\eea
%---------------  
where the factors $C_{fV}$ and $C_{fA}$ account for the final 
state mass and QCD corrections for quarks. 
Their numerical values are listed in Table~\ref{tab:cvca}. 
The $\alps$-dependence in $C_{qV}, C_{qA}$ is parametrized in 
terms of the parameter $\xs$~(\ref{eq:param_xs}). 
The last term proportional to $\alpha(\mzsq)/\pi$ in 
eq.~(\ref{eq:partial_width}) accounts for the final state QED 
correction. 
%---------------
\cvca_tab
%---------------
The total decay width $\Gamma_Z$ and the hadronic decay width 
$\Gamma_h$ are given in terms of the partial width $\Gamma_f$: 
%---------------  
\bsub
\begin{eqnarray}
\Gamma_Z &=& 3\Gamma_{\nu} + \Gamma_e 
	+ \Gamma_{\mu} + \Gamma_{\tau} + \Gamma_h, 
\label{eq:total_width}\\
\Gamma_h &=& \Gamma_u + \Gamma_c + \Gamma_d + \Gamma_s + \Gamma_b. 
\label{eq:hadron_width}
\end{eqnarray}
\esub
%---------------  
The ratios $R_l, R_c^{}, R_b^{}$ and the hadronic peak 
cross section $\sigma_h^0$ are given by: 
%------
\begin{equation}
R_l      = \frac{\Gamma_h}{\Gamma_l},\;
R_c      = \frac{\Gamma_c}{\Gamma_h},\;
R_b      = \frac{\Gamma_b}{\Gamma_h},\;
\sigma^0_h = \frac{12\pi}{\mzsq}
	\frac{\Gamma_e\Gamma_h}{\Gamma_Z^2}, 
\end{equation}
%%%-------------------
where $l=e,\mu$ or $\tau$. 
%%%----------------------------------
%%%	a new paragraph

%%%	a new paragraph
%%%----------------------------------
The left-right asymmetry parameter $A^f$ is also 
given in terms of the effective couplings $g_\alpha^f$ as  
%------ 
\bea
A^f 	= \frac{(g^{f}_L)^2-(g^{f}_R)^2}{(g^{f}_L)^2+(g^{f}_R)^2}. 
\eea
%------  
The forward-backward (FB) asymmetry $A^{0,f}_{FB}$ and the 
left-right (LR) asymmetry $A^{0}_{LR}$ are then expressed as follows: 
%------ 
\bsub
\bea
A^{0,f}_{FB} &=& \frac{3}{4}A^{e}A^{f}, \\
A^{0}_{LR} &=&  A^{e}. 
\eea
\esub
%%%------------ 
The effective parameter $\sin^2\theta_{\rm eff}^{\rm lept}$ 
measured from the jet-charge FB asymmetry is defined as 
%%%------------ 
\bea
\sin^2\theta_{\rm eff}^{\rm lept} &=& \half \frac{g^e_R}{g_R^e - g_L^e}. 
\eea
%%%----------------------------------
%%%	a new paragraph

%%%	a new paragraph
%%%----------------------------------
All the $Z$-boson parameters in Table~\ref{tab:ewdata98} are 
now calculable for arbitrary values of $\mt,\mh,\alpha(\mzsq)$ 
and $\alps(\mz)$, or $\xt,\xh,\xa$ and $\xs$, respectively, 
in the SM by using the parametrizations given in this section. 
The predictions of the MSSM are calculated by using the formulae 
in Appendices~\ref{section:two_point}, \ref{section:zdecay_amplitudes} 
and \ref{section:muon_decay}, by using the mixing matrix and the 
coupling notation~\cite{mssml99} of Appendix A.   
%%%%%%%%%%%%%%%%%%%%%%%%%%%%%%%%%%%%%%%%%%%%% 
%%%%%
%%%%%
%%%%%	Subsection: $W$-boson mass
%%%%%
%%%%%
%%%%%%%%%%%%%%%%%%%%%%%%%%%%%%%%%%%%%%%%%%%%%
\subsection{The $W$-boson mass}
%%%------------ 
The theoretical prediction of $\mw$ can be parametrized 
as~\cite{hhkm94,chu98} 
%-------------------------
\bsub
\bea
\mw({\rm GeV})  &=& 80.402 + \dmw, \\
\label{eq:mw}
\dmw({\rm GeV}) &=& - 0.288 \ds + 0.418 \dt +0.337 \du 
\nonumber \\
	&&~~~~~~~ + 0.012 \xa - 0.126 \frac{\ddelg}{\alpha}. 
\label{eq:dmw}
\eea
\esub
%------------------------- 
Here in addition to the $S$ and $T$ parameters, the 
$U$ parameter~\cite{stu90} is needed to calculate the effective 
charge $\gwbarsq(0)$~\cite{hhkm94} that determines the 
muon decay constant $\gf$; see eq.~(\ref{eq:gf}). 
The SM contribution to the shift $\dmw$ from the reference 
prediction can be parametrized as~\cite{chu98,arnps98}
%%%-------------
\bea
(\dmw)_{\smr} &=& 0.064 \xt - 0.060 \xh - 0.009 x_h^2  
	+ 0.001 \xt (\xt -\xh) \nonumber \\ 
	&& + 0.001 x_h^3  + 0.012 \xa, 
\eea
%%%-------------
which is obtained by inserting the SM contributions to 
$\ds,\dt$ and $\du$~\cite{chu98,arnps98} to eq.~(\ref{eq:mw}). 
The MSSM gives additional contributions to $S,T,U$ and 
to $\ddelg$: 
%%%-------------
\bea
\dmw &=& (\dmw)_{\smr} - 0.288 S_{\rm new} + 0.418 T_{\rm new} 
	+ 0.337 U_{\rm new} - 0.126 \frac{\ddelg}{\alpha}, 
\eea
%%%-------------
where 
%%%-------------
\bea
S_{\rm new} = \ds - (\ds)_{\rm SM}, ~~
T_{\rm new} = \dt - (\dt)_{\rm SM}, ~~
U_{\rm new} = \du - (\du)_{\rm SM}. 
\eea
%%%-------------
The MSSM contribution to $S_{\rm new}$, $T_{\rm new}$, $U_{\rm new}$ 
and $\ddelg$ are given in appendices~\ref{section:two_point} 
and \ref{section:muon_decay}. 
%%%%%%%%%%%%%%%%%%%%%%%%%%%%%%%%%%%%%%%%%%%%%
%%%%%
%%%%%
%%%%%	Subsection: Constraints on the SM parameters
%%%%%
%%%%%
%%%%%%%%%%%%%%%%%%%%%%%%%%%%%%%%%%%%%%%%%%%%%
\subsection{Constraints on $\mt,\alpha(\mzsq)$ and $\alps(\mz)$, 
	and the SM fit}
%%%-----------------
All the electroweak observables are now calculated in the SM 
as functions of the four parameters, $\mt, \mh, \alpha(\mzsq)$ 
and $\alps(\mz)$, or $\xt,\xh,\xa$ and $\xs$, respectively 
via~(\ref{eq:param_sm}). 
The Higgs boson has not been found yet and the lower mass 
bound 
%%%-----------------
\bea
\mh({\rm GeV}) \simgt 90 \hsp{1}(\xh \simgt -0.11), 
\label{eq:smhiggs}
\eea
%%%----------------- 
is obtained from the LEP2 experiment~\cite{higgs_moriond}. 
%%%----------------------------------
%%%	a new paragraph

%%%	a new paragraph
%%%----------------------------------
In our analysis we use the following constraints on the 
parameters $\mt~$\cite{PDG98}, $\alps(\mz)$~\cite{PDG98} 
and $\alpha(\mzsq)$~\cite{EJ} 
%%%----------------- 
\bsub
\bea
\mt({\rm GeV}) &=& 173.8 \pm 5.2 \hsp{0.65} (\xt= -0.12 \pm 0.52),  
\label{eq:mt_pdg}
\\
1/\alpha(\mzsq) &=& 128.90 \pm 0.09 \hsp{0.3}  (\xa= 0 \pm 1),  
\label{eq:ej_alpha}
 \\
\alps(\mz) &=& 0.119 \pm 0.002 \hsp{0.3} (\xs= 0.33 \pm 0.67),
\label{eq:alpha_s_pdg}
\eea
\label{eq:SMparams}
\esub
%%%----------------- 
as shown in the bottom of Table~\ref{tab:ewdata98}. 
As an alternative to the model-independent 
estimate~(\ref{eq:ej_alpha})~\cite{EJ}, 
we also examine the case with the estimate~\cite{DH}:  
%%%----------------- 
\bea
1/\alpha(\mzsq) &=& 128.94 \pm 0.04 \hsp{0.3} (\xa= 0.44 \pm 0.44), 
\label{eq:dh_alpha}
\eea
%%%----------------- 
which is obtained by using the perturbative QCD constraints 
down to the $\tau$-lepton mass scale. 
%%%----------------------------------
%%%	a new paragraph

%%%	a new paragraph
%%%----------------------------------
We find that the reference point of our analysis ($\xt=\xh=\xa=\xs=0$) 
is not far from the global minimum of the SM fit.  
With the external constraints of eq.~(\ref{eq:SMparams}), 
we find $\chi^2_{\rm min}/(\dof)=18.2/(22-4)$ at 
%%%----------------- 
\bsub
\bea
\mt({\rm GeV}) &=& 172.3 \pm 5.0, \\
\alps(\mz) &=& 0.119 \pm 0.002, \\
1/\alpha(\mzsq) &=& 128.90 \pm 0.09, \\
\mh({\rm GeV}) &=& 117^{+98}_{-64}.  
\label{eq:higgs_ej}
\eea
\label{eq:sm_fit_ej}
\esub
%%%----------------- 
By replacing the estimate~(\ref{eq:ej_alpha})~\cite{EJ} 
by (\ref{eq:dh_alpha})~\cite{DH}, we find 
$\chi^2_{\rm min}/(\dof)=18.3/(22-4)$ at  
%%%----------------- 
\bsub
\bea
\mt({\rm GeV}) &=& 172.6 \pm 4.9, \\
\alps(\mz) &=& 0.119 \pm 0.002, \\
1/\alpha(\mzsq) &=& 128.94 \pm 0.04, \\
\mh({\rm GeV}) &=& 144^{+88}_{-58}. 
\label{eq:higgs_dh}
\eea
\label{eq:sm_fit_dh}
\esub
%%%----------------- 
In summary, the SM gives a good fit to all the electroweak data if 
the Higgs boson mass is relatively light, with the mass below 
a few hundred GeV, as suggested by the ranges (\ref{eq:higgs_ej}) 
or (\ref{eq:higgs_dh}). 
%%%%%--------------------------------------------------------------
%%%%%
%%%%%
%%%%%	Section: The minimal supersymmetric standard model (MSSM)
%%%%%
%%%%%
%%%%%--------------------------------------------------------------
\clean
\section{The minimal supersymmetric standard model (MSSM)}
%%%-----------------
In subsection 3.1 we briefly summarize the MSSM and our coupling 
conventions. 
In subsection 3.2, we summarize the constraints on supersymmetric 
particle masses from direct search experiments. 
%%%%%%%%%%%%%%%%%%%%%%%%%%%%%%%%%%%%%%%%%%%%%
%%%%%
%%%%% subsection: The MSSM
%%%%%
%%%%%%%%%%%%%%%%%%%%%%%%%%%%%%%%%%%%%%%%%%%%%
\subsection{The MSSM Lagrangian}
\label{section:mssm_Lagrangian}
%%%----------------- 
We study consequences of the MSSM under the following constraints.  
%%%----------------- 
\begin{enumerate}
\item 
It has the minimal particle content, with the gauginos of the 
SU(3)$_C$, SU(2)$_L$ and U(1)$_Y$ groups, three generations of 
squarks and sleptons, and two pairs of Higgs doublets and their 
superpartners.  
\item 
It has the minimal superpotential, which is sufficient to give 
masses to all the quarks and leptons.  In particular, we do not consider 
$R$-parity non-conserving interactions.  
\item 
The three gaugino masses, $M_3, M_2, M_1$ are taken to be independent, 
while most of our numerical results are obtained under the `unification' 
condition, $M_3/M_2/M_1 = 
\hat{\alpha}_3(\mz)/\hat{\alpha}_2(\mz)/\hat{\alpha}_1(\mz)$, 
where $\hat{\alpha}_i(\mu)$ are the $\msbar$ couplings.
\item
The scalar masses are introduced in such a way that squark 
and slepton interactions with neutralinos and gluinos are flavor 
conserving in the basis where quarks and leptons have definite mass.  
\item 
The scalar masses of the first two generations are taken to be equal.  
We therefore have 5 scalar masses for the first two generations, 
and another 5 for the third generation squarks and sleptons.  
\item   
We neglect mixings between the left- and right-chirality sfermions 
in the first two generations.  Accordingly, we retain the soft SUSY 
breaking $A$ parameters only in the third generations, 
$A_t, A_b$ and $A_\tau$.  
\end{enumerate}
%%%-----------------
Under the above constraints, the MSSM interactions can be parameterized 
in terms of 19 parameters:  
the ratio of the two v.e.v.'s $\tan\beta$, the Higgs-mixing mass $\mu$, 
the pseudo-scalar Higgs-boson mass $\mpseudo$, the three gaugino 
masses $M_3, M_2, M_1$, the three $A$ parameters $A_t, A_b, A_\tau$, 
the five sfermion masses for the first two generations, 
$m_{\wt{Q}}$, $m_{\wt{U}}$, $m_{\wt{D}}$, $m_{\wt{L}}$, $m_{\wt{E}}$, 
and the five sfermion for the third generations, 
$m_{\wt{Q}_3}, m_{\wt{U}_3}, m_{\wt{D}_3}, m_{\wt{L}_3}, m_{\wt{E}_3}$.  
Summing up, the 19 parameters of our MSSM Lagrangian are 
%%%--------------------
\bsub\label{eq:mssm_para}
\bea
&& \label{eq:mssm_para1} \tan\beta, \mu, \mpseudo,  \\
&&\label{eq:mssm_para2} M_3, M_2, M_1,  \\
&&\label{eq:mssm_para3} A_t, A_b, A_\tau,  \\
&&\label{eq:mssm_para4} 
m_{\wt{Q}}, m_{\wt{U}}, m_{\wt{D}}, m_{\wt{L}}, m_{\wt{E}}, \\ 
&&\label{eq:mssm_para5} 
m_{\wt{Q}_3}, m_{\wt{U}_3}, m_{\wt{D}_3}, m_{\wt{L}_3}, 
m_{\wt{E}_3}.
\eea
\esub
%%%--------------------
All our analytic expressions are valid when the above 19 parameters of 
the MSSM are independently varied.  
Because the parameter space of the MSSM is too large even with the 
above restrictions, we present our numerical results often by varying  
only the few most relevant parameters while keeping the rest of the 
parameters fixed at some appropriate values.  
Systematic investigation of the parameter space of the two representative 
models, the supergravity mediated and the gauge-interaction mediated 
supersymmetry breaking models, will be reported elsewhere \cite{chk99}.  
%%%----------------------------------
%%%	a new paragraph

%%%	a new paragraph
%%%----------------------------------
The MSSM Lagrangian with the above constraints are adopted for the 
new Feynman amplitude generator MadGraph2~\cite{madgraph2}, and its 
explicit form can be found in ref.~\cite{mssml99}. 
A compact summary of our notation is given in Appendix A.  
All the physical masses and couplings of the supersymmetric particles 
are calculated in the tree-level by using the MSSM Lagrangian with 
the above restrictions. 
%%%----------------------------------
%%%	a new paragraph

%%%	a new paragraph
%%%----------------------------------
We denote the two chargino mass eigenstates as, $\chargino{i}$ with 
$m_{\chargino{1}} < m_{\chargino{2}}$, the four neutralinos as, 
$\neutralino{i}$, with $m_{\neutralino{1}} < \cdots < m_{\neutralino{4}}$, 
and the gluino as $\gluino$.  
The 7 sfermion mass eigenstates for each generation are denoted by, 
$\sup_L, \sdown_L, \sup_R, \sdown_R, \sneutrino_l, \wt{l}_L, \wt{l}_R$ 
for the first two generations, and $\stop_1, \stop_2, \sbottom_1, 
\sbottom_2, \sneutrino_\tau, \stau_1, \stau_2$ for the third generation.  
The four Higgs boson mass eigenstates, the light and heavy CP-even neutral 
Higgs bosons $h$ and $H$ ($\mhlight < \mhheavy$), the CP-odd neutral 
Higgs boson $A$, and the charged Higgs boson $H^\pm$ are obtained by using 
the improved effective potential of ref.~\cite{susymh91} 
that assumes CP invariance\footnote{
The radiative corrections beyond the 1-loop level on the lightest 
Higgs boson mass have been discussed in refs.~\cite{higgs_twoloop_ren,
higgs_twoloop_eff,higgs_twoloop_dia}. 
Those effects shift the theoretical prediction on $\mhlight$ 
in the leading order a few ${\rm GeV}$. 
Since the result of our analysis, however, is not affected 
quantitatively by such a small shift of $\mhlight$, we take into 
account the leading order correction on $\mhlight$ in our study 
for brevity. }.  
%%%---------------------------------- 
%%%	a new paragraph

%%%	a new paragraph
%%%----------------------------------
A few comments are in order.  
Although we do not consider models with non-minimal interactions, 
such as the $R$-parity violating models or models with gauge 
interaction mediated supersymmetry breaking where the very 
light gravitino becomes the lightest supersymmetric particle (LSP), 
all our results should be valid in those models with additional 
interactions, as long as their strengths are negligibly small 
as compared to the gauge interactions. 
Our analytic expressions allow for arbitrary CP violating phases 
in the mass parameters, $\mu$ and $A_f$, and hence some of 
the couplings have associated complex phases.  
Our numerical results are, however, given in the CP conserving limit 
of the MSSM interactions.  
This is mainly because we adopt the effective Higgs potential of 
ref.~\cite{susymh91} that assumes CP invariance in the scalar sector.  
Effects of CP violating interactions in the MSSM will be reported 
elsewhere. 
%%%%%%%%%%%%%%%%%%%%%%%%%%%%%%%%%%%%%%%%%%%%% 
%%%%%
%%%%% subsection: Bounds on SUSY particle masses
%%%%%
%%%%%%%%%%%%%%%%%%%%%%%%%%%%%%%%%%%%%%%%%%%%%
\subsection{Bounds on the masses of SUSY particles from 
	the direct search}
%--------------------------------------
The predictions of the MSSM for all the electroweak observables 
reduce to those of the SM with the restricted Higgs boson mass range, 
$\mhlight \simlt 135\gev$~\cite{susymh91,higgs_twoloop_ren,
higgs_twoloop_eff,higgs_twoloop_dia}, in the limit where all the 
supersymmetry breaking mass parameters are large.  
In the following analysis, we explicitly demonstrate this decoupling 
behavior quantitatively for all the electroweak observables.  
%%%----------------------------------
%%%	a new paragraph

%%%	a new paragraph
%%%----------------------------------
The differences between the predictions of the SM and those of the 
MSSM are hence largest when the supersymmetric particle masses 
are near their present lower bounds from the direct search experiments.  
We calculate the particle masses from the 19 parameters of 
the MSSM, eq.~(\ref{eq:mssm_para}), and confront them with those 
lower mass bounds.  
%%%----------------------------------
%%%	a new paragraph

%%%	a new paragraph
%%%----------------------------------
Current limits on the masses of scalar leptons are obtained from 
the LEP2 experiments as~\cite{susy_moriond} 
%%%--------------------
\bsub
\bea
m_{\selectron_R} &\simgt& 88\gev, \\
m_{\smuon_R} &\simgt& 80\gev, \\ 
m_{\stau_R} &\simgt& 69\gev, 
\eea
\esub
%%%--------------------
where $\stau_R$ has been assumed to be the mass eigenstate.  
The corresponding limits for the chirality-left sleptons are weaker 
and depend on the mass of the charginos.  
The lower mass bounds for scalar quarks and gluinos are obtained 
from the Tevatron search experiments as~\cite{squark_gluino_bounds}
%%%--------------------
\bsub
\bea
m_{\wt{q}} &\simgt& 212 \gev, \\
m_{\gluino} &\simgt& 173 \gev, 
\eea
\label{eq:tevatron_bound}
\esub
%%%--------------------
when the squark masses are common for the 5 light flavors and both 
chiralities, and when either the squarks or the gluino are much heavier.  
The bounds on squark masses depend, however, on details of their 
mass spectrum and on their decay patterns.  
The (almost) model-independent lower mass bounds are found from 
the LEP2 experiments~\cite{susy_moriond}:
%%%-------------------- 
\bsub
\bea
m_{\stop_1} &\simgt& 88\gev, \\
m_{\sbottom_1} &\simgt& 76\gev.
\eea
\label{eq:stop_bounds_lep2}
\esub
%%%--------------------
For charginos and neutralinos the following bounds are found from 
the LEP2 experiments~\cite{susy_moriond}: 
%%%--------------------
\bsub
\bea
m_{\neutralino{1}} &\simgt& 33\gev, \\
m_{\chargino{1}} &\simgt& 90\gev. 
\eea
\label{eq:inos_bounds_lep2}
\esub
%%%--------------------
Those on the Higgs particles are~\cite{higgs_moriond}
%%%--------------------
\bsub
\bea
\mh       & > & 95\gev, \\
\mhlight  &\simgt& 84\gev, \\
\label{eq:susy_higgs_bound}
\mpseudo &\simgt& 85\gev, \\
\mcharged &\simgt& 69\gev. 
\eea
\label{eq:bound_higgs}
\esub
%%%----------------------------------
In addition, for the neutral particles, we assume 
%%%-------------------- 
\bsub\label{mnbound}
\bea
m_{\sneutrino} &>& 45\gev, \\
m_{\neutralino{1}} + m_{\neutralino{2}} &>& \mz, 
\eea
\esub
%%%-------------------- 
so that they do not contribute to the total width of the 
$Z$-boson~\cite{lepewwg98}. 
%%%%%--------------------------------------------
%%%%%
%%%%%
%%%%%	Section: Oblique corrections in the MSSM
%%%%%
%%%%%
%%%%%--------------------------------------------
\clean
\section{Oblique corrections in the MSSM}
\label{section:oblique}
%%%-----------------
As explained in the introduction, the MSSM contributions to the 
vertex or box corrections vanish whenever either the sypersymmetric 
scalars (squarks and sleptons) or the supersymmetric 
fermions (charginos, neutralinos, and the gluino) are heavy enough.  
In those cases, the MSSM particles can still affect the electroweak 
observables via their contributions to the gauge-boson propagators, 
which are often called the {\it oblique } corrections 
\cite{stu90,ht90_gw91,mr90_kl90,ab91}. 
Because the oblique corrections affect all the electroweak observables 
in a flavor-independent manner (universality), and because their 
effects are found to be most significant under the present constraints 
on the new particle masses as summarized in the previous section, 
we study them in this section in great detail.  
%%%----------------------------------
%%%	a new paragraph

%%%	a new paragraph
%%%----------------------------------
The formalism presented in section 2 tells us that the precision 
electroweak experiments at the $Z$-boson pole constrain just two 
oblique parameters, $S_Z$ and $T_Z$, whereas the $W$-boson mass 
$\mw$ can be taken as the third oblique parameter.  
We favor $\mw$ over the $U$ parameter as our third oblique 
parameter, because we can avoid correlations among the three 
oblique parameters this way and because we could not gain  
insight by adopting the $U$ parameter in our MSSM analysis.  
%%%----------------------------------
%%%	a new paragraph

%%%	a new paragraph
%%%----------------------------------
In the following subsections, we examine the oblique corrections 
from each sector of the MSSM, since all the oblique corrections 
are expressed as linear sum of individual contributions~\cite{dh90}: 
squarks and sleptons (sec.~4.1), the MSSM Higgs bosons (sec.~4.2), 
and charginos and neutralinos (sec.~4.3).   
The effects of combining all the contributions are discussed 
in subsection 4.4.  
As remarked above, the dominance of the oblique contributions 
to the electroweak observables is justified only when either 
the sfermions or the supersymmetric (-ino) fermions are heavy.  
All the oblique contributions are of course relevant always 
as a part of the full MSSM contributions.  
%%%----------------------------------
%%%	a new paragraph

%%%	a new paragraph
%%%----------------------------------
\begin{figure}[t]
\begin{center}
\leavevmode\psfig{figure=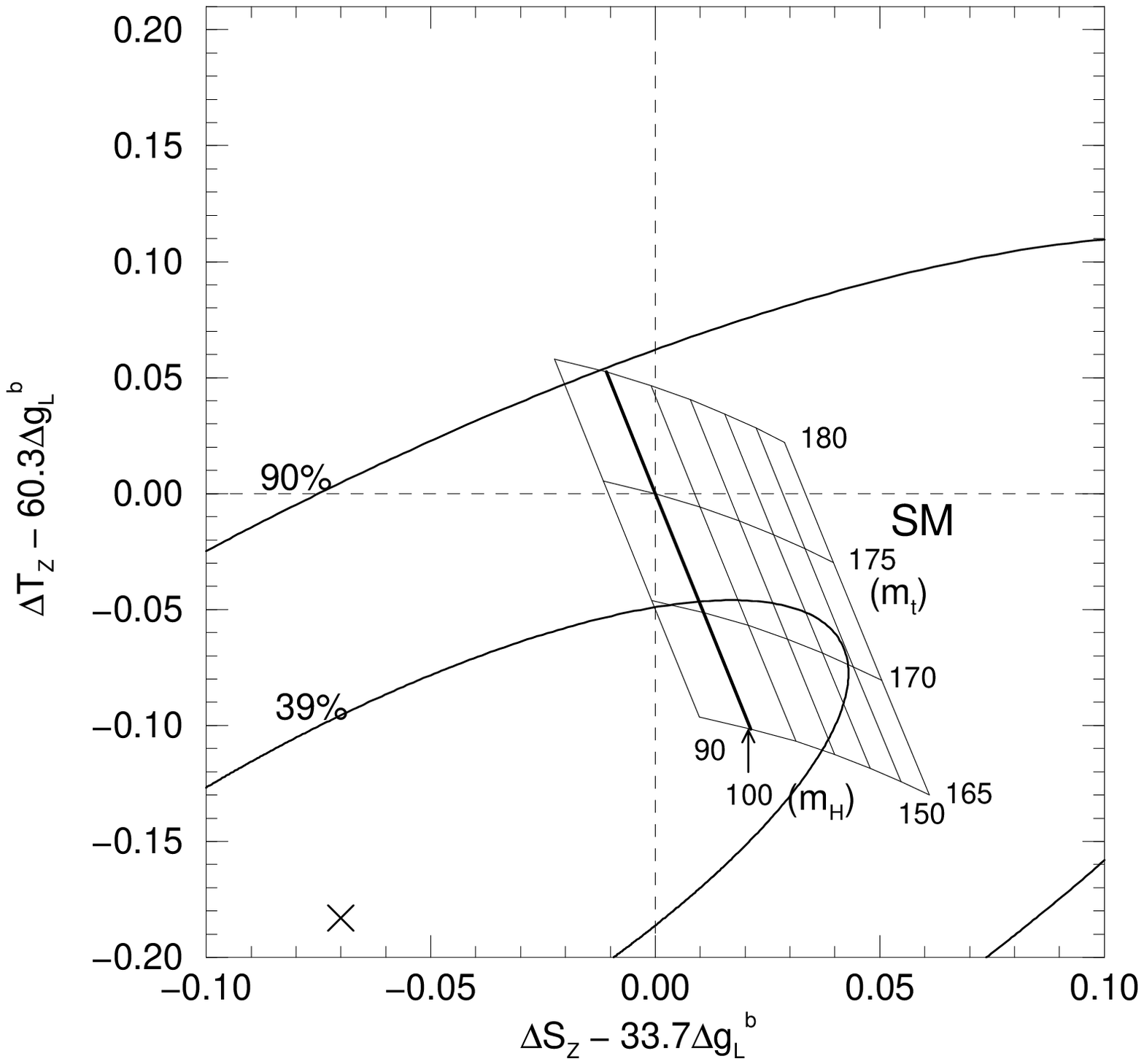,width=10cm}
\end{center}
\caption{Constraints on $\dsz - 33.7\dgbl$ and $\dtz - 60.3\dgbl$ 
from the electroweak precision measurements. 
The symbol ($\times$) denotes the best fit from the electroweak data. 
The 39\% ($\dx = 1$) and 90\% ($\dx = 4.61$) contours are shown. 
The SM predictions are given for $\mt = 165 \sim 180\gev$ and 
$\mh = 90 \sim 150\gev$. }
\label{fig:sm_oblique}
\end{figure}
%%%----------------------------------
In presenting our results, we find it most convenient to 
parameterize first the experimental constraints on the oblique 
parameters, and then confront the MSSM contributions to those 
parameters against the experimental constraints.  
We further observe that the uncertainty in the SM contributions 
to the oblique parameters which arise from the uncertainty in 
$\mh$~(\ref{eq:smhiggs}), $\mt$~(\ref{eq:mt_pdg}) and 
$\alpha(\mzsq)$, (\ref{eq:ej_alpha}) or (\ref{eq:dh_alpha}),  
is significant as compared to the magnitudes of the MSSM contributions.  
The uncertainty in $\alps(\mz)$~(\ref{eq:alpha_s_pdg}) 
has been found to affect our results little. 
We therefore obtain the constraints on the oblique parameters 
in such a way that we can examine the sum of the SM and the 
MSSM contributions for arbitrary values of $\mt, \mh$ and 
$\alpha (\mzsq)$ within their present constraints.  
This can be achieved by observing that the $\mt$ and $\mh$ dependences 
of the SM predictions appear only in the oblique parameters and 
in the $\zblbl$ vertex correction, $\dgbl$, and by observing that their 
$\alpha(\mzsq)$ dependences appear only through the combination 
$\sz$~(\ref{eq:dsz}) and $\dmw$~(\ref{eq:dmw}).
%%%----------------------------------
%%%	a new paragraph

%%%	a new paragraph
%%%----------------------------------
By adopting the 5 parameters, $\dsz, \dtz, \dgbl, \dmw$
and $\alps(\mz)$ as the free adjustable parameters, 
we find the following fit for all the electroweak observables 
of Table 1, under the constraint~(\ref{eq:alpha_s_pdg}): 
%%%-----------------
\bsub
\bea
&&
	\left. 
	\begin{array}{lcl}
	\dsz -33.7 \dgbl &=& -0.070 \pm 0.113 \\
	\dtz -60.3 \dgbl &=& -0.183 \pm 0.137 \\
	\end{array} \right \}, ~~~~
\rho = 0.89, 
\label{eq:bound_dsz_dtz}
\\
&&
\dmw(\gev) = \hph 0.008 \pm 0.046, 
\\
\vsk{0.2}
&&
\disp{
\chi^2_{\rm min} = 15.4 + \biggl( \frac{\dgbl + 0.00086}{0.00076}
	\biggr)^2, 
}
\eea
\label{eq:chisq_oblique}
\esub
%%%--------
where \dof $= 20-5 = 15$. 
%%%----------------------------------
\begin{figure}[t]
\begin{center}
\leavevmode\psfig{figure=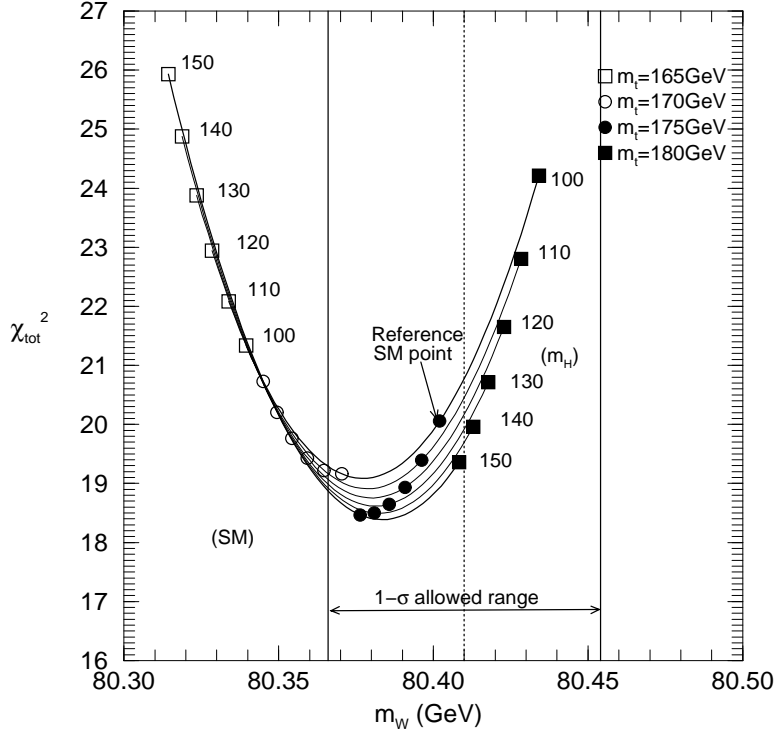,width=10cm}
\end{center}
\caption{
The total $\chi^2$ versus $\mw$. The 1-$\sigma$ experimental bound 
on $\mw$ is shown. The SM predictions are given for $\mt = 165 \sim 
180\gev$ and $\mh = 100\sim 150\gev$. The reference SM point at 
$\mt=175\gev, \mh=100\gev, 1/\alpha(\mzsq)=128.90$ is shown by 
the arrow. }
\label{fig:sm_mw}
\end{figure}
%%%----------------------------------
%%%	a new paragraph

%%%	a new paragraph
%%%---------------------------------- 
The results of the fit are shown in Figs.~\ref{fig:sm_oblique} 
and \ref{fig:sm_mw}, where the 
SM predictions for some representative values of $\mt$ and $\mh$
are given for $1/\alpha(\mzsq)=128.90~(\xa=0)$~\cite{EJ}. 
The predictions for different estimate of $1/\alpha(\mzsq)$ can 
easily be obtained by using the $\xa$-dependences of $\dsz$~(\ref{eq:dsz}) 
and $\dmw$ (\ref{eq:dmw}).  
Fig.~\ref{fig:sm_oblique} shows the constraint (\ref{eq:bound_dsz_dtz}) 
in the plane of $\dsz-33.7\dgbl$ and $\dtz-60.3\dgbl$. 
It is clearly seen from this figure that relatively 
low value of $\mt$ as compared to the present experimental mean 
value~(\ref{eq:mt_pdg}) and relatively light Higgs boson are favored 
within the SM fit.
In Fig.~\ref{fig:sm_mw}, the SM prediction for $\mw$ (at $\xa=0$) 
and the resulting total $\chi^2$ are shown.  
The $\mw$ data slightly favors $\mh=100\gev$ over $\mh=150\gev$ for 
$\mt \simlt 170\gev$, so does the total $\chi^2$.  
When $\mt \simgt 175\gev$, the trend is reversed and $\mh=150\gev$ is 
favored against $\mh=100\gev$ in the total 
$\chi^2$.\footnote{
The $\chi^2_{\rm tot}$ value in the figures~\ref{fig:sm_mw}, 
\ref{fig:mw:sfermion}, \ref{fig:mw:inos} for the SM reference 
point at $\mt=175\gev$ and $\mh=100\gev$ is about 0.25 larger 
than the quoted value of 19.8 in Table~\ref{tab:ewdata98}, 
because of the contribution from the $\alps(\mz)$ 
constraint~(\ref{eq:alpha_s_pdg}). }
%%%----------------------------------
%%%	a new paragraph

%%%	a new paragraph
%%%---------------------------------- 
The SM fits reported in section 2, eqs.~(\ref{eq:sm_fit_ej}) and 
(\ref{eq:sm_fit_dh}), are obtained from the above fit by combining 
it with the constraint on $\mt$~(\ref{eq:mt_pdg}), and that on 
$1/\alpha(\mzsq)$, (\ref{eq:ej_alpha}) or (\ref{eq:dh_alpha}), 
respectively.   
In both Figs.~\ref{fig:sm_oblique} and \ref{fig:sm_mw}, the theoretical 
predictions are given for the reference value of $1/\alpha(\mzsq)=128.90$, 
or $\xa=0$, in eq.~(\ref{eq:xa}).  
Effects of changing $1/\alpha(\mzsq)$ can easily be studied 
by shifting the theoretical predictions shown for $\xa=0$ by 
%---------- 
\bsub
\bea
\dsz (\xa=0)  &\to& \dsz (\xa=0) -0.064\xa \,, \label{dszvsxa} \\
\dmw (\xa=0)  &\to& \dmw (\xa=0) +0.012\xa \,, \label{dmwvsxa}
\eea
\esub
%----------- 
as required by eqs.(\ref{eq:dsz}) and (\ref{eq:dmw}).  
Change in $1/\alpha(\mzsq)$ of the full uncertainty of the conservative 
estimate (\ref{eq:ej_alpha}) makes $\xa=\pm 1$ in eq.~(\ref{eq:xa}). 
We can clearly see from Fig.~\ref{fig:sm_oblique} that the resulting 
horizontal shift of the SM prediction due to eq.(\ref{dszvsxa}) can 
affect significantly the preferred range of the Higgs boson mass in the SM.  
Inspection of Fig.~\ref{fig:sm_mw} with the shift due to 
eq.~(\ref{dmwvsxa}) tells us that the effect is not significant for 
$\mw$ with its present measurement error. 
We can verify the effect by comparing the SM fits (\ref{eq:sm_fit_ej}) 
and (\ref{eq:sm_fit_dh}), that were obtained by using the two estimates, 
(\ref{eq:ej_alpha}) and (\ref{eq:dh_alpha}), respectively. 
The mean value of the estimate (\ref{eq:dh_alpha}) is about $\xa=0.4$, 
and hence the SM prediction in Fig.~\ref{fig:sm_oblique} moves 
horizontally in the negative direction by about 0.03.  
This may affect the Higgs boson mass by about 30\%.  
Although naive, this simple estimate reproduces qualitatively 
the difference in the most favored value of $\mh$ in the two fits, 
that are found to be about 22\% between eqs.(\ref{eq:sm_fit_ej}) and 
(\ref{eq:sm_fit_dh}).  
%%%----------------------------------
%%%	a new paragraph

%%%	a new paragraph
%%%----------------------------------
In the following subsections, we show the MSSM contributions to 
the oblique parameters by superposing them on Figs.~\ref{fig:sm_oblique} 
and \ref{fig:sm_mw}, by choosing the reference SM point at $\mt=175\gev$ 
and $\mh=100\gev$ $(\xt=\xh=0)$. 
The MSSM predictions for other choices of $\mt$ and $\mh$ are then 
obtained simply by shifting the SM reference point.  
%%%---------------------------------------
%%% Subsection: 4.1 squarks and sleptons 
%%%---------------------------------------
\begin{figure}[t]
\begin{center}
\leavevmode\psfig{figure=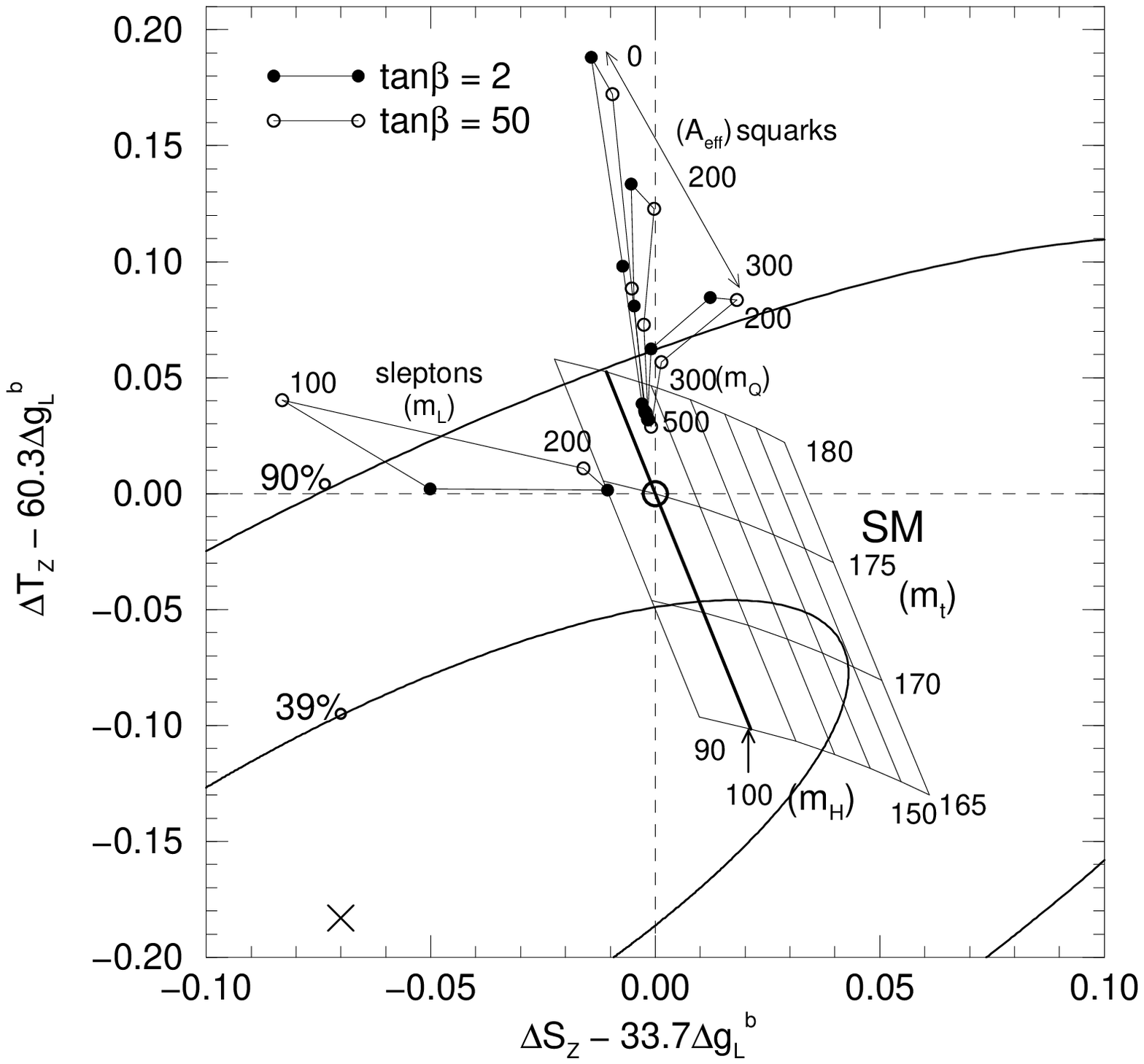,width=10cm}
\end{center}
\caption{The squark and slepton contributions to $(\dsz, \dtz)$ 
for $\tan\beta = 2$ and 50. 
The SUSY breaking scalar masses for the left-handed and right-handed 
squarks are assumed to be same, denoted by $m_{\wt{Q}}$. 
The $\stop_L$-$\stop_R$ and $\sbottom_L$-$\sbottom_R$ mixings are 
controlled by $A_\eff = A_\eff^t = A_\eff^b$. 
In the slepton sector, although sizable $\stau_L$-$\stau_R$ mixing 
may be induced for large $\tan\beta$, the mixing is neglected 
because it does not affect both $\dsz$ and $\dtz$ significantly. 
The reference SM point ($\dsz=\dtz=\dgbl=0$) is marked by the open 
circle at the origin. }
\label{fig:oblique:sfermion}
\end{figure}
%%%---------------------------------------
\subsection{Squarks and sleptons}
%%%---------------------------------------
In Fig.~\ref{fig:oblique:sfermion} we show the sfermion contributions 
to $\dsz$ and $\dtz$ that are superposed to the SM contribution of 
Fig.~\ref{fig:sm_oblique}. 
The origin of the plot $(\dsz=\dtz=\dgbl=0)$, marked by the big open 
circle, gives the SM prediction at $\mt=175\gev$ and $\mh=100\gev$  
for $1/\alpha(\mzsq)=128.90~(\xa=0)$.  
The contributions of the squarks and sleptons are shown separately.  
The net result should be obtained by adding the SM contribution, 
the squark contribution and the slepton contribution vectorially 
in the two-dimensional plot.  
%%%----------------------------------
%%%	a new paragraph

%%%	a new paragraph
%%%----------------------------------
It is clear from the figure that for $\mt\sim 175\gev$ and 
$\mh \simlt 130\gev$, both the squark and slepton contributions 
make the fit worse than the SM.  
The situation does not improve by changing our estimate for 
$1/\alpha(\mzsq)$ in the range (\ref{eq:ej_alpha}).  
The squark contribution makes the fit worse, because it always makes 
$\tz$ larger than the SM prediction, which is already larger 
than the preferred value of the data.  
The slepton contribution makes the fit worse, because it 
gives negative $\sz$ with either $\tz=0$ ($\tan\beta=2$) 
or with slightly positive $\tz$ ($\tan\beta=50$).  
The better fit to the data requires a contribution with 
{\it both} $\dsz<0$ {\it and} $\dtz<0$, which cannot be 
achieved by the contributions from squarks and sleptons.    
%%%----------------------------------
%%%	a new paragraph

%%%	a new paragraph
%%%----------------------------------
Let us examine the squark and slepton contributions to the 
$\dsz$ and $\dtz$ parameters in more detail.  We find that 
they do not contribute significantly to $R$, and hence we 
can understand the qualitative behavior by studying their 
contribution to $S$ and $T$.  
%-----------------
To fix our notation, we first give the squared mass matrix of 
the squarks and sleptons. 
Since the structure of the mass matrix of sfermions are similar 
among the different flavors, we give the mass-squared matrix 
of the stop as an example.  
In the $(\stop_L,\stop_R)$ basis it is given by 
%-----------------
\bsub
\bea
M_{\wt{t}}^2 &=& 
	\left( 
	\begin{array}{cc} 
	m_{\wt{t}_L}^2 & \mt (A_\eff^t)^*
	\\
\vsk{0.1}
	\mt A_\eff^t 	& m_{\wt{t}_R}^2 
	\end{array}
	\right), 
\label{eq:stopmass}
\\
\vsk{0.1}
m_{\wt{t}_L}^2 &=& m_{\wt{Q}_3}^2 + \mzsq \cos 2\beta 
	(I_{3u} - \shatsq Q_u) + \mtsq, 
\label{eq:stop_diagonal_left}
\\
\vsk{0.1}
m_{\wt{t}_R}^2 &=& m_{\wt{U}_3}^2 + \mzsq \cos 2\beta 
	\shatsq Q_u + \mtsq,  
\label{eq:stop_diagonal_right}
\eea
\esub
%-----------------
where $\shatsq=\sin^2\theta_W(\mz)_{\msbar}$.  
The diagonal elements of the mass-squared matrices for the other 
sfermions are obtained by replacing $(m_{\wt{Q}_3}, m_{\wt{U}_3})$ by 
$(m_{\wt{Q}_3}, m_{\wt{D}_3})$ for $\sbottom$, 
$(m_{\wt{Q}}, m_{\wt{U}})$ for $\sup$, 
$(m_{\wt{Q}}, m_{\wt{D}})$ for $\sdown$,  
$(m_{\wt{L}_3}, m_{\wt{E}_3})$ for $\stau$,  
$(m_{\wt{L}}, m_{\wt{E}})$ for $\selectron$, and by 
choosing appropriate values for 
the third component of the weak isospin $I_{3f}$ and 
the electric charge $Q_f$. 
The off-diagonal elements in (\ref{eq:stopmass}), $\mt A_\eff^t$, 
should be replaced by $m_f A_\eff^f$   
%-----------------  
\bea
A_\eff^f &=& \left\{ 
	\begin{array}{l}
	A_f - \mu^*\cot\beta ~~~(f=t)\\
	A_f - \mu^*\tan\beta ~~~(f=b,\tau)
	\end{array}
	\right. 
\label{eq:aeff}
\eea
%-----------------
for $\sbottom$ and $\stau$, while it is set to zero for the other sfermions.  
The sfermion mass-squared matrix can be diagonalized by using the unitary 
matrix $U^{\wt{f}}$ for $\wt{f} = \stop, \sbottom, \stau$:    
%----------------- 
\bea
(U^{\wt{f}})^\dagger M_{\wt{f}}^2 U^{\wt{f}} &=& 
	\diag(m_{\wt{f}_1}^2, m_{\wt{f}_2}^2), 
~~~~~~~~~~(m_{\wt{f}_1} < m_{\wt{f}_2}). 
\eea
%----------------- 
In all our numerical examples, we put real values for $A_f$ and $\mu$.  
Since we neglect the left-right mixing in the first two generations, 
their mass eigenvalues are given by the diagonal elements. 
%%%----------------------------------
%%%	a new paragraph

%%%	a new paragraph
%%%----------------------------------
It is useful to examine analytic expressions of the $S$ and 
$T$ parameters under some approximations. 
If the mixing between the left- and right-handed states is negligible, 
the sfermion contribution to the $S$ parameter, $\ds$, can be given by 
the following simple form: 
%%%----------------------------------
\bea
\ds &=& -\frac{1}{\pi \mzsq}\sum_f C_f I_{3f} 
	Y_{f_L} F_5(\mzsq:m_{\wt{f}_L}, m_{\wt{f}_L}), 
\label{eq:ds_sfermion}
\eea
%%%----------------------------------
where $C_f$ is 3 for the squarks and 1 for the sleptons. 
The symbol $f$ runs over the flavor space. 
In this limit, only the SU(2)$_L$ doublets contributes to the 
$S$ parameter. 
The explicit form of the function $F_5(\mzsq:m_{\wt{f}}, m_{\wt{f}})$ 
in the first line has been given in ref.~\cite{hhkm94}. 
Since $\ds$ is proportional to the hypercharge $Y_f$, the relative 
sign of $\ds$ is opposite between squarks $(Y_q=1/6)$ and sleptons 
$(Y_L=-1/2)$ in the first two generations. 
Moreover, since $\ds$ is proportional to $I_{3f}$, it receives 
contribution only 
when there is a mass splitting among SU(2)$_L$ multiplet members. 
It takes a particularly simple form in the large sfermion mass limit. 
For example, the $\sup_L$-$\sdown_L$ and $\sneutrino_l$-$\slepton_L$ 
contributions can be expressed as  
%%%----------------------------------
\bsub\label{eq:ds_sfermion_approx}
\bea
(\ds)_{\squark} &\approx& 
	-\frac{1}{12 \pi} 
\frac{m_{\sup_L}^2 - m_{\sdown_L}^2}{m_{\sup_L}^2 + m_{\sdown_L}^2}
	\biggl[
	1+ O\biggl(
\frac{m_{\sup_L}^2 - m_{\sdown_L}^2}{m_{\sup_L}^2 + m_{\sdown_L}^2}
	\biggr)
	\biggr], 
\\
(\ds)_{\slepton} &\approx& 
	+\frac{1}{12 \pi} 
\frac{m_{\sneutrino_l}^2-m_{\slepton_L}^2}{m_{\sneutrino_l}^2+m_{\slepton_L}^2}
	\biggl[
	1+ O\biggl(
\frac{m_{\sneutrino_l}^2-m_{\slepton_L}^2}{m_{\sneutrino_l}^2+m_{\slepton_L}^2}
	\biggr)
	\biggr], 
\eea
\esub
%%%----------------------------------
respectively.  
The magnitude of $\ds$ is determined by the mass difference between 
the up- and down-type states in the SU(2)$_L$ doublet. 
Except for the $\stop_L$-$\sbottom_L$ case, the mass difference 
of the up- and down-type components of the SU(2) doublets is 
determined by the $D$-terms in the mass-squared matrix by neglecting 
the tiny contribution from the fermion masses 
%%%---------------------------------- 
\bea
m_{\sneutrino_l}^2 - m_{\slepton_L}^2 = 
m_{\sup_L}^2 - m_{\sdown_L}^2 = (1-\shatsq)\mzsq\cos2\beta. 
\label{eq:mass_diff_slepton}
\eea
%%%----------------------------------
Since the r.h.s.\ of eq.~(\ref{eq:mass_diff_slepton}) is negative 
for $\tan\beta > 1$ and its magnitude grows as $\tan\beta$ increases, 
we can understand the qualitative behavior of the squark and slepton 
contributions in Fig.~\ref{fig:oblique:sfermion}. 
%%%----------------------------------
%%%	a new paragraph

%%%	a new paragraph
%%%----------------------------------
The analytic form of the $T$ parameter in the zero left-right mixing 
limit of squarks and sleptons are also simple. 
It also receives contribution from particles which carry the SU(2)$_L$ 
quantum number. 
For example, the $\stop_L$-$\sbottom_L$ contribution is given by 
%%%----------------------------------
\bea
\dt &=& \frac{\gf}{\rttwo}\frac{1}{4\pi^2\alpha} C_f
	\biggl[ 
	\half (m_{\stop_L}^2 + m_{\sbottom_L}^2) 
	+ \frac{m_{\stop_L}^2 m_{\sbottom_L}^2}
	{m_{\stop_L}^2 - m_{\sbottom_L}^2}
	\ln \frac{m_{\sbottom_L}^2}{m_{\stop_L}^2}
	\biggr] 
\nonumber \\
&\approx& 
	\frac{\gf}{24 \rttwo \pi^2\alpha} C_f
	\frac{(m_{\stop_L}^2 - m_{\sbottom_L}^2)^2 }
	{m_{\stop_L}^2 + m_{\sbottom_L}^2 }
	\biggl[ 1 + O \biggl( 
	\frac{m_{\stop_L}^2 - m_{\sbottom_L}^2 }
	{m_{\stop_L}^2 + m_{\sbottom_L}^2 } 
	\biggr)	\biggr]. 
\label{eq:dt_approx}
\eea
%%%----------------------------------
\begin{figure}[t]
\begin{center}
\leavevmode\psfig{figure=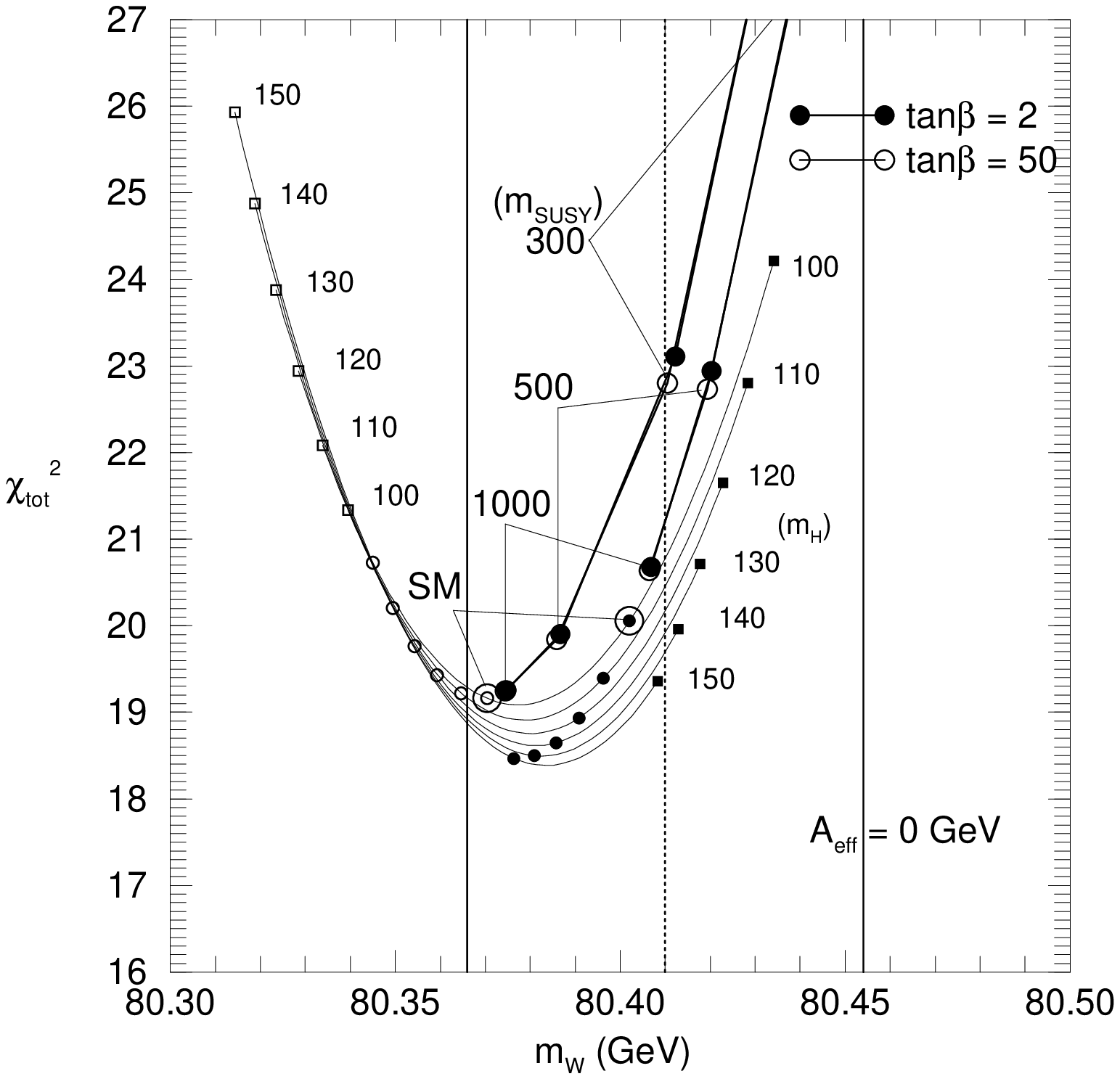,width=7cm}
\leavevmode\psfig{figure=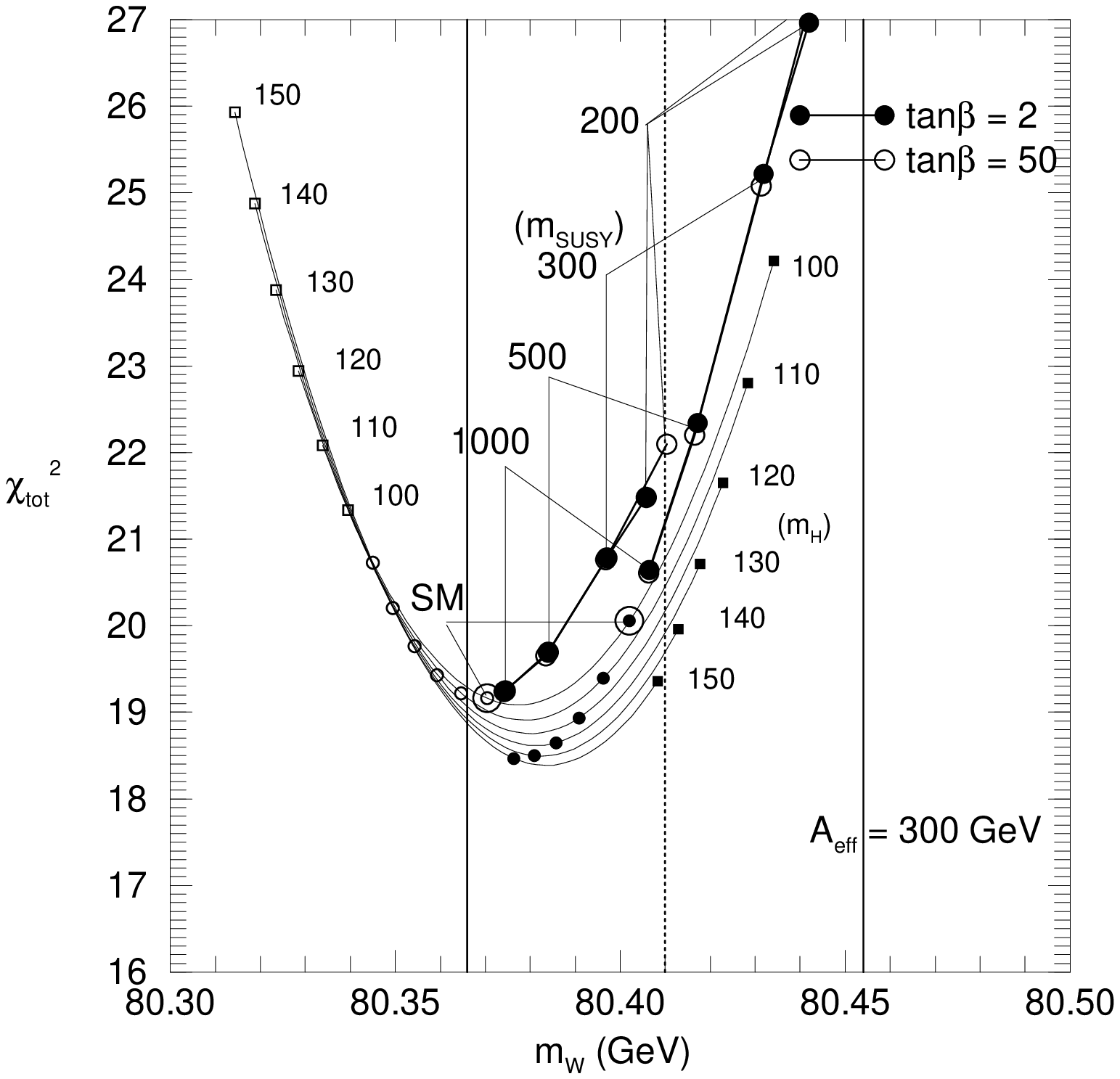,width=7cm}
\end{center}
\caption{The squark and slepton contributions to $\mw$ 
for $\tan\beta = 2$ and 50. 
The SUSY breaking scalar masses for the left- and right-handed 
squarks/sleptons are taken to be common, which are denoted by 
$m_{\rm SUSY}$. 
Two cases of the left-right mixing, $A_{\rm eff} = 0\gev$ (left) 
and $A_{\rm eff} = 300\gev$ (right) are examined. 
In each figure, two cases of the SM reference point, 
$(\mt({\rm GeV}),\mh({\rm GeV}))=(175,100)$ and $(170,100)$, are 
marked by circles.} 
\label{fig:mw:sfermion}
\end{figure}
%%%----------------------------------
The second line in eq.~(\ref{eq:dt_approx}) is an expression 
in the large $\stop_L$, $\sbottom_L$ mass limit. 
Contributions from the other squarks and sleptons are obtained by 
replacing the masses and $C_f$ factors appropriately.   
The squarks in the first two generations and sleptons contributions 
can be found by replacing the mass parameters and $C_f$ appropriately. 
Since the function in the parenthesis of eq.~(\ref{eq:dt_approx}) 
is positive for any values of $(m_{\sup_L}, m_{\sdown_L})$ or 
$(m_{\sneutrino_l},m_{\slepton_L} )$, $\dt$ always receives 
positive contributions from the squarks and sleptons. 
%%%----------------------------------
%%%	a new paragraph

%%%	a new paragraph
%%%----------------------------------
The special attention should be payed for the stop-sbottom 
contribution. 
There is a large mass difference between $\stop_L$ and $\sbottom_L$, 
which is dominated by $\mtsq-m_b^2$. 
It is rather easy to understand that the effect of the left-right 
mixing of the stop and sbottom qualitatively. 
Since the off-diagonal elements of the mass-squared matrices of 
the stop and sbottom are multiplied by the top and bottom mass 
squared, respectively, the effect of the left-right mixing in 
the stop sector is much larger than that in the sbottom sector. 
Let us focus on the left-right mixing in the stop mass matrix for brevity. 
Suppose that only $\stop_1$ is light and the other squarks are 
heavy and degenerate. 
Suppose also that there is no mixing between $\sbottom_L$ and $\sbottom_R$. 
With these assumptions, the squark contribution to $\dt$ in the third 
generation may be parametrized by $m_{\stop_1}$ and $m_{\sbottom_L}$ 
($\approx m_{\stop_2} \approx m_{\sbottom_R}$) and the unitary 
matrix $U^{\wt{t}}$. 
Then, the stop-sbottom contribution to $\dt$ is give by 
%%%----------------------------------
\bea
\dt &\approx& \frac{\gf}{\rttwo}\frac{1}{4\pi^2\alpha} C_q
	\biggl[ 
	|U^{\wt{t}}_{11}|^2 F_5(0:m_{\stop_1}, m_{\sbottom_L}) 
-	|U^{\wt{t}}_{11}|^2 |U^{\wt{t}}_{21}|^2 
	F_5(0:m_{\stop_1}, m_{\sbottom_L}) 
	\biggr]
\nonumber \\
\vsk{0.2}
	&\approx& 
	\frac{\gf}{\rttwo}\frac{1}{4\pi^2\alpha} C_q
	|U^{\wt{t}}_{11}|^4 F_5(0:m_{\stop_1}, m_{\sbottom_L})
\nonumber \\
\vsk{0.2}
	&\approx& 
	\frac{\gf}{\rttwo}\frac{1}{4\pi^2\alpha} C_q
	|U^{\wt{t}}_{11}|^4 
	\biggl[
	\frac{1}{2} (m_{\stop_1}^2 + m_{\sbottom_L}^2) 
	+ \frac{m_{\stop_1}^2 m_{\sbottom_L}^2 }
	{m_{\stop_1}^2 - m_{\sbottom_L}^2 } 
	\ln \frac{m_{\sbottom_L}^2}{m_{\stop_1}^2} 
	\biggr].
\label{eq:dt_stop}
\eea
%%%----------------------------------
Let us recall that the left-right mixing is induced by $A_\eff^t$ 
in eq.~(\ref{eq:aeff}). 
The factor $|U^{\wt{t}}_{11}|^4$ decreases ($\le 1$) as $A_\eff^t$ 
increases and it suppresses $\dt$. 
The behavior of the squark contribution in Fig.~\ref{fig:oblique:sfermion} 
may be interpreted in this way.  
On the other hand, if only $m_{\stop_1}$ is kept small in 
eq.~(\ref{eq:dt_stop}) while keeping $|U^{\wt{t}}_{11}|$ finite, 
the $T$ parameter grows with $m_{\sbottom_L}^2$.   
This reflects the growth of the $A_\eff^t$ parameter as 
$(m_{\sbottom_L}^2/\mt)|U^{\wt{t}}_{11}|^2/(1-|U^{\wt{t}}_{11}|^2)$, 
that breaks the custodial SU(2) symmetry \cite{dh90}.   
%%%----------------------------------
%%%	a new paragraph

%%%	a new paragraph
%%%----------------------------------
Let us give the analytic expression of $\dr$ in the large 
sfermion mass limit. 
Because $\dr$ is a linear combination of the $Z$-boson propagator 
corrections between two different momentum transfer scales, 
there are contributions not only from the left-handed sfermions 
but also from the right-handed sfermions. 
By taking $m_{\wt{f}_L}$ and $m_{\wt{f}_R}$ as the left- and 
right-handed sfermion masses, respectively, 
$\dr$ in the heavy mass limit is given by 
%%%----------------------------------
\bea
\dr \approx -\frac{1}{30\pi} C_f 
	\biggl[
	(I_{3f}-\shatsq Q_f)^2 \frac{\mzsq}{m_{\sfermi_L}^2}
	+      (\shatsq Q_f)^2 \frac{\mzsq}{m_{\sfermi_R}^2}
	\biggr]. 
\label{eq:dr_sfermion}
\eea
%%%----------------------------------
Although the negative sign of $\dr$ leads to negative contribution 
to $\dtz$ in (\ref{eq:dtz}), its magnitude is so tiny that the net 
contribution of sfermions to $\dtz$ is found to be always positive.  
%%%----------------------------------
%%%	a new paragraph

%%%	a new paragraph
%%%----------------------------------
In Fig.~\ref{fig:mw:sfermion}, we superimposed the sfermion contribution 
to $\mw$ and to the total $\chi^2$.  
$A_\eff=0$ and $A_\eff=300\gev$ cases are shown side by side.  
Because the total $\chi^2$ is not a linear sum of the SM contribution 
and the new physics contribution, we show two representative cases 
of the reference SM point, one for $(\mt,\mh)=(175,100)$ and the other 
for $(\mt,\mh)=(170,100)$ in $\gev$ units.  
For each representative $(\mt,\mh)$ case, we show the MSSM predictions 
of $\mw$ and the total $\chi^2$ as a function of the common sfermion 
mass $m_\susy$.  
Here, for simplicity, we set all the 10 sfermion mass parameters 
in eqs.~(\ref{eq:mssm_para4}) and (\ref{eq:mssm_para5}) to have a 
common value $m_\susy$, and the 3 $A_\eff^f$ parameters in 
eq.~(\ref{eq:aeff}) to have a common value $A_\eff$.    
We find that the sfermion contributions always make 
$\mw$ larger than the SM prediction.  
Because the SM prediction for $\mw$ is smaller than the experimental 
mean value at $\mt \simlt 170\gev$, the sfermion contribution can 
improve the fit for $\mw$.  
As may be seen from the four examples in Fig.~\ref{fig:mw:sfermion}, 
however, the sfermion contribution always make the overall fit worse 
than the SM.  
By comparing the $A_\eff=0$ case (left) and the $A_\eff=300\gev$ case 
(right), we find that the unfavorable sfermion contribution can be 
made small by introducing $A_\eff$ for the same value of $m_\susy$.  
The trend can be understood qualitatively by using the analytic 
expressions (\ref{eq:ds_sfermion_approx}), (\ref{eq:dt_approx}) and 
(\ref{eq:dt_stop}) for the sfermion contributions to the $S$ and $T$ 
parameters. 
The sfermion contribution to the $U$ parameter is found numerically small.  
%%%----------------------------------
%%%	a new paragraph

%%%	a new paragraph
%%%----------------------------------
As is clear from the figures, the present $\mw$ data gives little 
contribution to the total $\chi^2$ of the fit.  
Although the positive contribution to $\dmw$ can improve the fit 
to the $\mw$ data if $\mt \sim 170\gev$, the overall $\chi^2$ 
always gets worse because of the $Z$-parameter constraints as 
summarized in Fig.~\ref{fig:oblique:sfermion}. 
%%%%%----------------------------------------------
%%%%% Subsection: 4.2 Higgs
%%%%%----------------------------------------------
\subsection{MSSM Higgs bosons}
\label{section:higgs}
%%%----------------------------------
The MSSM has an extended Higgs boson sector with three neutral 
Higgs bosons, $h$, $H$, and $A$, and one charged Higgs boson $H^\pm$.  
All their masses and couplings are calculated in terms of the MSSM 
Lagrangian parameters, by using the improved one-loop 
potential~\cite{susymh91}. 
%%%%%----------------------------------------------
\begin{figure}[b]
\begin{center}
\leavevmode\psfig{figure=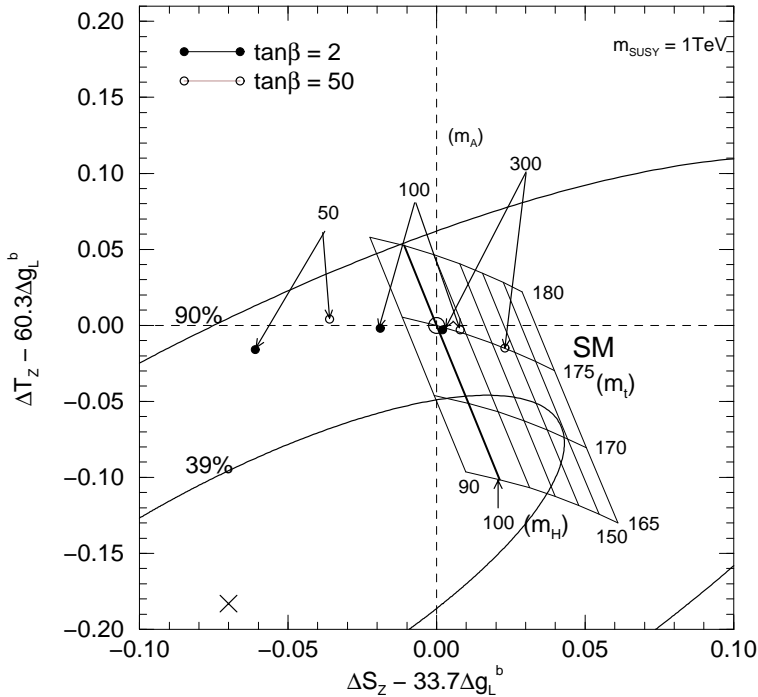,width=10cm}
\end{center}
\caption{Contributions to $(\dsz,\dtz)$ from the Higgs bosons 
for $\tan\beta = 2$ and 50. 
The 1-loop improved Higgs potential, in which only the top 
and bottom Yukawa couplings are retained, are used to 
find the mass eigenvalues. 
The inputs are the pseudo-scalar Higgs-boson mass $\mpseudo$ and 
$\tan\beta$. Three values of $\mpseudo$, $\mpseudo=50,100$ and 
$300\gev$ are examined. 
The scale parameter $m_{\rm SUSY}$ which appears in the 
1-loop potential is set at 1\tev. }
\label{fig:oblique:higgs}
\end{figure}
%%%%%----------------------------------------------
Their contribution to the oblique parameters are evaluated carefully, 
because parts of the SM corrections contain the $\mh$ dependence 
from the two-loop corrections.  They appear in the $T$ parameter and 
in the $\dgbl$ parameter, although the $\mh$ dependence of the latter 
turned out to be negligibly small.  
In order to avoid discontinuity in the theoretical predictions 
because of the lack of the corresponding two-loop results in 
the MSSM, we make the following arrangement in our actual calculation.  
For each MSSM parameters, we first calculate the lightest CP-even 
neutral Higgs boson mass, $\mhlight$.  We then evaluate all the SM 
corrections for $\mh=\mhlight$.  The contributions of the MSSM Higgs 
sector to the radiative corrections are then evaluated as the sum of 
the SM contributions, that partially contain the two-loop correction,  
and the difference between the MSSM and the SM Higgs sector 
contributions that is calculated strictly in the one-loop order 
by using the formulae in the Appendices.  In this way, we can 
test numerically the decoupling of the MSSM Higgs sector in the 
large $\mpseudo$ limit.  
%%%----------------------------------
%%%	a new paragraph

%%%	a new paragraph
%%%----------------------------------
In Fig.~\ref{fig:oblique:higgs}, we show the contributions of the 
MSSM Higgs sector to the $\dsz$ and $\dtz$ parameters when 
$\mpseudo = 50\gev$, $100\gev$ and $300\gev$.  
As always, the results for $\tan\beta=2$ are shown 
by solid blobs, and those for $\tan\beta=50$ are shown by open blobs.  
We find that the MSSM prediction is remarkably near to the SM 
prediction at $\mh=\mhlight$ already at $\mpseudo=300\gev$.  
The lightest Higgs boson mass in ref.\cite{susymh91} can be approximately 
expressed as 
%-------
\bea
\mhlight &\approx& 
	102 - 6.7\ln\frac{\tan\beta}{2} 
	+ \frac{\tan\beta - 2}{\tan\beta}
	(0.13 \tan\beta + 43.5)
\nonumber \\
&&
+ 20 x_\susy \biggl[ 1 - 0.08 x_\susy
	\biggl\{ 1 + 1.5\biggl( 1 - \frac{2}{\tan\beta} \biggr) 
	\biggr\} \biggr], 
\label{eq:higgs_approx}
\eea
%-------
when $\mpseudo = 300\gev$. The parameter $x_\susy$ is defined by 
$x_\susy \equiv \ln(m_\susy/1\tev)$. 
Eq.~(\ref{eq:higgs_approx}) is valid for $2 \le \tan\beta \le 50$ and 
$1\tev \le m_\susy \le 2\tev$ in which the error is smaller than 
$0.6\gev$. 
%%%--------------------------------------------------
%%%	a new paragraph

%%%	a new paragraph
%%%--------------------------------------------------
%%%
%%%  Table: Higgs contributions to the oblique parameters for 
%%%         \tan\beta = 2, 50 and m_SUSY = 1 TeV
%%%
%%%-------------------------------------------------- 
\begin{table}[t]%[htbp]
\begin{center}
\begin{tabular}{ccccccccc}  \hline \hline
$\tan\beta$ & $\mpseudo$ & $\mhlight$ & $\mhheavy$ & 
$\mcharged$ & $\dsz$ & $\dtz$ & $\dmw$ &$\chi^2_{\rm tot}$
\\ \hline \hline 
2 &$\hpz\hpz$50 & 49.1 &$\hpz$137 &94.7 & $-0.061$ & $-0.016$ & $\hph0.021$
	& 23.2 \\
  &$\hpz$100 & 75.0 & $\hpz$152 & \,128 & $-0.019$ & $-0.002$ & $\hph0.007$
	& 21.0 \\
  &$\hpz$200 & 97.3&$\hpz$222 & \,216 &$-0.003$&$-0.003$&$\hph$0.000 
	& 20.0 \\
  &$\hpz$300 &\,102&$\hpz$313 & \,311 & $\hph$0.002 & $-0.003$ & $-0.002$ 
	& 19.8 \\
  &1000 &\,106  &1004 & 1003 & $\hph$0.006 & $-0.003$ & $-0.003$ 
	& 19.7 \\
\hline
50 &$\hpz\hpz$50 &50.3 & $\hpz$129 & 94.7& $-0.036$& $\hph0.004$ &$\hph0.023$ 
	& 22.7 \\
   &$\hpz$100   &\,100  & $\hpz$129 &\,128 &$\hph$ 0.008 & $-0.003$ &$-0.003$ 
	& 19.6 \\
   &$\hpz$200   &\,128  & $\hpz$200 &\,216 &$\hph$ 0.020 & $-0.014$ &$-0.012$ 
	& 18.8 \\
   &$\hpz$300   &\,129  & $\hpz$300 &\,311 &$\hph$ 0.023 & $-0.015$ &$-0.014$ 
	& 18.7 \\
   & 1000    & \,129  & 1000& 1003&$\hph$ 0.025 & $-0.017$ &$-0.015$
	& 18.7 \\
\hline
\end{tabular} 
\caption{Oblique parameters, $\dsz, \dtz$ and 
$\dmw$ in the MSSM Higgs sector for $\tan\beta = 2$ and 50. 
The mass spectrum of the Higgs bosons are calculated by using the 
1-loop improved scalar potential which is approximated by retaining 
only the top- and bottom-quark Yukawa couplings \cite{susymh91}. 
The scale parameter $m_{\rm SUSY}$ which appears in the 
1-loop potential is set at 1~TeV. }
\end{center}
\label{tab:higgs}
\end{table}
%%%----------------------------------
We find $\mhlight=102\gev$ and $129\gev$ for $\tan\beta=2$ and 50, 
respectively, at $\mpseudo=300\gev$ for $m_\susy=1\tev$, 
and the predictions are remarkably near to the SM predictions at 
$\mh=\mhlight$.  
The $\mpseudo$ dependence of the MSSM predictions are shown in 
Table~3, together with the masses of all the MSSM Higgs bosons 
for $\tan\beta=2$ and 50 and at $m_\susy=1\tev$.  
From the Table, we can see that the predictions of the MSSM 
reduces essentially to those of the SM at $\mh=\mhlight$ when 
$\mpseudo\simgt 200\gev$.  
%%%%%--------------------------------------
%%%%% Subsection: 4.3 Inos
%%%%%--------------------------------------
\subsection{Supersymmetric fermions: charginos and neutralinos}
\label{section:inos}
%%%%%--------------------------------------
In this subsection, we study the contributions of the charginos and 
neutralinos.  The chargino and neutralino mass matrices depend on four 
parameters, the two gaugino masses $M_1$ and $M_2$, the supersymmetric 
Higgs-mixing mass $\mu$ and $\tan\beta$.  
Because the gaugino masses and the Higgs-mixing mass  are invariant 
under the electroweak gauge symmetry, they contribute neither to the 
$S$ nor the $T$ parameter.  
%%%---------------------------------- 
%%%	a new paragraph

%%%	a new paragraph
%%%---------------------------------- 
On the other hand, because they are fermions, their virtual creation 
can affect the running of the gauge-boson propagator, the $R$ parameter 
(\ref{eq:running_of_gzbar}), strongly when the pair-creation threshold 
is near the $Z$-boson pole \cite{hhkm94,zpair}. 
%%%--------------------------------------------------
\begin{figure}[t]
\begin{center}
%%%\leavevmode\psfig{figure=fig_inos02.eps,width=10cm}
\leavevmode\psfig{figure=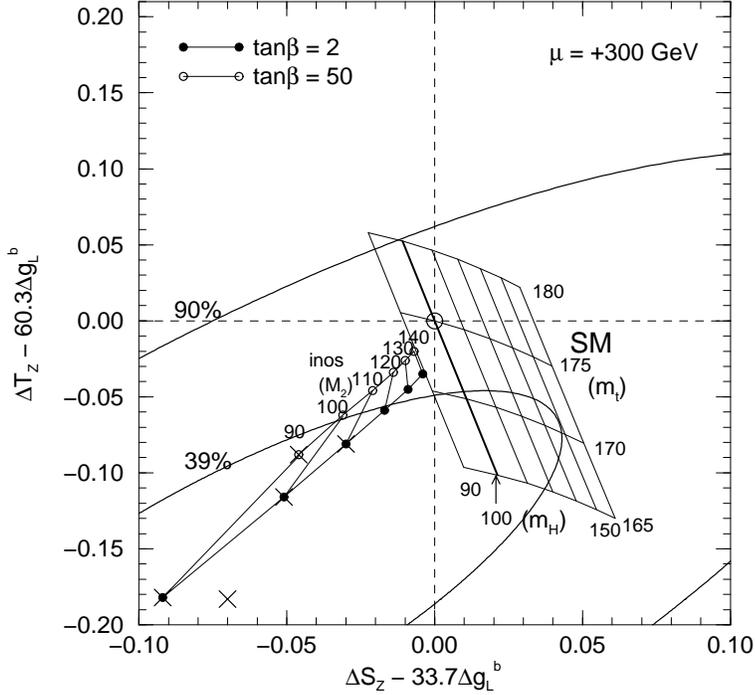,width=10cm}
\end{center}
\caption{The chargino and neutralino contributions to $(\dsz,\dtz)$ 
for $\tan\beta = 2$ and 50. 
The supersymmetric Higgs-mixing mass $\mu$ is fixed at 
$\mu =300\gev$. 
The GUT relation for the gaugino masses, 
$M_2/\hat{\alpha}_2 = M_1/\hat{\alpha}_1$ is assumed, 
and cases for $M_2 = 140 \sim 90\gev$ are shown by using the marked 
SM reference point at $(\mt,\mh)=(175,100)$ in GeV unit as the origin.  
Those points with the cross ($\times$) symbols give the lighter 
chargino mass $m_{\chargino{1}}$ below $90\gev$. 
}
\label{fig:oblique:inos}
\end{figure}
%%%--------------------------------------------------
\begin{figure}[t]
\begin{center}
\leavevmode\psfig{figure=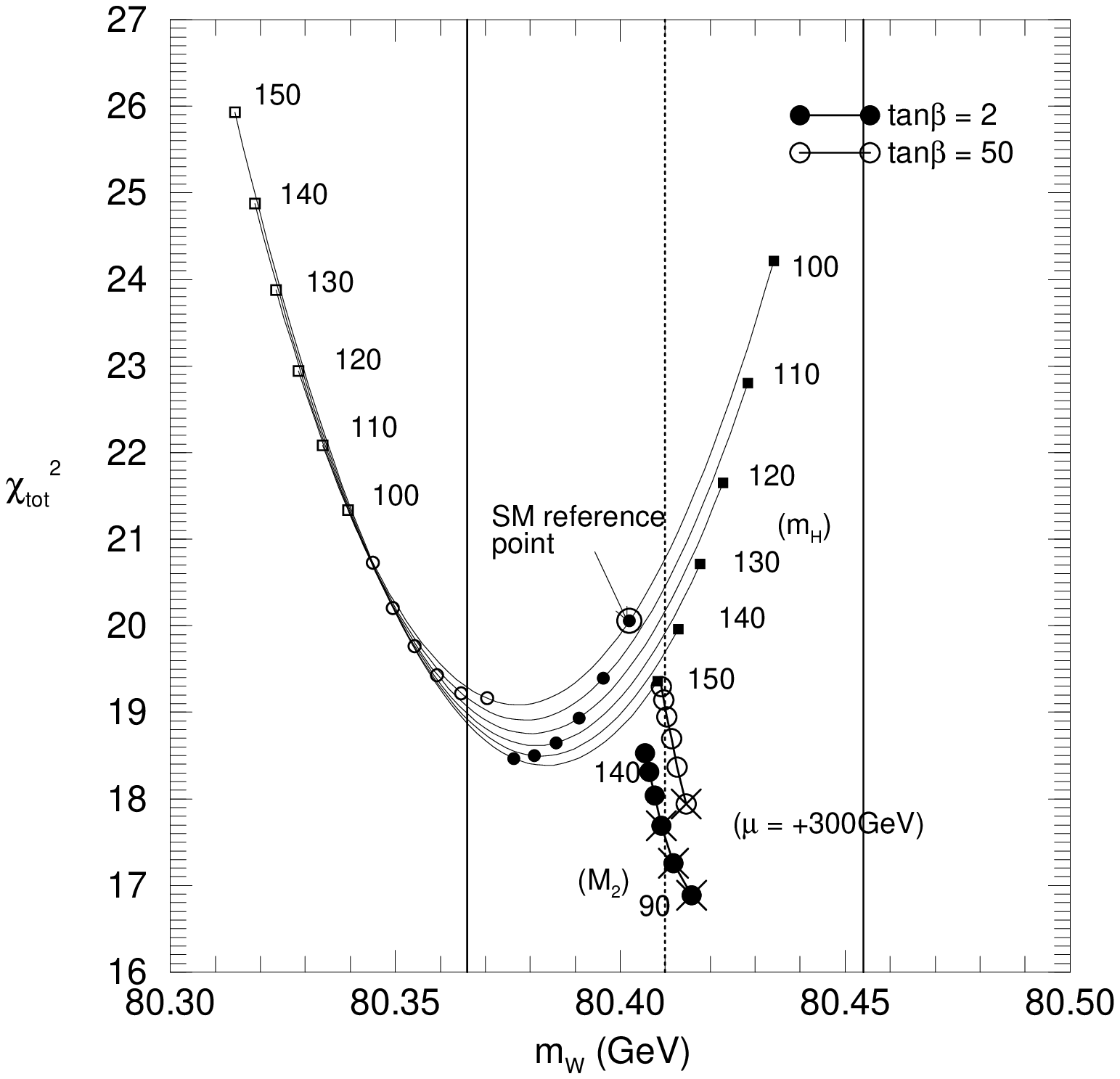,width=10cm}
\end{center}
\caption{The chargino and neutralino contributions to 
$(\mw,\chi^2_{\rm tot})$ for $\tan\beta = 2$ and 50. 
The supersymmetric Higgs-mixing mass $\mu$ is fixed by 
$\mu = 300\gev$. 
The GUT relation for the gaugino masses, 
$M_2/\hat{\alpha}_2 = M_1/\hat{\alpha}_1$ 
is assumed. 
The 6 blobs are for $M_2({\rm GeV})=140,130,120,110,100,90$ 
at $\tan\beta=2$ (solid blobs) and at $\tan\beta=50$ (open blobs). 
Those points with the cross $(\times)$ symbols indicate that the 
lightest chargino mass $m_{\chargino{1}}$ is smaller that 
$90\gev$. 
}\label{fig:mw:inos}
\end{figure}
%%%---------------------------------- 
%%%	a new paragraph

%%%	a new paragraph
%%%---------------------------------- 
%%%--------------------------------------------------
%%%
%%%  Table: chargino/neutralino contributions 
%%%         to the oblique parameters for 
%%%         \tan\beta = 2 and \mu = 200,300 GeV
%%%
%%%--------------------------------------------------
\begin{table}[t]%[htbp]
\begin{center}
\begin{tabular}{rcccccccc}  \hline \hline
$\mu$ & $M_2$ & $m_{\chargino{1}}$ 
& $m_{\neutralino{1}}$ & $\dr $ & $\dsz$& $\dtz$& $\dmw$ 
& $\chi^2_{\rm tot}$
\\ \hline \hline 
\vspace{0.1cm}
+300 & 110 & 85.0 & 45.9 & $-0.070$ & $-0.030$ & $-0.081$ &$\hph 
0.007$ & 17.7\\
\vspace{0.1cm}
     & 120 & 93.8 & 50.9 & $-0.054$ & $-0.017$ & $-0.059$ &$\hph 
0.006$ & 18.0\\
\vspace{0.1cm}
     & 130 & 103  & 55.9 & $-0.044$ &$-0.009$  & $-0.045$ &$\hph
0.004$ & 18.3\\
\vspace{0.1cm}
     & 140 & 111  & 60.8 & $-0.036$ &$-0.004$  & $-0.035$ &$\hph
0.004$ & 18.5\\
\hline
\vspace{0.1cm}
$-300$ &\hpz70& 82.2 & 38.1 & $-0.082$ &$-0.058$&$-0.104$&$\hph 
0.010$ & 17.8\\
\vspace{0.1cm}
       &\hpz80& 91.7 & 43.2 & $-0.062$ &$-0.041$&$-0.076$&$\hph 
0.008$ & 18.2\\
\vspace{0.1cm}
       &\hpz90& 101  & 48.3 & $-0.049$ &$-0.030$&$-0.058$&$\hph 
0.007$ & 18.5\\
\vspace{0.1cm}
       & 100  & 111  & 53.3 & $-0.040$ &$-0.023$&$-0.046$&$\hph 
0.006$ & 18.8\\
\hline \hline
\vspace{0.1cm}
+105    & 300  & 80.5 & 62.4 & $-0.045$ &$\hph0.009$&$-0.057$&$-0.007$& 17.7\\
\vspace{0.1cm}
+115    & 300  & 89.4 & 69.8 & $-0.036$ &$\hph0.013$&$-0.043$&$-0.006$& 18.0\\
\vspace{0.1cm}
+127    & 300  & 100  & 78.7 & $-0.028$ &$\hph0.014$&$-0.032$&$-0.005$& 18.2\\
\vspace{0.1cm}
+137    & 300  & 109  & 85.4 & $-0.024$ &$\hph0.015$&$-0.026$&$-0.004$& 18.4\\
\hline 
\vspace{0.1cm}
$-68$  & 300  & 80.2 & 64.2 & $-0.056$ &$-0.041$&$-0.071$&$\hph 
0.007$& 18.4\\
\vspace{0.1cm}
$-78$  & 300  & 89.8 & 73.9 & $-0.042$ &$-0.029$&$-0.051$&$\hph 
0.006$& 18.8\\
\vspace{0.1cm}
$-89$  & 300  & 101  & 84.7 & $-0.032$ &$-0.021$&$-0.037$&$\hph 
0.005$& 19.1\\
\vspace{0.1cm}
$-99$  & 300  & 110  & 94.4 & $-0.027$ &$-0.016$&$-0.028$&$\hph 
0.005$& 19.3\\
\hline
\end{tabular}
\caption{  
Chargino and neutralino contributions to the oblique parameters, 
$\dr, \dsz, \dtz, \dmw$ and the total $\chi^2$ of the fit 
at $\tan\beta = 2$. }
\end{center}
\label{table:oblique:inos:one}
\end{table}
%%%%----------------------------------------------------
Although the present lower mass bound on the charginos is as 
large as the $Z$-boson mass rather than its half, we find that 
the effect can still be significant.  
We show in Fig.~\ref{fig:oblique:inos} the contributions of 
the charginos and neutralinos to the $\dsz$ and $\dtz$ parameters.  
In the figure, the supersymmetric Higgs mass is fixed at 
$\mu=300\gev$, and the SU(2)$_L$ gaugino mass $M_2$ has been varied.  
The U(1)$_Y$ gaugino mass is scaled by the relation 
$M_1/M_2=\hat{\alpha}_1/\hat{\alpha}_2$ for definiteness.  
The case for $\tan\beta=2$ are shown by solid blobs, and 
those for $\tan\beta=50$ are shown by open blobs.  
As the $M_2$ decreases, the lightest chargino mass decreases, 
and those points with the cross symbol give the chargino mass  
below the direct search bound (\ref{eq:inos_bounds_lep2}).  
%%%---------------------------------- 
%%%	a new paragraph

%%%	a new paragraph
%%%---------------------------------- 
Most importantly, we find that the effect of relatively light 
charginos and neutralinos gives {\it both } $\dsz$ {\it and } 
$\dtz$ negative, which improves the fit over the SM. 
This is essentially because of the $R$ parameter that reside 
both in the $\dsz$ and the $\dtz$ parameters (\ref{eq:sztzdr}).  
For instance, if the wino-like chargino mass is near to half 
the $Z$-boson mass, its contribution to the parameter $R$ behaves as 
%%%---------
\bea
\dr &\approx& -\chat^4 
	\biggl( \frac{1}{\beta} -\frac{16}{3\pi} \biggr),
\label{eq:dr_wino}
\eea
%--------- 
where $\chatsq$ is given by $\chatsq = 1-\shatsq$. 
$\beta = \sqrt{4M_2^2/\mzsq-1}$ is the analytic 
continuation of the chargino velocity below the threshold.  
When $4M_2^2/\mzsq \gg 1$, the effect is suppressed as 
%---------
\bea
\dr &\approx& 
	-\frac{4\chat^4}{15\pi}\, \frac{\mzsq}{M_2^2}
	\biggl[1 +O\biggl( \frac{\mzsq}{M_2^2} \biggr) \biggr] . 
\label{eq:dr_wino_approx}
\eea
%---------
The chargino contribution to $\dr$ is negative just like the sfermion 
contribution to $\dr$ in eq.~(\ref{eq:dr_sfermion}).  
The coefficient of the wino contribution in eq.~(\ref{eq:dr_wino_approx}) 
is found to be about 90 times larger than that of the $\slepton_R$ 
contribution.   
This large negative contribution to $\dr$ makes both 
$\dsz$ and $\dtz$ significantly negative when a relatively light 
chargino of mass about $100\gev$ exists.  
%%%----------------------------------
%%%	a new paragraph

%%%	a new paragraph
%%%----------------------------------
%%%--------------------------------------------------
%%%
%%%  Table: chargino/neutralino contributions 
%%%         to the oblique parameters for 
%%%         \tan\beta = 50 and \mu = 200,300 GeV
%%%
%%%--------------------------------------------------
\begin{table}[ht]%[htbp]
\begin{center}
\begin{tabular}{rcccccccc}  \hline \hline
$\mu$ & $M_2$ & $m_{\chargino{1}}$
& $m_{\neutralino{1}}$ & $\dr$ & $\dsz$& $\dtz$& $\dmw$ 
& $\chi^2_{\rm tot}$
\\ \hline \hline 
\vspace{0.1cm}
+300 &\hpz90& 82.9 & 43.8 & $-0.077$&$-0.046$&$-0.089$& 0.013 & 17.9\\
\vspace{0.1cm}
     & 100 & 92.1 & 48.7 & $-0.059$&$-0.031$& $-0.063$ & 0.011 & 18.4 \\
\vspace{0.1cm}
     & 110 & 101  & 53.6 & $-0.047$&$-0.021$&$-0.046$& 0.009 & 18.7 \\
\vspace{0.1cm}
     & 120 & 110  & 58.4 & $-0.039$&$-0.014$&$-0.034$& 0.008 & 18.9 \\
\hline
$-300$ &\hpz90& 84.6 & 44.4 & $-0.074$&$-0.044$&$-0.084$& 0.012 & 18.0 \\
\vspace{0.1cm}
       & 100 & 93.8 & 49.3 & $-0.057$& $-0.030$ & $-0.060$ & 0.011 & 18.4 \\
\vspace{0.1cm}
       & 110 & 103  & 54.1 & $-0.045$&$-0.020$&$-0.044$& 0.009 & 18.8 \\
\vspace{0.1cm}
       & 120 & 112  & 59.0 & $-0.038$&$-0.014$&$-0.033$& 0.008 & 19.0 \\
\hline \hline 
+87  & 300 & 80.2 & 64.7 & $-0.049$&$-0.011$&$-0.054$& 0.003 & 18.1\\
\vspace{0.1cm} 
+98  & 300 & 90.5 & 74.0 & $-0.037$&$-0.004$&$-0.036$& 0.003 & 18.5\\
\vspace{0.1cm} 
+108 & 300 & 99.7 & 82.2 & $-0.030$&$-0.001$&$-0.026$& 0.003 & 18.8\\
\vspace{0.1cm} 
+120 & 300 & 110  & 91.0 & $-0.025$&$+0.001$&$-0.018$& 0.004 & 19.1\\
\hline 
\vspace{0.1cm} 
$-85$  & 300 & 79.8 & 64.5 & $-0.050$&$-0.013$&$-0.056$& 0.003 & 18.1\\
\vspace{0.1cm} 
$-96$  & 300 & 90.2 & 74.0 & $-0.038$&$-0.006$&$-0.037$& 0.004 & 18.5\\
\vspace{0.1cm} 
$-107$  & 300 &100  & 83.2 & $-0.030$&$-0.002$&$-0.026$& 0.004 & 18.9\\
\vspace{0.1cm} 
$-117$  & 300 &110  & 91.2 & $-0.025$&$+0.000$&$-0.019$& 0.004 & 19.1\\
\hline
\end{tabular}
\caption{  
Chargino and neutralino contributions to the oblique parameters, 
$\dr, \dsz, \dtz, \dmw$ and the total $\chi^2$ of the fit 
at $\tan\beta = 50$. }
\end{center}
\label{table:oblique:inos:two}
\end{table}
%%%--------------------------------
More detailed numerical results are summarized in 
Table~4 for $\tan\beta=2$ and in Table~5 for $\tan\beta=50$. 
From both tables, we can see that $\dtz$ is always negative for these 
inputs and its magnitude is dominated by $\dr$, see 
eq.~(\ref{eq:dtz}). 
On the other hand, $\dsz$ and $\dmw$ can be positive or negative. 
As compared to the total $\chi^2$ value of 20.1 which is obtained 
at the SM reference value $(\mt({\rm GeV}),\mh({\rm GeV}),\alpha(\mzsq) 
=(175,100,129.90)$ under the $\alps(\mz)$ constraint 
(\ref{eq:alpha_s_pdg}), see eq.~(\ref{eq:chisq_oblique}), 
we find an improvement of $\chi^2_{\rm tot}$ in the tables 
by between 1.0 and 1.9, by including the contribution of a chargino 
if its mass is around $100\gev$, just above the present mass bound.  
%%%%%-----------------------------------
%%%%% Subsection: 4.4. Summary 
%%%%%-----------------------------------
\begin{figure}[t]
\begin{center}
%%%\leavevmode\psfig{figure=fig_st_total01.eps,width=10cm}
\leavevmode\psfig{figure=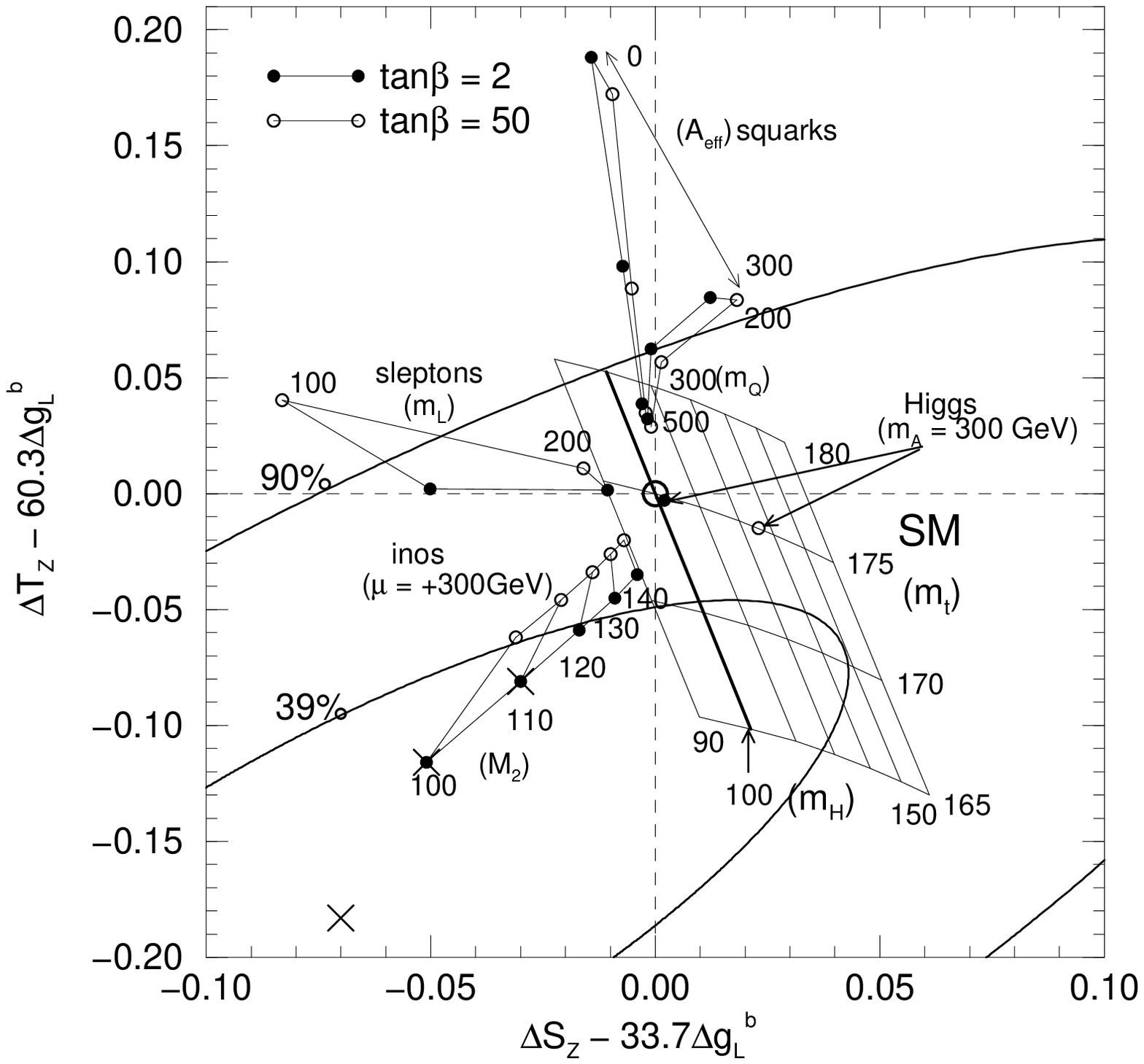,width=10cm}
\end{center}
\caption{Summary of the supersymmetric contributions to 
($\dsz,\dtz$). 
}
\label{fig:oblique:summary}
\end{figure}
%%%%%-----------------------------------
\subsection{Summary of oblique corrections} 
%%%%%-----------------------------------
We show in Fig.~\ref{fig:oblique:summary} the contributions 
of all the sectors of the MSSM together.  
The total oblique corrections are obtained by summing 
the SM contribution for one's favorite $\mt$ at $\mh=100\gev$, 
the MSSM Higgs sector contribution, the squark contribution, 
the slepton contribution, and the chargino-neutralino contribution 
vectorially in the plane.  
%%%---------------------------------- 
%%%	a new paragraph

%%%	a new paragraph
%%%----------------------------------
It is surprising that each sector of the MSSM gives distinctive 
contributions in the $\dsz$ and $\dtz$ plane.  
The squark sector contributes essentially to positive $\tz$ direction, 
which is strongly disfavored by the electroweak data.  
The slepton sector contributes negatively to $\sz$, but $\tz$ remains 
constant or slightly positive for large $\tan\beta$.  
This is also found to be disfavored by the data.  
Therefore, if only the squarks and sleptons are light, their 
contributions would make the fit to the electroweak data 
significantly worse than the SM.  
%%%---------------------------------- 
%%%	a new paragraph

%%%	a new paragraph
%%%----------------------------------
On the other hand, the MSSM Higgs sector is found to give the 
contribution very similar to that of the SM when the lightest 
CP-even Higgs-boson mass $\mhlight$ is taken to be the SM Higgs 
boson mass.  Therefore, if $\mhlight$ is significantly larger 
than the reference value of $100\gev$, either with large 
$\tan\beta$, or by having very heavy squark masses that affect 
the effective scalar potential, the fit improves slightly.  
%%%---------------------------------- 
%%%	a new paragraph

%%%	a new paragraph
%%%----------------------------------
Finally, the contribution of charginos and neutralinos make both 
$\dsz$ and $\dtz$ negative, and hence the fit improves more 
significantly if the lightest chargino mass is near the 
present direct search limit of around $100\gev$.  
%%%---------------------------------- 
%%%	a new paragraph

%%%	a new paragraph
%%%----------------------------------
As emphasized in section 2, the oblique corrections dominate the 
whole radiative effects if either squarks and sleptons, or 
the charginos and neutralinos are very heavy.  Otherwise, 
in addition to the oblique corrections summarized in this Figure, 
there will be additional vertex and box corrections.  
%%%%%--------------------------------------
%%%%%
%%%%%
%%%%%	Section: Non-oblique corrections
%%%%%
%%%%%
%%%%%--------------------------------------
\clean
\section{Non-oblique corrections}
\label{section:nonoblique}
%%%---------------------------------- 
We found in the previous section that the existence of relatively 
light charginos (of mass $\sim 100\gev$) improves the SM fit to 
the electroweak data, whereas the existence of relatively light 
squarks and sleptons (of mass $\sim$ $100\gev$) makes the fit 
significantly worse.  
On the other hand, if both supersymmetric fermions (charginos 
and neutralinos) and supersymmetric scalars (squarks and sleptons) 
are light, then in addition to the universal gauge-boson-propagator 
corrections (the oblique corrections), we should expect process 
specific vertex and box corrections to be non-negligible.  
We would like to address this problem in this section.  
%%%---------------------------------- 
%%%	a new paragraph

%%%	a new paragraph
%%%---------------------------------- 
There are just two types of process specific corrections which 
are relevant for the precision electroweak experiments on the 
$Z$-boson properties and the $W$-boson mass.  
One is the vertex corrections to various $Z \to f\ov{f}$ 
amplitudes, the parameters $\dgfa$, in eqs.(\ref{eq:amp_sm}). 
Even if we assume the universality of the first two generation 
squarks and sleptons, there are 14 independent $Zff$ vertex 
corrections in the MSSM. 
The other is the vertex and box corrections to the muon-decay amplitude, 
the parameter $\ddelg$, that affect the $\dtz$ parameter (\ref{eq:dtz}) 
in the $Z$-boson experiments and the $\dmw$ parameter (\ref{eq:dmw}) 
for the $W$-boson mass.  Although this is just a single correction 
factor, it affects all the electroweak predictions because we use 
the observed magnitude of the muon decay constant $G_F$ as one of 
the basic inputs of our calculation.  
%%%---------------------------------- 
%%%	a new paragraph

%%%	a new paragraph
%%%----------------------------------
We present our studies on these non-oblique corrections step by 
step in the increasing order of the number of observables that 
will be affected.   
In section \ref{section:zff_higgs} we study the MSSM Higgs boson 
contribution to the $\zbb,\ztautau$ and $Z \nu_\tau \nu_\tau$ 
vertices.  
In section \ref{section:susy_qcd}, we study the order $\alps$ 
vertex correction when gluinos and squarks are both light.  
This will affect the $\zqq$ vertices only, and we find it useful 
to study its quantitative significance independently of the other 
electroweak corrections.  
In section \ref{section:zqq_squark}, we study the electroweak 
corrections to the 
$\zqq$ vertices when both squarks and charginos/neutralinos 
are light.  Effects on the $Zbb$ vertices are carefully studied. 
In section \ref{section:zll_slepton}, 
we study the $\zll$ vertices when both sleptons 
and charginos/neutralinos are light.  
We notice, however, if both the supersymmetric fermions 
(charginos and neutralinos) and the sleptons are light enough 
to affect the $\zll$ vertices significantly, they should also affect 
the muon-decay amplitude, $\ddelg$, and hence all the remaining 
electroweak observables.  
All the numerical results in the following subsections are found 
at $(\mt({\rm GeV}),\alps(\mz),\alpha(\mzsq))=(175, 0.118, 128.90)$. 
%%%%--------------------------------------------------------------
\subsection{$Zff$-vertex corrections by the MSSM Higgs bosons }
\label{section:zff_higgs}
%%%%--------------------------------------------------------------
\begin{figure}[t] 
\begin{center}
\leavevmode\psfig{figure=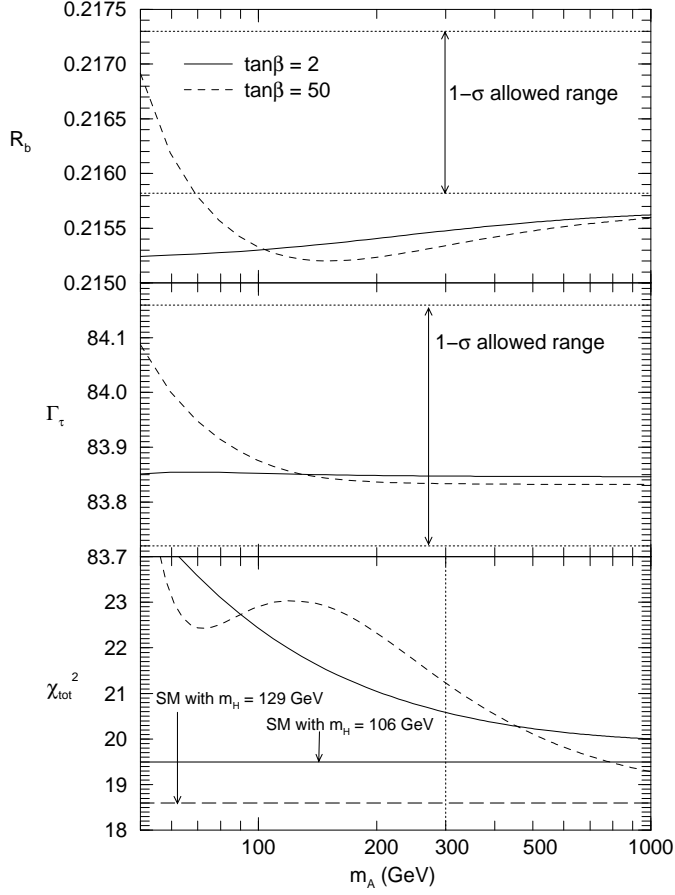,height=12cm}
\end{center}
\caption{{\small
	The Higgs boson contributions to $R_b$ (top), $\Gamma_\tau$ 
	(middle) and the total $\chi^2$ (bottom) as functions of 
	$\mpseudo$ for $\tan\beta = 2$ (solid line) and 50 (dashed line). 
	The squarks, sleptons and lighter chargino masses are 
	fixed by 1\tev. 
	The other parameters, $\mt, \alps(\mz)$ and $\alpha(\mzsq)$ 
	are fixed by $175\gev, 0.118$ and $128.90$, respectively. 
	The 1-$\sigma$ allowed range of $R_b$ and $\Gamma_\tau$	
	($83.94 \pm 0.22 \mev$~\cite{lepewwg98}) are shown in 
	the top and middle graphs, respectively. 
	Two thick lines in the bottom graph denote the SM fit for 
	$\mh=106\gev$(solid) and $\mh=129\gev$(dashed) which are 
	the lightest Higgs boson mass in the MSSM at 
	$\mpseudo=1\tev$ for $\tan\beta =2$ and 50, respectively. 
	}}
\label{fig:higgs_vtx}
\end{figure}
%%%%--------------------------------------------------------------
When $\tan\beta$ is very large, the Yukawa couplings of the $b$-quark 
and the $\tau$-lepton get large, and we expect significant vertex 
corrections for both $\zbb$ and $\ztautau$ vertices.  
The most significant contributions are found to arise from the 
vertices with the Higgs bosons.  
In Fig.~\ref{fig:higgs_vtx}, we show $R_b$, $\Gamma_\tau$ and 
the total $\chi^2$ as functions of $\mpseudo$, the pseudo-scalar 
Higgs-boson mass in our Higgs potential.  
The $\tan\beta=2$ case is given by the solid lines, whereas 
the $\tan\beta=50$ case is given by the dashed lines.  
For simplicity, we set all squarks and slepton masses as well 
as the lightest chargino mass to be 1~TeV so that their contribution 
can be neglected.  
%%%---------------------------------- 
%%%	a new paragraph

%%%	a new paragraph
%%%----------------------------------
We find that for sufficiently small $\mpseudo$ at large $\tan\beta$, 
there is a significant positive contribution to both $\Gamma_b$ 
and $\Gamma_\tau$.  These effects have been 
noted~\cite{hisano_kiyoura_murayama} when the 
experimental data on $R_b$ larger than the SM prediction was 
reported~\cite{rb1995}. 
We find that the fit to the latest $R_b$ data is slightly better with 
larger $R_b$ predicted in the MSSM with $\tan\beta=50$ and 
$m_A\sim 60\gev$, but as we can see from the total $\chi^2$ behavior, 
the scenario is not favored over the SM.  
This is because the MSSM predicts light Higgs particles whose 
contributions to the oblique parameters worsen the fit, as shown 
in Table~3 in Section~\ref{section:oblique}. 
Also the pseudo-scalar Higgs-boson mass $\mpseudo$ significantly 
below $100\gev$ is not acceptable because it gives the lightest 
CP-even Higgs boson mass ($\mhlight$) below the direct search 
bound (\ref{eq:bound_higgs}).
%%%---------------------------------- 
%%%	a new paragraph

%%%	a new paragraph
%%%---------------------------------- 
In Fig.~\ref{fig:higgs_vtx} we show by horizontal lines the 
$\chi^2_{\rm tot}$ of the SM prediction at $\mt=175\gev$ and 
$1/\alpha(\mzsq)=128.90$ for $\mh=106\gev$ and $129\gev$. 
These Higgs boson masses correspond to the masses of the lightest 
CP-even Higgs boson ($h$) in the MSSM at $\tan\beta=2$ and 50, 
respectively. 
Table~3 shows that the oblique corrections from 
the MSSM Higgs sector reduces to the SM predictions at the corresponding 
Higgs boson mass, $\mh=\mhlight$, already at $\mpseudo=300\gev$. 
In contrast, Fig.~\ref{fig:higgs_vtx} shows that the vertex corrections 
can affect the total $\chi^2$ significantly even at $\mpseudo=300\gev$, 
especially for $\tan\beta=50$. 
This is mainly because of the enhanced $bbH$ and $\tau\tau H$ couplings 
at large $\tan\beta$, where $H$ is the heavier of the CP-even Higgs 
bosons with mass $m_H^{} \approx m_A^{}$.  
%%%%%---------------------------------------------
%%%%% Subsection: 5.1. SUSY QCD corrections
%%%%%---------------------------------------------
\begin{figure}[t]%[p]
\begin{center}
\leavevmode\psfig{figure=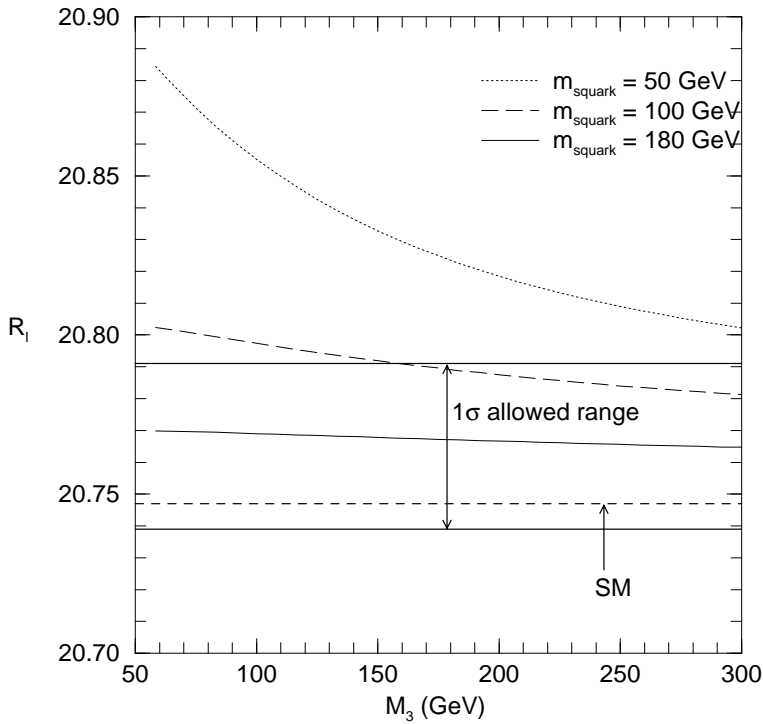,width=10cm}
\end{center}
\caption{The SUSY-QCD correction to $R_\ell$ as a function of the 
gluino mass $M_3=M_{\gluino}$. All the squark masses are assumed 
to be given by the universal scalar mass, $m_{\rm squark}$. 
}
\label{fig:vertex:sqcd}
\end{figure}
%%%----------------------------------
\subsection{SUSY-QCD corrections }
\label{section:susy_qcd}
%%%----------------------------------
Supersymmetric QCD corrections to the $Z$-boson hadronic width, 
$\Gamma_h$, or its ratio $R_\ell$ to the leptonic width $\Gamma_\ell$ 
is one of the first vertex corrections to the $Z$-boson decay 
amplitudes calculated \cite{hm90}. 
They contribute at the order $\alps$ level, and hence could 
be significant if both squarks and gluinos are light.  
We would like to first examine their quantitative significance 
in view of the present lower mass bounds on squarks and 
gluinos~(\ref{eq:tevatron_bound}). 
%%%%%-------------------------------------
%%%%%
%%%%%	SUSY QCD correction to R_\ell
%%%%%
%%%%%-------------------------------------
	\begin{table}[t]%[htbp]
	\begin{center}
	\begin{tabular}{c|ccc|cc|cc}
	\hline	\hline
%%%------------------
	& \multicolumn{3}{c|}{Inputs (GeV)}
	& \multicolumn{2}{c|}{$O(\alps)$ vertex} 
	& \multicolumn{2}{c}{vertex + obl.} \\
\hline 
	& $m_{\sup_L}$ & $m_{\sup_R}$ & $M_{\gluino}$
	& $R_\ell/R_\ell(\smr)$ & $\chi^2_{\rm SQCD}$
	& $R_\ell/R_\ell(\smr)$ & $\chi^2_{\rm tot}$
\\ \hline 
%%%-------------------------------------------
I	& 100 & 100 & 100 & 1.00152 &  24.1 & 1.00241 &   129.1 \\
II 	& 180 & 180 & 180 & 1.00048 &  20.5 & 1.00095 &\hpz52.9 \\
III 	& 1000& 180 & 180 & 1.00003 &  19.9 & 1.00005 &\hpz20.5 \\
IV 	& 1000& 180 & 421 & 1.00001 &  19.8 & 1.00004 &\hpz20.4 \\
\hline \hline 
%%%------------------
	\end{tabular} 
	\end{center}
	\caption{The SUSY-QCD corrections to $R_\ell/R_\ell(\smr)$ 
	and $\chi^2$ for $\tan\beta = 2$. 
	In all cases, $m_{\sup_R} = m_{\sdown_R}$ is assumed. 
	The universality of the SUSY breaking scalar masses among 
	the different generations, 
	$m_{\wt{Q}}=m_{\wt{Q}_3}$, $m_{\wt{U}}=m_{\wt{U}_3}$ and 
	$m_{\wt{D}}=m_{\wt{D}_3}$ is also assumed. 
	The left-right mixings in the stop and sbottom 
	sector are taken to be zero by using an appropriate values of 
	$A_t,A_b$ and $\mu$. 
	The effects of the SUSY-QCD vertex correction for four cases 
	are summarized in the second column while 
	the SUSY-QCD vertex correction and the oblique correction by 
	squarks are given in the last column. 
	Only in IV, the GUT relation $M_{\gluino} = M_3 = 
	(\hat{\alpha}_3/\hat{\alpha}_2) M_2$ has been 
	respected. 
	The value of $M_{\gluino}$ in IV leads to $M_2 = 120\gev$, and 
	$m_{\chargino{1}} = 94\gev$ for $\mu= 300\gev$. 
	}	
	\label{table:susyqcd}
	\end{table}
%%%---------------------------------- 
%%%	a new paragraph

%%%	a new paragraph
%%%----------------------------------
The one-loop $Zff$ amplitudes of Appendix C, which is 
presented by using a generic notation that apply for all the 
vertices and for all the MSSM corrections, take a particularly 
simple form when only the order $\alps $ corrections are 
considered.  For instance, if the mixing between the left and 
right chirality squarks is negligible, which we assume to be 
the case for the first two family squarks and sleptons, we find 
%%%---------------------------------- 
\bea
\Del g_\alpha^q &=& 
	\frac{g^{qqZ}_\alpha \alps }{\sqrt{4\rttwo \gf \mzsq}} 
	\frac{32\pi}{3} 
	\biggl[
	\biggl( B_0 + B_1 \biggr) (0:m_{\wt{q}_\alpha}, M_{\gluino}) 
\nonumber \\
&&~~~~~~~~~~~~~~
	- 2C_{24} (\mzsq: m_{\wt{q}_\alpha}, M_{\gluino}, 
	m_{\wt{q}_\alpha})
	\biggr], 
\eea
%%%---------------------------------- 
where the two- and three-point functions, $B_0, B_1$ and $C_{24}$ 
are given in ref.~\cite{hhkm94}. 
%------------
We reproduced the results of refs.~\cite{hm90}. 
%%%---------------------------------- 
%%%	a new paragraph

%%%	a new paragraph
%%%----------------------------------
In Fig.~\ref{fig:vertex:sqcd} we show the SUSY-QCD correction to 
the ratio $R_\ell$ as functions of the gluino mass 
$M_3=M_{\gluino}$ when all the squark masses are taken to be the 
same for brevity.  
Significantly larger magnitude of $R_\ell$ as compared to the 
SM prediction at the same $\alps$ is found only when the common 
squark mass is as light as $50\gev$ and the gluino is lighter than 
$100\gev$, consistent with the early observation~\cite{hm90}.  
Once we take into account the present lower mass bound of squarks 
and gluinos~(\ref{eq:tevatron_bound}), the quantitative significance 
of the SUSY-QCD correction diminishes.  
%%%---------------------------------- 
%%%	a new paragraph

%%%	a new paragraph
%%%----------------------------------
We show in Table~\ref{table:susyqcd} the magnitude of the SUSY-QCD 
correction to $R_\ell$ for several squark and gluino mass inputs. 
We should compare the percentage corrections with the present 
experimental accuracy of $R_\ell$, which is 0.125\% from 
Table~\ref{tab:ewdata98}. 
Only when all the squarks and gluino masses are around $100\gev$,
we can expect non-negligible effect from this sector.  
When all the masses are moved up to $180\gev$, consistent with the present 
direct search limits~(\ref{eq:tevatron_bound}), the effect on $R_\ell$ 
is about $0.05\%$ and is already insignificant.  
Our studies on the oblique correction suggest that the left-handed 
squarks should be rather heavy not to worsen the electroweak fit.  
If only the right-handed squarks are kept light at $180\gev$, together 
with the gluino of the same mass, the effect essentially goes away, 
as shown for the case III in Table~\ref{table:susyqcd}. 
In the last line of the Table, we study the case IV where  
$M_{\gluino}=M_3=421\gev$, which is the mass obtained by using the 
`unification' relation $M_3/M_2=\hat{\alpha}_3/\hat{\alpha}_2$ 
in one of our selected MSSM points that accommodate a chargino of 
mass around $100\gev$.  The effect diminishes to $10^{-4}$. 
%%%----------------------------------  
%%%	a new paragraph

%%%	a new paragraph
%%%----------------------------------
In the following analysis, we assume that the left-handed squarks 
are relatively heavy ($m_{\wt{q}_L}\simgt 1\tev$) in order not to 
spoil the good fit to the electroweak data.  
We also adopt the `unification' relation for the three gaugino masses 
$M_3/\hat{\alpha}_3= M_2/\hat{\alpha}_2= M_1/\hat{\alpha}_1$ 
in all the subsequent numerical examples.  
Under such conditions, we find that the SUSY-QCD corrections 
cannot have significant effect on our electroweak studies.  
%%%%%-------------------------------------------------------------------
%%%%% Subsection: 5.2. Electroweak $Zqq$ vertex corrections with squarks
%%%%%-------------------------------------------------------------------
\subsection{Electroweak $Zqq$ vertex corrections with squarks }
\label{section:zqq_squark}
%%%%%-------------------------
In this subsection we explore the possibility that the electroweak 
SUSY corrections to the $\zqq$ vertices improve the overall fit to 
the data.  
As in the case of the SUSY-QCD corrections with a gluino-squark loop, 
we find that $m_{\wt{q}_L} \approx m_{\stop_L}$ should be kept 
high ($\simgt 1\tev$) in order for their positive $T$ contribution not 
to spoil the good overall fit of the SM.  
%%%---------------------------------- 
%%%	a new paragraph

\begin{figure}[t]
\begin{center}
\leavevmode\psfig{figure=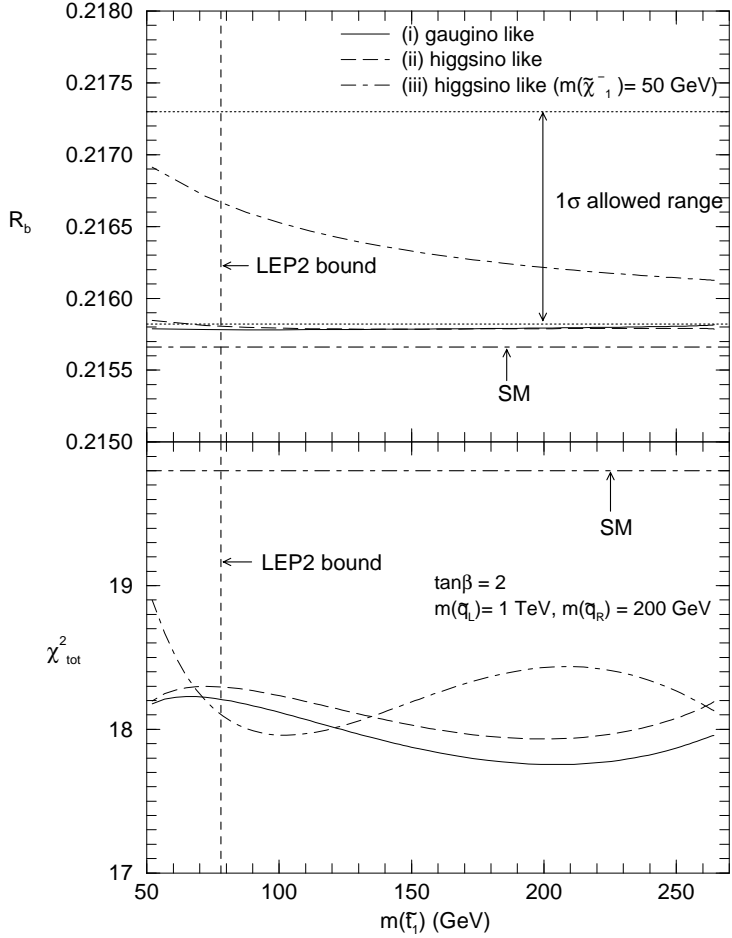,width=10cm}
\end{center}
\caption{
	The stop contribution to $R_b$ (top) and the total $\chi^2$ 
	(bottom) as functions of $m_{\stop_1}$ for 
	$m_{\wt{Q}} = 1\tev$ and $m_{\wt{U}} = m_{\wt{D}}= 200\gev$ 
	where the universality in the generation space of the squark 
	sector is assumed. 
	Three different inputs for the chargino mass 
	are studied: 
	(i) $(\mu,M_2)=(300,120)$, 
	(ii) $(\mu,M_2)=(100,800)$, 
	(iii) $(\mu,M_2)=(90,200)$ in the GeV unit. 
	All the other SUSY particle masses are fixed by $1\tev$. 
	The SM prediction, $R_b = 0.21566$ and $\chi^2 = 19.8$,  
	are also shown.  
	}
\label{fig:zbb}
\end{figure}

%%%	a new paragraph
%%%----------------------------------
The only case that SUSY electroweak correction to the $\zqq$ vertices  
are found to give nontrivial effects is when loops of light charginos 
and a light $\stop_1$ affect the $\zbb$ vertices. 
We show in Fig.~\ref{fig:zbb} the effect of supersymmetric vertex 
corrections to the ratio $R_b$. 
In order to avoid large unfavorable oblique corrections to the $T$ 
parameter, we set all the $\wt{q}_L$ masses large by setting  
$m_{\wt{Q}}=m_{\wt{Q}_3}=1\tev$.  
We take the singlet squark masses at 
$m_{\wt{U}}=m_{\wt{U}_3}=m_{\wt{D}}=m_{\wt{D}_3}=200\gev$ near the 
Tevatron search limit (\ref{eq:tevatron_bound}).  
The lighter stop, $\stop_1$, can still be light enough to affect the 
$\zblbl$ vertex by having large $\stop_L$-$\stop_R$ mixing.  
%%%---------------------------------- 
%%%	a new paragraph

%%%	a new paragraph
%%%----------------------------------
In the figure we show the present 1-$\sigma$ experimental bound on 
$R_b$, and the MSSM predictions as functions of the $\stop_1$ mass.  
The lower part of the figure shows the total $\chi^2$ of the fit 
that includes contributions from all the electroweak data 
in Table~\ref{tab:ewdata98}. 
Three cases with a light chargino $(m_{\chargino{1}} \simlt 100\gev)$ 
have been examined:
%%%----------------------------------
\begin{center}
\begin{tabular}{rll}
(i) & $\chargino{1}$ is almost wino & 
	($\mu=300\gev,M_2=120\gev, m_{\chargino{1}}=94\gev)$ 
\\
(ii)&	$\chargino{1}$ is almost higgsino & 
	($\mu=100\gev,M_2=800\gev, m_{\chargino{1}}=93\gev)$
\\
(iii)&	$\chargino{1}$ is almost higgsino &
	($\mu=90\gev,M_2=200\gev, m_{\chargino{1}}=54\gev$ )
\end{tabular}
\end{center}
%%%--------------------------------------------------------------
%%%--------------------------------------------------------------
The case (iii) violates the LEP chargino mass bound 
(\ref{eq:inos_bounds_lep2}).  
As is clear from these figures, the MSSM effects on the $\zbb$ vertices 
are not significant under the present constraints on the 
$\stop_1$ and $\chargino{1}$ masses.  
The scale of the total $\chi^2$, which is significantly 
below that of the SM, is set by the light chargino-neutralino contribution 
to the oblique parameters as we explained in the last section, and the 
slight decrease that we observe for the cases (i) and (ii) for lighter
$m_{\stop_1}$ is not significant.  
Only when both the $\stop_1$ and $\chargino{1}$ are as light as 
$50\gev$ and when the $\chargino{1}$ is almost higgsino and $\stop_1$ 
is almost $\stop_R$, we could obtain significant positive contribution 
to $R_b$ as discussed in the literatures a couple of years ago 
\cite{zbb_susy,yhm95}. 
We reproduced all the numerical results presented in ref.~\cite{yhm95}.  
%%%---------------------------------- 
%%%	a new paragraph

%%%	a new paragraph
%%%----------------------------------
With the above choice of squark parameters, $m_{\wt{Q}}=1\tev$ and 
$m_{\wt{U}}=m_{\wt{D}}=200\gev$, and with our neglect of all the 
left-right mixing in the first two generations, the electroweak vertex 
corrections to the four light-quark flavors are found to be negligibly 
small.  
%%%-------------------------
\subsection{$\zll$-vertex and $\mu$-decay corrections with sleptons }
\label{section:zll_slepton}
%%%-----------------------------
\begin{figure}[t]
\begin{center}
\leavevmode\psfig{figure=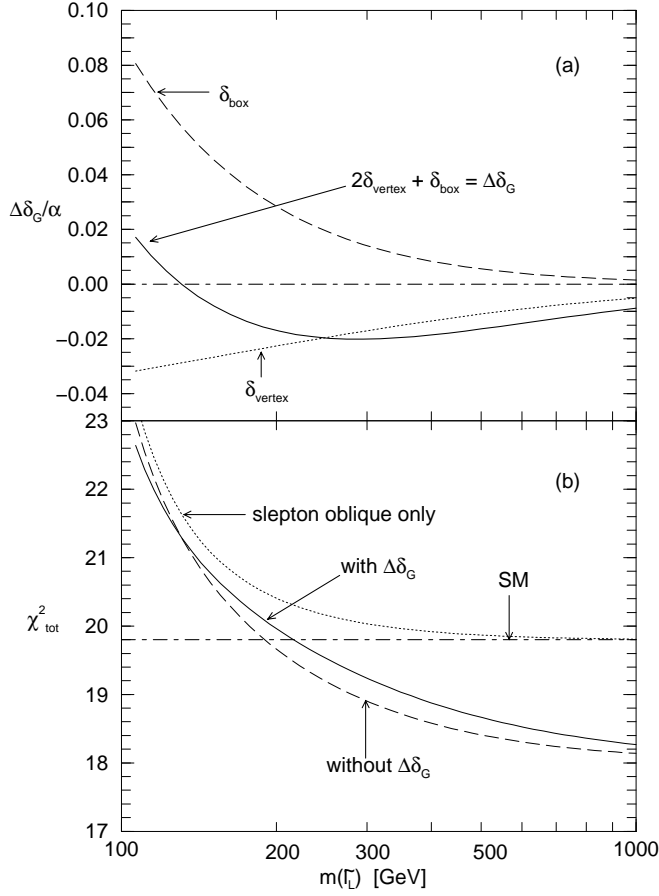,height=12cm}
\end{center}
\caption{{\small
	The MSSM contribution to $\ddelg/\alpha$
	(a) and the total $\chi^2$ (b) as a function of 
	the left-handed slepton mass, $m_{\wt{l}_L}$ 
	for $\tan\beta = 2$. 
	In the figures, $\mu$ and $M_2$ are fixed at $300\gev$ 
	and $120\gev$, respectively.
	The GUT relation for the gaugino masses are assumed. 
	In (a), the vertex and box corrections are shown by 
	dotted and dashed lines, respectively. The total 
	contribution, $2\delta_{\rm vertex} + \delta_{\rm box}$,  
	is given by the solid line. 
	In (b), the right-handed slepton mass is fixed by 
	$m_{\wt{l}_R} = 100\gev$. The masses of squarks and Higgs bosons 
	besides the lightest Higgs are fixed by 1\tev. 
	The solid and dashed lines show the total $\chi^2$ with 
	and without $\ddelg$, respectively. 
	}}
\label{fig:slepton-one}
\end{figure}
%%%---------------------------------- 
\begin{figure}[t]
\begin{center}
\leavevmode\psfig{figure=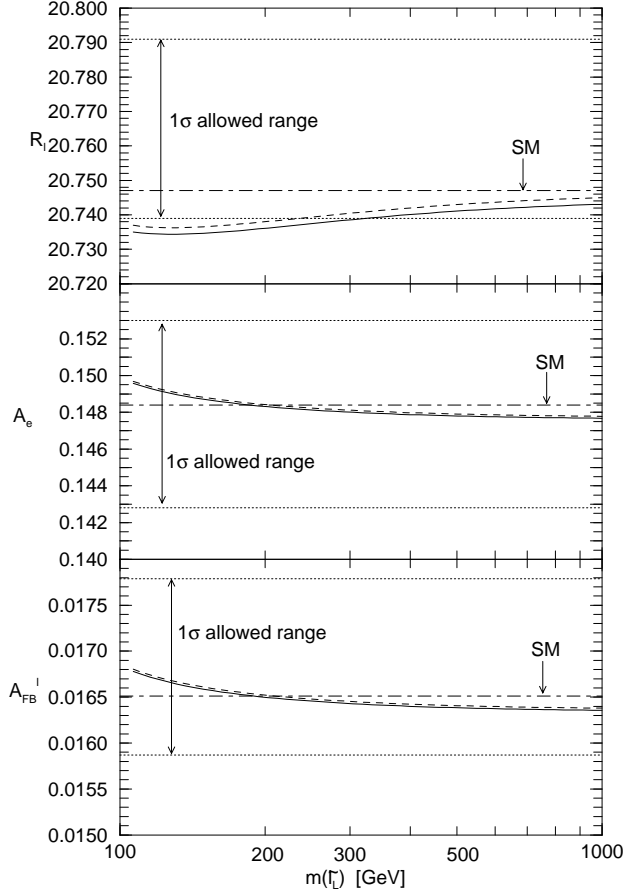,height=12cm}
\end{center}
\caption{{\small
	The scalar-lepton contributions to $R_\ell$ (top), $A_e$ (middle) 
	and $A_{\rm FB}^\ell$ (bottom) as functions of $m_{\wt{l}_L}$ 
	for $\tan\beta=2$. In each figure, the impact of the right-handed 
	slepton mass is shown by solid ($m_{\wt{l}_R}=100\gev$) 
	and dashed ($m_{\wt{l}_R}=1\tev$) lines, respectively.  
	The other parameters are the same with those in 
	Fig.~\ref{fig:slepton-one}. 
	}}
\label{fig:slepton-two}
\end{figure}
%%%-------------------------
So far, we set all the slepton masses to infinity so that they 
do not participate in the MSSM corrections.  
When sleptons are relatively light, together with relatively light 
charginos and neutralinos as motivated by their favorable contributions 
to the oblique parameters $\dsz$ and $\dtz$, we expect two distinctive 
effects in the electroweak predictions of the MSSM.  
First, they affect the muon-decay amplitude and hence the parameter 
$\ddelg/\alpha$, in eq.~(\ref{eq:gf}), 
%$\gf = \frac{\gwbarsq(0) + \ghatsq \delg}{4\sqrt{2}\mwsq}$, 
that appear in the basic expression for the oblique 
parameter $\dtz$ (\ref{eq:dtz}) and the $W$-boson mass (\ref{eq:dmw}). 
Second, they affect the $\zll$ vertices, and hence the parameters, 
$\dgll, \dglr, \dgnll$ in eqs.(\ref{eq:amp_sm}). 
%%%---------------------------------- 
%%%	a new paragraph

%%%	a new paragraph
%%%----------------------------------
We first examine the MSSM contributions to the muon-decay amplitude 
carefully, because if they can make $\ddelg/\alpha$ positive 
$\dtz$ can also become negative: see eq.~(\ref{eq:dtz}).  
A combination of this negative $\dtz$ from the muon-decay correction 
and the negative $\dsz$ from the oblique corrections (see Fig.~3) 
could improve the overall fit to the $Z$-pole data.  
If the muon-decay amplitude receives positive contribution from 
new physics, then it is equivalent to make the effective charged 
current strength larger, and the $T$ parameter, which is a measure 
of the neutral-current to charged-current strengths, is effectively 
reduced.  On the figures of the oblique parameters $\dsz$ vs. $\dtz$, 
the positive $\ddelg/\alpha$ would be regarded as the negative 
contribution to the $\dtz$ parameter, which is favored by the 
electroweak data.  
On the other hand, as soon as the muon-decay amplitude is affected, 
we should expect non-oblique $\zll$ vertex corrections to affect the 
MSSM predictions for the $Z$-boson parameters.  
We therefore present in Fig.~\ref{fig:slepton-one} the MSSM 
contribution to $\ddelg/\alpha$ and that to the total $\chi^2$, 
as functions of the $\slepton_L$ mass by assuming 
$m_{\wt{L}}=m_{\wt{L}_3}$.  
%%%----------------------------------
%%%	a new paragraph

%%%	a new paragraph
%%%----------------------------------
By using the analytic expressions for the muon-decay amplitude in 
Appendix C, we find that box corrections are always positive, while 
the vertex corrections are negative.  After summing up, we find 
that the net MSSM correction to $\ddelg$ can become positive
only when $m_{\slepton_L} < 130\gev$, when the lighter chargino 
mass is around 90\gev. 
Fig.~\ref{fig:slepton-one} shows that the total $\chi^2$ improves 
slightly by including the $\ddelg/\alpha$ correction to the muon 
decay, $\chi^2_{\rm tot}~~({\rm with} \ddelg) < 
\chi^2_{\rm tot}~~({\rm without} \ddelg)$, when 
$m_{\slepton_L} \simlt 130\gev$ and $\ddelg<0$. 
%%%----------------------------------
%%%	a new paragraph

%%%	a new paragraph
%%%----------------------------------
Fig.~\ref{fig:slepton-two} shows the slepton contributions to 
$R_\ell, A_e, A_\fb^\ell$ when $\tan\beta = 2$ and the lighter chargino 
mass is around $90\gev~(\mu=300\gev, M_2=120\gev)$. 
In order to find the impact of the right-handed slepton on the observables, 
we show the cases of $m_{\slepton_R} = 100\gev$ and 
$m_{\slepton_R} = 1\tev$ by solid and dotted lines, respectively. 
We find that the non-oblique corrections from the slepton-ino loops 
are generally small and that they cannot remedy the unfavorable 
oblique effects of light left-handed sleptons. 
On the other hand, the right-handed slepton mass $m_{\slepton_R}$ is 
not constrained significantly by the electroweak data.  
%%%%%------------------------------------------ 
%%%%%
%%%%%
%%%%%	Section: Summary and Discussions
%%%%%
%%%%%
%%%%%------------------------------------------
\clean
\section{Summary and Discussions}
%%%%------------------------------
In this paper, we studied radiative corrections to the electroweak 
observables due to new particles in the MSSM systematically, 
confronting them against the latest data on the $Z$-boson parameters 
and the $W$-boson mass. 
After introducing our analytic framework and giving parametrizations 
of the SM predictions, we first studied the effects of the oblique 
(gauge-boson-propagator) corrections from squarks and sleptons 
(sec.~4.1), the MSSM Higgs bosons (sec.~4.2), charginos and
neutralinos (sec.~4.3), separately. 
The effects of non-oblique corrections are then studied systematically; 
the MSSM Higgs-boson contributions to the $\zbb$ and $\ztautau$ vertices 
(sec.~\ref{section:zff_higgs}), 
the gluino-squark-loop corrections to the $\zqq$ vertices 
(sec.~\ref{section:susy_qcd}), 
the squark-chargino/neutralino-loop corrections to the $\zqq$ vertices 
(sec.~\ref{section:zqq_squark}), 
and the slepton-chargino/neutralino-loop corrections to the $\zll$ 
vertices and the muon-decay amplitude 
(sec.~\ref{section:zll_slepton}). 
When studying contribution of each sector of the MSSM, we set all the 
irrelevant new particle masses sufficiently high so that their contributions 
can be neglected. 
%%%----------------------------------
%%%	a new paragraph

%%%	a new paragraph
%%%---------------------------------- 
Our observations can be summarized as follows. 
The agreement between the SM predictions, and the electroweak 
data of Table~1 is excellent: 
for instance, the four-parameter fit~(\ref{eq:sm_fit_ej}) to the 
22 data points of Table~\ref{tab:ewdata98} (18 $Z$-parameters, 
$\mw, \mt, \alps(\mz)$ and $\alpha(\mzsq)$~\cite{EJ}) finds  
$\chi^2_{\rm min}/(\dof) = 18.2/(22-4)$, 
with the probability 44\%. 
However, close examination of the fit shows that the data favor 
additional small negative contributions to the two oblique parameters, 
$\sz$ and $\tz$: see Fig.~\ref{fig:sm_oblique}. 
Small positive contribution to $\mw$ can also improve the fit slightly 
if $\mt<175\gev$: see Fig.~\ref{fig:sm_mw}. 
%%%----------------------------------
%%%	a new paragraph

%%%	a new paragraph
%%%---------------------------------- 
Our study of the oblique corrections in the MSSM can be 
summarized as follows by using, as a measure of the goodness 
of the MSSM fit as compared to the SM fit, the quantity
%%%-------------------
\bea
\dx \equiv ( \chi^2_{\rm tot} )_{\rm MSSM} 
	-  ( \chi^2_{\rm tot} )_{\rm SM}, 
\eea
%%%-------------------
where we compare the two fits by using the common parameter set 
($\mt({\rm GeV})$, $1/\alpha(\mzsq)$, $\alps(\mz)$ )=($175$, 
$128.90$, $0.118$) 
at $\mh=\mhlight=100\gev$. 
%%%-------------------
\begin{itemize}
\item 
	Small left-handed slepton mass ($m_{\slepton_L}$) makes 
	$\dsz$ negative while keeping $\dtz$ essentially unchanged. 
	It contribute positively to $\dmw$. 
	The fit worsens slightly $(\dx \simgt 1)$ 
	if $m_{\slepton_L} \simlt 170\gev$ for $\tan\beta=2$ 
	or $m_{\slepton_L} \simlt 250\gev$ for $\tan\beta=50$,  
	and significantly $(\dx \simgt 4)$ 
	if $m_{\slepton_L} \simlt 110\gev$ for $\tan\beta=2$ 
	of $m_{\slepton_L} \simlt 150\gev$ for $\tan\beta=50$. 

\item
	Small left-handed squark mass parameters ($m_{\wt{Q}}, 
	m_{\wt{Q}_3}$) make $\dtz$ and $\dmw$ positive while 
	keeping $\dsz$ essentially unchanged. 
	The effective $\stop_L$-$\stop_R$ mixing parameter $A_{\rm eff}^t$ 
	of the magnitude of the order of $m_{\wt{Q}_3}$ tame 
	the unfavorable positive $\dtz$ contribution, but not completely. 
	The fit worsens significantly $(\dx \simgt 4)$ 
	if $m_{\wt{Q}}(=m_{\wt{Q}_3}) \simlt 340\gev$ for $\tan\beta=2$ 
	or $m_{\wt{Q}}(=m_{\wt{Q}_3}) \simlt 300\gev$ for $\tan\beta=50$, 
	even for $A_\eff=300\gev$, which correspond to 
	$m_{\stop_1} \simlt 300\gev$ ($\tan\beta=2$) and 
	$m_{\stop_1} \simlt 260\gev$ ($\tan\beta=50$).

\item
	The MSSM Higgs boson contribution can be approximated by the 
	SM Higgs boson contribution at $\mh = \mhlight$ if the 
	pseudo-scalar Higgs-boson mass satisfies $\mpseudo \simgt 
	200\gev$. 
	The effects on the electroweak observables are small,  
	$|\dx| \simlt 1$ for $\tan\beta=2$ or 
	$|\dx| \simlt 0.5$ for $\tan\beta=50$,  
	when the direct search bound of $\mhlight > 75\gev$ 
	(\ref{eq:susy_higgs_bound}) is taken into account. 
	See Fig.~\ref{fig:oblique:higgs} and Table~3. 
\item
	In contrast, the light charginos and neutralinos are found to 
	contribute negatively to both $\dsz$ and $\dtz$, while they 
	contribute negligibly to $\dmw$. 
	Their contribution can hence improve the SM fit	
	$(\dx \sim -1)$: 
	see Figs.~\ref{fig:oblique:inos} and \ref{fig:mw:inos}. 
\end{itemize}
%%%-------------------
We find that the best fit is obtained when the lightest chargino 
mass ($m_{\chargino{1}}$) is near its experimental lower bound, 
$m_{\chargino{1}} \sim 90\gev$, see Tables~4 and 5.
Overall picture of the oblique corrections in the MSSM is summarized 
in Fig.~\ref{fig:oblique:summary}. 
%%%----------------------------------
%%%	a new paragraph

%%%	a new paragraph
%%%---------------------------------- 
Non-oblique corrections in the MSSM are then studied systematically 
in section~\ref{section:nonoblique} as follows. 
The MSSM Higgs bosons can affect the $\zbb,\ztautau$ and 
$Z\nu_\tau\nu_\tau$ vertices through their Yukawa couplings to 
quarks and leptons of the third generation (sec.~\ref{section:zff_higgs}). 
The SUSY-QCD corrections to the $\zqq$ vertices are studied 
when both squarks and gluinos are light (sec.~\ref{section:susy_qcd}). 
The contributions of the light squarks and light charginos/neutralinos 
to the $\zqq$ vertices are then studied (sec.~\ref{section:zqq_squark}). 
When both sleptons and charginos/neutralinos are light, 
not only the $\zll$ vertices but also the muon-decay amplitude 
is affected (sec.~\ref{section:zll_slepton}). 
We find: 
%%%---------------
\begin{itemize}
\item
	The vertex corrections due to the MSSM Higgs bosons can  
	be significant $(\dx \simgt 4)$ 
	if the pseudo-scalar Higgs-boson mass $\mpseudo$ 
	is smaller than about $100\gev$ for $\tan\beta\sim 1$. 
	For $\tan\beta\sim 50$, the effects are significant when 
	$\mpseudo \simlt 300\gev$.  
	The overall fit to the electroweak observables can only be 
	worse than the SM: see Fig.~\ref{fig:higgs_vtx}. 
\item
	The order $\alps$ SUSY-QCD corrections to the $\zqq$ 
	vertices are found to be negligibly small 
	$(|\dx| \simlt 0.5)$ when squarks 
	and gluino masses are bigger than about $200\gev$: 
	see Fig.~\ref{fig:vertex:sqcd} and Table.~\ref{table:susyqcd}. 
\item
	The $\zqq$ vertex corrections due to light squarks and 
	light charginos/neutralinos are found to be insignificant 
	if we take into account the direct search mass bounds on the 
	lightest squarks (\ref{eq:stop_bounds_lep2}) and 
	the lightest chargino (\ref{eq:inos_bounds_lep2}). 
	We also found that good overall fit to the electroweak data 
	with $m_{\chargino{1}}\sim 100\gev$ can be maintained with 
	$m_{\stop_1} \sim 100\gev$ if the left-handed squark mass 
	is kept large, $m_{\wt{Q}}(= m_{\wt{Q}_3}) \simgt 1\tev$: 
	see Fig.~\ref{fig:zbb}. 
\item
	When the left-handed slepton mass is small
	($m_{\slepton_L} \simlt 200\gev$) and 
	charginos/neutralinos are 
	light ($m_{\chargino{1}}\sim 100\gev$), the $\zll$ vertices 
	as well as the muon-decay amplitude are affected significantly. 
	Although we find that their contribution to the muon-decay 
	amplitude can improve the fit when $m_{\slepton_L} \sim 
	100\gev$, its unfavorable oblique contributions make the 
	overall fit worse than the SM ($\dx \simgt 1$) for 
	$m_{\slepton_L} \simlt 150\gev$. 
\end{itemize}
%%%----------------------------------
%%%	a new paragraph

%%%	a new paragraph
%%%---------------------------------- 
Summing up, we find that no contribution of the MSSM particles 
can improve the fit to the electroweak data by taking into account 
the non-oblique corrections to the $\zff$ vertices and the muon-decay 
amplitude. 
Best fit to the data is found when the supersymmetric fermions 
(charginos and neutralinos) are light, $m_{\chargino{1}} \sim 100\gev$, 
and the five scalar mass parameters $\mpseudo, m_{\wt{L}}, 
m_{\wt{L}_3}, m_{\wt{Q}}$ and $m_{\wt{Q}_3}$ are all large; 
$\mpseudo \simgt 300\gev, m_{\wt{L}} = m_{\wt{L}_3} \simgt 500\gev, 
m_{\wt{Q}} = m_{\wt{Q}_3} \simgt 1\tev$. 
The singlet scalar-mass parameters, $m_{\wt{U}}, m_{\wt{U}_3}, 
m_{\wt{D}}, m_{\wt{D}_3}, m_{\wt{E}}$ and $m_{\wt{E}_3}$ do not 
affect the electroweak fit significantly, and hence light 
SU(2)$_L$-singlet scalar bosons are allowed. 
%%%----------------------------------
%%%	a new paragraph

%%%	a new paragraph
%%%---------------------------------- 
Before closing, we give a few examples of the MSSM predictions for 
all the 19 electroweak observables of Table~\ref{tab:ewdata98} 
when we have light charginos/neutralinos ($m_{\chargino{1}}\sim 100\gev$) 
and relatively light singlet squarks and sleptons. 
We examine the four cases I to IV of Table~\ref{table:inputstwo}, 
all of which have a light chargino of about $100\gev$ and 
the lightest squark at around $200\gev$, and the lightest 
slepton at around $150\gev$. 
For brevity, we assumed that the SUSY breaking scalar masses, 
$m_{\wt{Q}}, m_{\wt{U}}, m_{\wt{D}}, m_{\wt{L}}, m_{\wt{E}}$ are 
universal among different generations, 
and the effective left-right mixing mass parameters
of $\stop,\sbottom$ and $\stau$, 
$A_{\rm eff}^t, A_{\rm eff}^b, A_{\rm eff}^\tau$, are also common. 
The doublet squarks and sleptons are kept heavy at around $1\tev$ 
and the pseudo-scalar Higgs-boson mass $\mpseudo$ is set at 
around $500\gev$  in all four cases. 
The scenarios I and II are for $\tan\beta=2$, III and IV are for 
$\tan\beta=50$. 
$\mu=300\gev$ for I and III while $\mu=-300\gev$ for II and IV. 
Masses of most relevant physical states are also shown in 
Table~\ref{table:inputstwo}. 
Because $M_2 < |\mu|$ in all four cases, the lightest charginos/neutralinos 
are dominantly gauginos. 
The lightest squarks and sleptons of the third generation squarks and 
sleptons all have predominantly right-handed component. 
The gluino mass ($M_{\gluino} = M_3$) and the U(1)$_Y$ gaugino 
mass $M_1$ are fixed by the `unification' condition $M_3/M_2/M_1 
= \hat{\alpha}_3/\hat{\alpha}_2/\hat{\alpha}_1$. 
%%%----------------------------------
%%%	a new paragraph

%%%	a new paragraph
%%%---------------------------------- 
The predictions are compared with the data in Table~\ref{table:pulltwoa} 
for the scenarios I and II ($\tan\beta=2$) and in Table~\ref{table:pulltwob} 
for the scenarios III and IV ($\tan\beta=50$). 
In each Table we give the reference SM predictions at $\mt=175\gev$ and 
$\mh=100\gev$ for comparison. 
The first column in each scenario shows the contributions of 
charginos/neutralinos and the MSSM Higgs bosons only. 
The results of this column represent the predictions of the MSSM 
when all new particles except the charginos/neutralinos and the 
Higgs bosons are heavy. 
In the second column, we show the predictions when only the oblique 
corrections from the squarks and sleptons are added, and in the last 
column we show the complete MSSM predictions including all the vertex 
and box corrections. 
We show in the second column the predictions of the oblique
corrections only in order to show the quantitative importance 
of non-oblique corrections. 
%%%----------------------------------
%%%	a new paragraph

%%%	a new paragraph
%%%---------------------------------- 
From Table~\ref{table:pulltwoa}, we find that the MSSM predictions 
are almost completely determined by the oblique contributions of 
charginos and neutralinos in the scenarios I and II with 
$\tan\beta=2$, where the three scalar-mass 
parameters, $\mpseudo,m_{\wt{Q}}$ and $m_{\wt{L}}$ 
are all kept large. 
The relatively light squarks and sleptons of a few hundred GeV 
barely contribute to oblique or non-oblique corrections. 
In contrast, the $\tan\beta=50$ cases of III and IV given in 
Table~\ref{table:pulltwob} show that the non-oblique corrections can 
have significant effects even though the physical mass spectrum in 
all four cases are similar in Table~\ref{table:inputstwo}. 
This is because the enhanced Yukawa coupling of the $b$-quark 
allows the relatively light $\sbottom_1$ of predominantly 
right-handed component to contribute to the $\zbb$ vertices. 
This affects $\Gamma_b$ (and hence $R_b$ in Table~\ref{table:pulltwob}) 
and $\Gamma_h$ (and hence $R_\ell$). 
Similar contributions to the $Z \tau_L \tau_L$ vertex are found 
not to affect the fit significantly. 
%%%%%%%%%%%%%%%%%%%%%%%%%%%%%%%%%%%%%%%%%%%%%%%%%%%%%%%%%%%%%
%%%%%
%%%%%   Table: Input parameters for case I--IV
%%%%%          (sbottom lighter than stop is allowed)
%%%%%%%%%%%%%%%%%%%%%%%%%%%%%%%%%%%%%%%%%%%%%%%%%%%%%%%%%%%%%
\begin{table}[t]
\begin{center}
\begin{tabular}{c|cccccccccc} \hline \hline 
%%%--------------------
	& \multicolumn{10}{c}{Inputs}\\ \hline 
	& $\tan\beta$ & $\mu$ & $M_2$ & $A_t$ 
	& $\mpseudo$ & $m_{\wt{Q}}$ & $m_{\wt{U}}$ & $m_{\wt{D}}$ 
	& $m_{\wt{L}}$ & $m_{\wt{E}}$  \\ \hline
I   &  2 & $\hph$300 & 125      & $-657$     & 486 & $\hpz$956  & 196 & 181 & 894 & 129 \\
II  &  2 & $-300$    & $\hpz$90 & $\hph 568$ & 434 & 1273       & 183 & 176 & 639 & 159 \\
III & 50 & $\hph$300 & 110      & $\hph 812$ & 481 & 1091       & 221 & 204 & 925 & 165 \\
IV  & 50 & $-300$    & 105      & $-657$     & 486 & $\hpz$956  & 196 & 181 & 894 & 129 \\
\hline 
	&\multicolumn{10}{c}{Outputs}\\ \hline 
	& $\mhlight$   	& $\mch{1}$ & $\mn{1}$ & $M_{\gluino}$ 	&
	$m_{\stau_1}$ & $m_{\sneutrino}$ & $m_{\stop_1}$ &
	$m_{\sbottom_1}$ & $m_{\sup_R}$ &  $m_{\sdown_R}$ \\ \hline 
I	& 104 & 98.2 & 53.4 & 439 & 134 & 893 & 214 & 182 & 194 & 182 \\
II	& 109 & 101  & 48.3 & 316 & 162 & 637 & 231 & 177 & 180 & 177 \\
III	& 130 & 101  & 53.6 & 386 & 169 & 922 & 247 & 196 & 219 & 206 \\
IV	& 128 & 98.4 & 51.7 & 369 & 133 & 892 & 230 & 168 & 193 & 183 \\
\hline \hline 
%%%--------------------
\end{tabular}
\caption{Inputs and outputs parameters of the MSSM scenarios I to IV 
with light charginos/neutralinos and light squarks and sleptons. }
\label{table:inputstwo}
\end{center}
\end{table}
%%%----------------------------------
%%%	a new paragraph

\begin{figure}[t]
\begin{center}
\leavevmode\psfig{figure=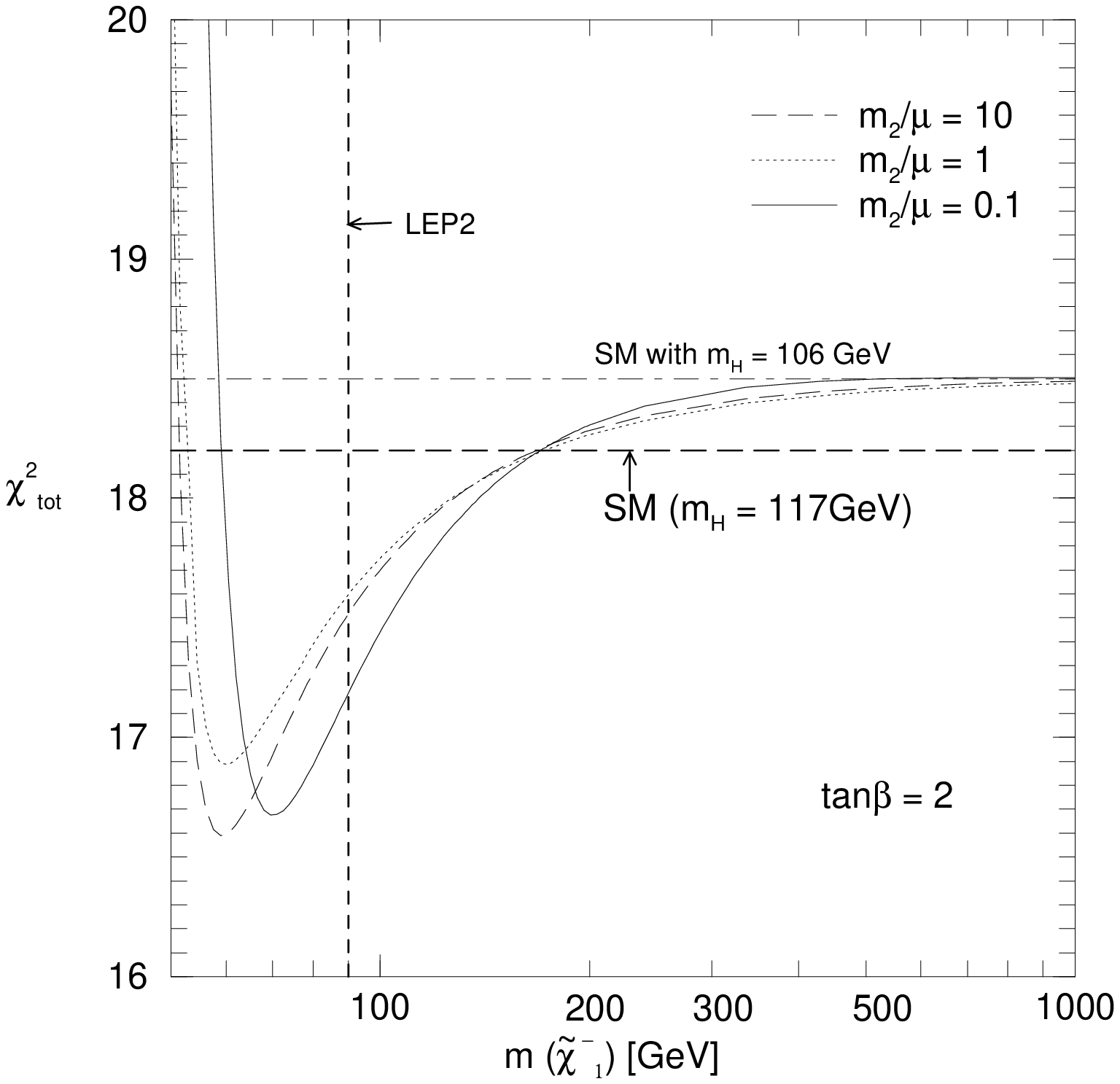,width=10cm}
\end{center}
\caption{ The total $\chi^2$ in the MSSM as a function of the lighter 
chargino mass $m_{\chargino{1}}$ for $\tan\beta=2$. 
The SM best fit ($\chi^2 = 18.2$) is shown by the dashed horizontal line. 
The dot-dashed horizontal line shows the SM fit using $\mh = 106\gev$ 
which is the lightest Higgs boson mass predicted in the MSSM.  
Three different $M_2$-$\mu$ ratio (10, 1, 0.1) are studied. 
The bound on $m_{\chargino{1}}$ from the LEP2 experiment is shown by 
the dashed vertical line. }
\label{fig:global_tnb02}
\end{figure}

\begin{figure}[t]
\begin{center}
\leavevmode\psfig{figure=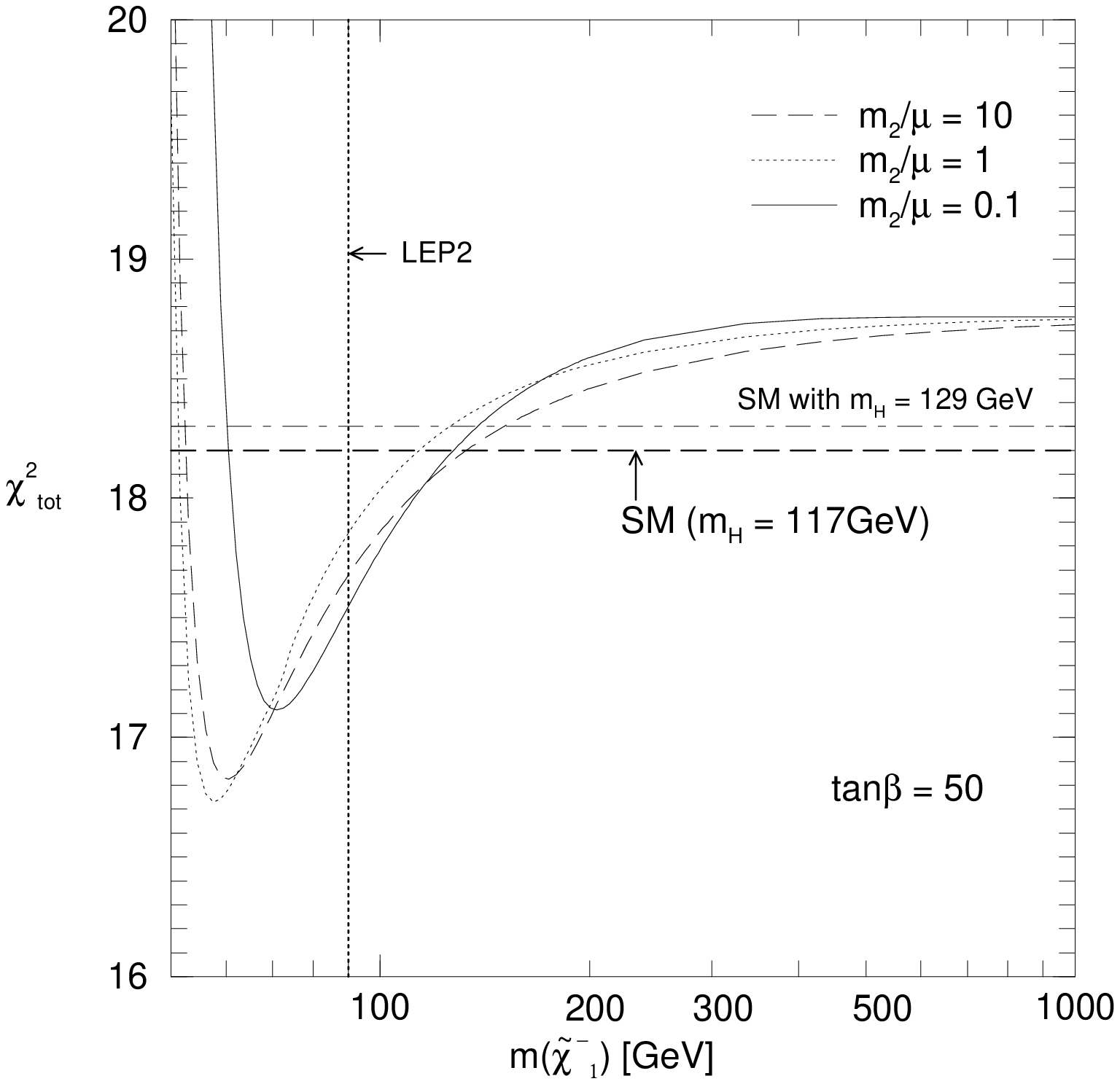,width=10cm}
\end{center}
\caption{ The total $\chi^2$ in the MSSM as a function of the lighter 
chargino mass $m_{\chargino{1}}$ for $\tan\beta=50$. 
The SM best fit ($\chi^2 = 18.2$) is shown by the dashed horizontal line. 
The dot-dashed horizontal line shows the SM fit using $\mh = 129\gev$ 
which is the lightest Higgs boson mass predicted in the MSSM.  
Three different $M_2$-$\mu$ ratio (10, 1, 0.1) are studied. 
The bound on $m_{\chargino{1}}$ from the LEP2 experiment is shown 
by the dashed vertical line. }
\label{fig:global_tnb50}
\end{figure}

%%%	a new paragraph
%%%---------------------------------- 
Finally, we would like to perform the `{\it global}' fit to all 
the electroweak data of Table~\ref{tab:ewdata98} in the MSSM, 
by taking into account the present data on 
$\mt$~(\ref{eq:mt_pdg}), $1/\alpha(\mzsq)$~(\ref{eq:ej_alpha}) 
and $\alps(\mz)$~(\ref{eq:alpha_s_pdg}). 
%%%%%%%%%%%%%%%%%%%%%%%%%%%%%%%%%%%%%%%%%%%%%%%%%%%%%%%%%%%%%%%%%%%%%%%%%
%%%%%
%%%%%	Table B (sbottom lighter than stop is allowed)
%%%%%   I & II 
%%%%%%%%%%%%%%%%%%%%%%%%%%%%%%%%%%%%%%%%%%%%%%%%%%%%%%%%%%%%%%%%%%%%%%%%%
	\begin{table}[p]%[htbp]
	\begin{center}
	\begin{tabular}{r|c||c|c|c||c|c|c}
	\hline	\hline
	 & \multicolumn{7}{|c}{Pull}\\ \hline
	 & SM$^*$ & \multicolumn{3}{c||}{I} 
	      & \multicolumn{3}{c}{II}  \\ \hline
	&   & inos/Higgs &obl. & all & inos/Higgs &obl. & all \\
	\hline
%%%------------------
$\Gamma_Z^{}$    & $-1.4$     & $-0.9$     & $-1.0$     & $-1.1$     
		& $-0.9$     & $-1.0$     & $-1.1$     \\
$\sigma^0_h$     & $\hph 0.3$ & $\hph 0.3$ & $\hph 0.3$ & $\hph 0.3$ 
		& $\hph 0.3$ & $\hph 0.3$ & $\hph 0.2$ \\
$R_{\ell}$       & $\hph 0.7$ & $\hph 0.8$ & $\hph 0.8$ & $\hph 0.7$ 
		& $\hph 0.8$ & $\hph 0.8$ & $\hph0.7$  \\
$A^{0,\ell}_\fb$ & $\hph 0.3$ & $\hph 0.5$ & $\hph 0.5$ & $\hph 0.5$ 
		& $\hph 0.5$ & $\hph 0.5$ & $\hph 0.4$ \\
	&&&&&&& \\
%---------
	$\left\{ \begin{array}{r}
        ~R_e^{}\; \\ ~R_\mu^{}\; \\ ~R_\tau^{}\; \end{array} \right.$ & 
        $\begin{array}{c} 
      	                  \hph 0.7 \\ \hph 1.3 \\ -0.7 \end{array}$ &
%%%---------- 
        $\begin{array}{c} \hph 0.7 \\ \hph 1.3 \\ -0.6 \end{array}$ & 
        $\begin{array}{c} \hph 0.7 \\ \hph 1.3 \\ -0.6 \end{array}$ & 
        $\begin{array}{c} \hph 0.7 \\ \hph 1.3 \\ -0.6 \end{array}$ & 
%%%----------
        $\begin{array}{c} \hph 0.7 \\ \hph 1.3 \\ -0.6 \end{array}$ & 
        $\begin{array}{c} \hph 0.7 \\ \hph 1.3 \\ -0.6 \end{array}$ & 
        $\begin{array}{c} \hph 0.7 \\ \hph 1.2 \\ -0.7 \end{array}$ \\
%%%----------
	&&&&&&& \\
%---------
	$\left\{ \begin{array}{r}
        A^{0,e}_\fb \\ A^{0,\mu}_\fb \\ A^{0,\tau}_\fb \end{array} \right.$ & 
        $\begin{array}{c} -0.5 \\-0.1 \\ \hph 1.1 \end{array}$ & 
%%%----------
        $\begin{array}{c} -0.4\\ \hph0.0 \\ \hph 1.2 \end{array}$ & 
        $\begin{array}{c} -0.4\\ \hph0.0 \\ \hph 1.1 \end{array}$ & 
        $\begin{array}{c} -0.4\\ \hph0.0 \\ \hph 1.1 \end{array}$ & 
%%%----------
        $\begin{array}{c} -0.4\\ \hph 0.0\\ \hph 1.1 \end{array}$ & 
        $\begin{array}{c} -0.4\\ \hph 0.0\\ \hph 1.1 \end{array}$ & 
        $\begin{array}{c} -0.4\\ \hph 0.0\\ \hph 1.1 \end{array}$ \\
%%%----------
	&&&&&&& \\
%---------
$A_{\tau}$&$-1.2$ 
	& $-1.0$ & $-1.0$ & $-1.0$ 
	& $-1.0$ & $-1.1$ & $-1.1$ \\
$A_{e} $  &$-0.1$ 
	& $\hph 0.1$ & $\hph0.0$ & $\hph0.1$ 
	& $\hph 0.0$ & $\hph0.0$ & $\hph0.0$ \\
	&&&&&&& \\
%---------
$R_b$ & $\hph 1.2$ 
	& $\hph 1.2$ & $\hph 1.2$ & $\hph 1.1$ 
	& $\hph 1.2$ & $\hph 1.2$ & $\hph 1.0$ \\
$R_c$ & $\hph 0.3$ 
	& $\hph 0.3$ & $\hph 0.3$ & $\hph 0.3$ 
	& $\hph 0.3$ & $\hph 0.3$ & $\hph 0.3$ \\
$A^{0,b}_{FB}$ & $-2.4$ 
	& $-2.1$ & $-2.2$ & $-2.1$ 
	& $-2.2$ & $-2.2$ & $-2.2$ \\
$A^{0,c}_{FB}$ & $-0.8$ 
	& $-0.7$ & $-0.7$ & $-0.7$ 
	& $-0.7$ & $-0.7$ & $-0.7$ \\
	&&&&&&& \\
%---------
$\sin^2\theta_{\rm eff}^{\rm lept}$ & $\hph 0.7$ 
	& $\hph 0.7$ & $\hph 0.7$ & $\hph 0.7$ 
	& $\hph 0.7$ & $\hph 0.7$ & $\hph 0.7$ \\
	&&&&&&& \\
%---------
$A^0_{LR}$&$\hph 1.0$
	& $\hph 1.4$ & $\hph 1.3$ & $\hph 1.3$
	& $\hph 1.3$ & $\hph 1.3$ & $\hph 1.2$\\
$A_b$ & $-1.9$ 
	& $-1.9$ & $-1.9$ & $-1.9$ 
	& $-1.9$ & $-1.9$ & $-1.9$ \\
$A_c$ & $-0.5$ 
	& $-0.5$ & $-0.5$ & $-0.5$ 
	& $-0.5$ & $-0.5$ & $-0.5$ \\
	&&&&&&& \\
%---------
$\mw$ & $\hph 0.2$ 
	& $\hph 0.1$ & $\hph0.1$ & $\hph0.1$ 
	& $\hph 0.1$ & $\hph0.1$ & $\hph0.1$ \\
	\hline
%---------
$\chi^2_{\rm tot}$ & 19.8
	& 18.1 & 18.2 & 18.1 & 18.2 & 18.3 & 18.2 \\
	\hline \hline
%%%---------------------------
	\end{tabular} 
	\end{center}
	\caption{The pull factors of the predictions of the MSSM 
	scenarios I and II. 
	Those of the reference SM predictions (see
	Table.~\cite{lepewwg98}) are given for comparison. 	
	}
	\label{table:pulltwoa}
	\end{table}
%%%%%%%%%%%%%%%%%%%%%%%%%%%%%%%%%%%%%%%%%%%%%%%%%%%%%%%%%%%%%%%%%%%%%%%
%%%%%%%%%%%%%%%%%%%%%%%%%%%%%%%%%%%%%%%%%%%%%%%%%%%%%%%%%%%%%%%%%%%%%%%%%
%%%%%
%%%%%	Table B (sbottom lighter than stop is allowed)
%%%%%   III & IV
%%%%%%%%%%%%%%%%%%%%%%%%%%%%%%%%%%%%%%%%%%%%%%%%%%%%%%%%%%%%%%%%%%%%%%%%%
	\begin{table}[p]%[htbp]
	\begin{center}
	\begin{tabular}{r|c||c|c|c||c|c|c}
	\hline	\hline
	 & \multicolumn{7}{|c}{Pull}\\ \hline
	 & SM$^*$ & \multicolumn{3}{c||}{III} 
	      & \multicolumn{3}{c}{IV}  \\ \hline
	&   & inos/Higgs &obl. & all & inos/Higgs &obl. & all \\
	\hline
%%%------------------
$\Gamma_Z^{}$ & $-1.4$ 
	& $-0.8$ & $-0.8$ & $-0.8$ 
	& $-0.8$ & $-0.8$ & $-0.7$  \\
$\sigma^0_h$ & $\hph 0.3$ 
	& $\hph 0.3$ & $\hph 0.3$& $\hph 0.2$
	& $\hph 0.3$ & $\hph 0.3$& $\hph 0.1$\\ 
$R_{\ell}$ & $\hph 0.7$ 
	& $\hph 0.8$ & $\hph 0.8$ & $\hph 1.1$
	& $\hph 0.8$ & $\hph 0.8$ & $\hph 1.1$\\
$A^{0,\ell}_\fb$ & $\hph 0.3$ 
	& $\hph 0.6$ & $\hph 0.6$ & $\hph 0.6$
	& $\hph 0.6$ & $\hph 0.6$ & $\hph 0.6$\\
	&&&&&&& \\
%---------
	$\left\{ \begin{array}{r}
        ~R_e^{}\; \\ ~R_\mu^{}\; \\ ~R_\tau^{}\; \end{array} \right.$ & 
        $\begin{array}{c} 
      	                  \hph 0.7 \\ \hph 1.3 \\ -0.7 \end{array}$ &
%%%----------
        $\begin{array}{c} \hph 0.8 \\ \hph 1.3 \\ -0.6 \end{array}$ & 
        $\begin{array}{c} \hph 0.8 \\ \hph 1.3 \\ -0.6 \end{array}$ & 
        $\begin{array}{c} \hph 0.9 \\ \hph 1.5 \\ -0.5 \end{array}$ & 
%%%----------
        $\begin{array}{c} \hph 0.8 \\ \hph 1.3 \\ -0.6 \end{array}$ & 
        $\begin{array}{c} \hph 0.8 \\ \hph 1.3 \\ -0.6 \end{array}$ & 
        $\begin{array}{c} \hph 0.9 \\ \hph 1.6 \\ -0.5 \end{array}$ \\
%%%----------
	&&&&&&& \\
%---------
	$\left\{ \begin{array}{r}
        A^{0,e}_\fb \\ A^{0,\mu}_\fb \\ A^{0,\tau}_\fb \end{array} \right.$ & 
        $\begin{array}{c} -0.5 \\-0.1 \\ \hph 1.1 \end{array}$ & 
%%%----------
        $\begin{array}{c} -0.4\\ \hph 0.1\\ \hph 1.2 \end{array}$ & 
        $\begin{array}{c} -0.4\\ \hph 0.1\\ \hph 1.2 \end{array}$ & 
        $\begin{array}{c} -0.4\\ \hph 0.1\\ \hph 1.2 \end{array}$ & 
%%%----------
	$\begin{array}{c} -0.4\\ \hph 0.1\\ \hph 1.2 \end{array}$ & 
	$\begin{array}{c} -0.4\\ \hph 0.1\\ \hph 1.2 \end{array}$ & 
	$\begin{array}{c} -0.4\\ \hph 0.1\\ \hph 1.2 \end{array}$ \\
%%%----------
	&&&&&&& \\
%---------
$A_{\tau}$&$-1.2$ 
	& $-0.9$ & $-0.9$ & $-0.8$ 
	& $-0.9$ & $-0.9$ & $-0.9$\\
$A_{e} $& $-0.1$ 
	& $\hph 0.2$ & $\hph0.2$ & $\hph0.2$ 
	& $\hph 0.2$ & $\hph0.1$& $\hph 0.1$\\
	&&&&&&& \\
%---------
$R_b$ & $\hph 1.2$ 
	& $\hph 1.2$ & $\hph 1.2$ & $\hph 1.4$ 
	& $\hph 1.2$ & $\hph 1.2$ & $\hph 1.5$\\
$R_c$ & $\hph 0.3$ 
	& $\hph 0.3$ & $\hph 0.3$ & $\hph 0.3$
	& $\hph 0.3$ & $\hph 0.3$ & $\hph 0.3$ \\
$A^{0,b}_{FB}$ & $-2.4$ 
	& $-1.9$ & $-2.0$ & $-2.1$ 
	& $-2.0$ & $-2.0$ & $-2.1$\\
$A^{0,c}_{FB}$ & $-0.8$ 
	& $-0.6$ & $-0.6$ & $-0.6$ 
	& $-0.6$ & $-0.6$ & $-0.6$\\
	&&&&&&& \\
%---------
$\sin^2\theta_{\rm eff}^{\rm lept}$ & $\hph 0.7$ 
	& $\hph 0.6$ & $\hph 0.6$ & $\hph 0.6$
	& $\hph 0.6$ & $\hph 0.6$ & $\hph 0.6$ \\
	&&&&&&& \\
%---------
$A^0_{LR}$&$\hph 1.0$
	& $\hph 1.6$ & $\hph 1.6$ & $\hph 1.6$
	& $\hph 1.5$ & $\hph 1.5$ & $\hph 1.5$\\
$A_b$ & $-1.9$ 
	& $-1.9$ & $-1.9$ & $-2.0$
	& $-1.9$ & $-1.9$ & $-2.0$ \\
$A_c$ & $-0.5$ 
	& $-0.5$ & $-0.5$ & $-0.5$
	& $-0.5$ & $-0.5$ & $-0.5$ \\
	&&&&&&& \\
%---------
$\mw$ & $\hph 0.2$ 
	& $\hph 0.3$ & $\hph0.3$ & $\hph0.3$ 
	& $\hph 0.3$ & $\hph0.2$ & $\hph0.2$ \\ 
	\hline
%---------
$\chi^2_{\rm tot}$ & 19.8
	& 17.8 & 17.8 & 19.4 & 17.7 & 17.8 & 19.7\\
	\hline \hline
%%%---------------------------
	\end{tabular} 
	\end{center}
	\caption{The pull factors of the predictions of the MSSM 
	scenarios III and IV. 
	Those of the reference SM predictions (see
	Table.~\cite{lepewwg98}) are given for comparison. 	
	}
	\label{table:pulltwob}
	\end{table}
%---------
From the previous comprehensive studies, it is clear that the 
improvement of the total $\chi^2$ over the SM can be realized only 
when a light ($\sim 100\gev$) chargino exists and when 
the left-handed squark 
and slepton masses ($m_{\wt{Q}}$ and $m_{\wt{L}}$) and 
the pseudo-scalar Higgs-boson mass ($\mpseudo$) are all 
sufficiently large. 
In the following study, 
we assume that the universality of the scalar mass parameters 
among the different generations, 
$m_{\wt{Q}}=m_{\wt{Q}_3}$, $m_{\wt{U}}=m_{\wt{U}_3}$, 
$m_{\wt{D}}=m_{\wt{D}_3}$, $m_{\wt{L}}=m_{\wt{L}_3}$ and 
$m_{\wt{E}}=m_{\wt{E}_3}$. 
Also, we fix all the scalar mass parameters at 
$1\tev$ 
($m_{\wt{Q}}=m_{\wt{L}}=m_{\wt{U}}=m_{\wt{D}}=m_{\wt{E}}=\mpseudo
=1\tev$) and $A_{\rm eff}^f = 0$ for brevity. 
Then, there are three parameters left in the MSSM, $\tan\beta$, $M_2$ 
and $\mu$. 
We find the total $\chi^2$ as a function of the lighter chargino mass 
$m_{\chargino{1}}$ for $\tan\beta = 2$ (Fig.~\ref{fig:global_tnb02}) 
and $\tan\beta=50$ (Fig.~\ref{fig:global_tnb50}). 
Since it is worth studying how the mixing between the gaugino and 
the higgsino affects the fit, we show the following three cases 
separately: $M_2/\mu = 0.1$ (solid lines), 1 (dotted lines) 
and 10 (dashed lines). 
The small number of $M_2/\mu$ implies that the lighter chargino is 
dominantly the wino while the large number of $M_2/\mu$ implies that 
it is dominantly the higgsino. 
%%%%--------------------------------------
%%%----------------------------------
%%%----------------------------------
The behavior of $\chi^2_{\rm tot}$ in the small $m_{\chargino{1}}$ region 
($\simlt 60\gev$) can be understood from the chargino contribution to 
the $R$ parameter, eq.~(\ref{eq:dr_wino}), which behaves as $1/\beta$ 
when $m_{\chargino{1}}$ is close to a half of $\mz$. 
%%%----------------------------------
%%%	a new paragraph

%%%	a new paragraph
%%%----------------------------------
In our assumption on the mass spectrum of the MSSM, the lightest 
Higgs boson mass is predicted as $\mhlight = 106\gev$ for $\tan\beta =2$ 
and $\mhlight = 129\gev$ for $\tan\beta =50$. 
The decoupling in the large SUSY mass limit is examined 
by comparing the MSSM fit with the SM fit at $\mh=\mhlight$ rather than 
the SM best fit at $\mh = 117\gev$. 
We show by horizontal lines the corresponding SM results 
in Figs.~\ref{fig:global_tnb02} and \ref{fig:global_tnb50}. 
The decoupling seems to hold at $m_{\chargino{1}} = 1\tev$ for 
$\tan\beta =2$. 
On the other hand, the MSSM prediction at $\tan\beta=50$ 
differs slightly with the SM prediction at $\mh=\mhlight$ 
even when $m_{\chargino{1}} = 1\tev$, as seen from 
Fig.~\ref{fig:global_tnb50}. 
%%%------------------------------
\begin{table}[t]
\begin{center}
\begin{tabular}{l|c|c|c|c} \hline \hline 
	& \multicolumn{2}{c|} {$\tan\beta=2$}
	& \multicolumn{2}{c} {$\tan\beta=50$} \\ \hline
	& obl. & obl. + vtx. & obl. & obl. + vtx. \\ \hline
$\chi^2_{\rm min}$ & 18.5 & 18.5 & 18.3 & 18.7 \\
$\chi^2_{\smr}$ at $\mh=\mhlight$ &---& 18.5  &---& 18.3 
\\ \hline \hline
\end{tabular}
\caption{
$\chi^2_{\rm min}$ in the MSSM for $\tan\beta=2$ and 50 
which are estimated by using the oblique corrections only 
(obl.) and all the MSSM corrections (obl. + vtx.). 
All the scalar mass parameters are the same with those in  
Figs.~\ref{fig:global_tnb02} and \ref{fig:global_tnb50}. 
We take $M_2/\mu = 0.1$ as an example. 
As a reference, $\chi^2$ in the SM ($\chi^2_{\smr}$) are shown 
for $\mh = 106\gev$ and $\mh = 129\gev$, which should each 
correspond to the decoupling limit of 
the $\tan\beta=2$ and $\tan\beta=50$ cases, respectively. 
}
\label{tab:chisq}
\end{center}
\end{table}
%%%------------------------------
We looked for the origin of this small discrepancy and found that it 
comes from the non-oblique corrections as shown in Table~\ref{tab:chisq}. 
Among the 14 distinct $Zff$ vertices of the MSSM that appear in 
eq.~(\ref{eq:dgfa}), we find that the $\zbrbr$ vertex receives the 
non-negligible corrections at $\tan\beta\sim 50$.  
Its origin is the enhanced Yukawa coupling among the $b$-quark, 
the $t$-quark and the heavy Higgs bosons ($A,H,H^\pm$) of 1~TeV mass.  
In particular, the charged Higgs boson contributes to the $\zbrbr$ 
vertex has an additional enhancement factor due to the exchanged 
top-quark mass. 
The Yukawa couplings among the charged Higgs boson, the right-handed 
$b$-quark and the left-handed $t$-quark is proportional to 
$\mb \tan\beta$, which gives the largest coupling in the MSSM when 
$\tan\beta > \sqrt{m_t/m_b} \approx 7$. 
The combined effects make the effective $\zbrbr$ coupling smaller 
than the SM by about 0.6\% even when $m_{H^-}^{} \approx 1\tev$.  
This 0.6\% decrease in $g^b_R$ slightly worsen the fit to the  
$R_b$ and $A_\fb^{0,b}$ data.  
%%%----------------------------------
%%%	a new paragraph

%%%	a new paragraph
%%%----------------------------------
Taking account of the direct search limit on the chargino mass from 
LEP2, we can conclude that $m_{\chargino{1}}\sim 100\gev$ gives the best fit 
point of the MSSM in both small and large $\tan\beta$. 
Among the three ratios of $M_2/\mu$, the gaugino dominant case 
($M_2/\mu=0.1$) gives slightly better fit at $m_{\chargino{1}}\sim 100\gev$ 
rather than the others. 
The improvement of the MSSM fit over the SM is not very significant, 
however, since the SM fit to the data is already good. 
%%%%%----------------------------------
%%%%%
%%%%%
%%%%%	Section: Conclusion
%%%%%
%%%%%
%%%%%----------------------------------
\clean
\section{Conclusion}
%%%----------------------------------
We have presented the results of our comprehensive study 
of the constraints on the MSSM parameters from the electroweak 
precision measurements. 
The gauge-boson-propagator (oblique) corrections are parametrized 
in terms of three parameters, $\dsz,\dtz$ and $\dmw$. 
The $\sz$ parameter is obtained from the standard $S$ parameter 
by taking into account the running effect of the $Z$-boson propagator 
correction between $q^2=0$ and $q^2=\mzsq$, which we denoted by $R$, 
while the $\tz$ parameter is obtained from the standard $T$ parameter 
by taking account of the $R$ parameter and $\ddelg$, 
which is the correction to the muon-decay amplitude. 
We found that the left-handed sfermion contributions always make 
the fit worse than the SM due to the positive contributions to 
the $\tz$ parameter (squarks) or the negative contributions to 
the $\sz$ parameter (sleptons). 
The contributions from the right-handed sfermions to the oblique 
parameters are negligible. 
The contributions of the MSSM Higgs bosons to the oblique parameters 
behave like the SM Higgs boson contribution at $\mh=\mhlight$ 
when the pseudo-scalar Higgs-boson mass $\mpseudo$ is 
large $(\mpseudo \simgt 300\gev)$. 
The most important finding of our study is that the light chargino 
contribution can make the fit better than that of the SM due to 
its effect on the $R$ parameter. 
We therefore studied vertex and box corrections that contain 
light charginos/neutralinos and find that no combination of 
squarks and sleptons can improve the fit further. 
%%%----------------------------------
%%%	a new paragraph

%%%	a new paragraph
%%%----------------------------------
The `{\it global}' fit of the MSSM has been performed by assuming the 
heavy mass limit of all the sfermions and the pseudo-scalar Higgs boson. 
We found that the best fit is obtained when the chargino mass is at 
the current lower bound from the LEP2 experiments. 
%--------------------------------
%
%
%         acknowledgements
%
%
%--------------------------------
\section*{Acknowledgment}
%%%%%
The work of G.C.C.\ is supported in part by Grant-in-Aid for Scientific 
Research from the Ministry of Education, Science and Culture of Japan.
K.H.\ thanks the JSPS-NSF Joint Research Project and the Phenomenology 
Institute of University of Wisconsin-Madison, for their hospitality 
while part of this work was done.  
%--------------------------------
%
%
%         Note added
%
%
%--------------------------------
\section*{Note added}
%%%%%
After submission of this paper, we learned that 
Djouadi \etal~\cite{djouadi97} calculated the two-loop 
QCD corrections to the gauge boson two-point functions.  
They found that the QCD correction enhances the one-loop 
squark contributions to the $\rho$-parameter (the $T$-parameter 
in our paper) by 10\%--30\%.  
This effect may shift effectively the stop mass in our results 
slightly higher.  We would like to thank S.~Heinemeyer for calling 
our attention to Ref.~\cite{djouadi97}.  
%%%----------------------------------
%%%	a new paragraph

%%%	a new paragraph
%%%----------------------------------
We also learned that it is not appropriate to use the experimental 
data on the effective weak mixing angle, 
$\sin^2\theta_{\rm eff}^{\rm lept}$ 
(see Table 1) which is measured from the jet-charge asymmetry,  
in our analysis when the non-universal vertex corrections are 
significant.  This is because the parameter is obtained from the 
asymmetry data by assuming that the $Z$-boson decay branching 
fractions to each quark flavor obey the SM prediction.  We confirm
that our results presented in this paper do not change significantly 
when we remove the jet-charge asymmetry data in the fit.  We are 
grateful to K.\ Moenig for clarifying the problem for us.

%--------------------------------
\newpage
%--------------------------------
%
%
%         Appendix 
%
%
%--------------------------------
\begin{flushleft}
{\Large\bf Appendix}
\end{flushleft}
\begin{appendix}
%--------------------------------------------
%
%         Appendix: Mass eigenstates and couplings 
%		    in the MSSM	
%
%--------------------------------------------
\section{Mass eigenstates and couplings in the MSSM}
\label{section:masseigenstates}
\clean
%%%----------------------------------
All the analytic expressions for the 1-loop contributions 
of the MSSM particles are expressed in terms of the masses 
of the mass-eigenstates and their couplings. 
We summarize the notation of ref.~\cite{mssml99} here for 
completeness. 
%%%----------------------------------
%%%	a new paragraph

%%%	a new paragraph
%%%----------------------------------
The mixing matrix elements between the current-eigenstates 
and the mass-eigenstates appear in the expression for 
the gauge-boson-propagator corrections. 
There are four types of the mixing matrices in our 
restricted MSSM. 
For squarks and sleptons, we consider only the chirality mixing 
among the third generation squarks and sleptons:   
%-----------------
\bsub\label{exp_sf}
\bea
\tilde{f}_L &=& \sum_{j=1}^{2} (U^{\tilde{f}})_{1j}   \tilde{f}_j  \,,\quad 
\tilde{f}_L^* = \sum_{j=1}^{2} (U^{\tilde{f}})_{1j}^* \tilde{f}_j^*\,,\\ 
\tilde{f}_R &=& \sum_{j=1}^{2} (U^{\tilde{f}})_{2j}   \tilde{f}_j  \,,\quad 
\tilde{f}_R^* = \sum_{j=1}^{2} (U^{\tilde{f}})_{2j}^* \tilde{f}_j^*\,,  
\eea
\esub
%-----------------
for $\tilde{f}=\tilde{t}$, $\tilde{b}$ and $\tilde{\tau}$.  
For the charginos we have 
%-----------------
\bsub\label{exp_cino}
\bea
\wt{W}^-_L   &=&\sum_{j=1}^{2}(U^C_L)_{1j}  \wt{\chi}^-_{jL}\,,\quad 
\wt{W}^+_R     =\sum_{j=1}^{2}(U^C_L)_{1j}^*\wt{\chi}^+_{jR}\,,\\ 
\wt{H}^-_{dL}&=&\sum_{j=1}^{2}(U^C_L)_{2j}  \wt{\chi}^-_{jL}\,,\quad 
\wt{H}^+_{dR}  =\sum_{j=1}^{2}(U^C_L)_{2j}^*\wt{\chi}^+_{jR}\,,\\ 
\wt{W}^-_R   &=&\sum_{j=1}^{2}(U^C_R)_{1j}  \wt{\chi}^-_{jR}\,,\quad 
\wt{W}^+_L     =\sum_{j=1}^{2}(U^C_R)_{1j}^*\wt{\chi}^+_{jL}\,,\\ 
\wt{H}^-_{uR}&=&\sum_{j=1}^{2}(U^C_R)_{2j}  \wt{\chi}^-_{jR}\,,\quad 
\wt{H}^+_{uL}  =\sum_{j=1}^{2}(U^C_R)_{2j}^*\wt{\chi}^+_{jL}\,, 
\eea
\esub
%-----------------
and for the neutralinos we have 
%-----------------
\bsub\label{exp_nino}
\bea
\wt{B}_L     &=&\sum_{j=1}^4(U^N_L)_{1j}\wt{\chi}^0_{jL}\,,\quad 
\wt{B}_R       =\sum_{j=1}^4(U^N_R)_{1j}\wt{\chi}^0_{jR}\,,\\ 
\wt{W}^3_L   &=&\sum_{j=1}^4(U^N_L)_{2j}\wt{\chi}^0_{jL}\,,\quad 
\wt{W}^3_R     =\sum_{j=1}^4(U^N_R)_{2j}\wt{\chi}^0_{jR}\,,\\ 
\wt{H}^0_{dL}&=&\sum_{j=1}^4(U^N_L)_{3j}\wt{\chi}^0_{jL}\,,\quad 
\wt{H}^0_{dR}  =\sum_{j=1}^4(U^N_R)_{3j}\wt{\chi}^0_{jR}\,,\\ 
\wt{H}^0_{uL}&=&\sum_{j=1}^4(U^N_L)_{4j}\wt{\chi}^0_{jL}\,,\quad 
\wt{H}^0_{uR}  =\sum_{j=1}^4(U^N_R)_{4j}\wt{\chi}^0_{jR}\,. 
\eea
\esub
%-----------------
Finally, the Higgs bosons are expressed as follows:  
%-----------------
\bsub\label{exp_higgs}
\bea
H_u^+ &=& (H_u^-)^* = i \chi^+ \sin\beta + H^+ \cos\beta \,,\\
H_d^- &=& (H_d^+)^* = i \chi^- \cos\beta + H^- \sin\beta \,,\\ 
H_u^0 &=& \halfsq( v\sin\beta + h \cos\alpha 
	+ H^0 \sin\alpha 
	- i \chi^3 \sin\beta + i A \cos\beta  ) \,,\\  
H_d^0 &=& \halfsq( v\cos\beta - h \sin\alpha 
	+ H^0 \cos\alpha 
	+ i \chi^3 \cos\beta + i A \sin\beta ) \,.
\eea
\esub
%-----------------
Here the phases $\alpha$ and $\beta$ are fixed by requiring 
$\cos\beta,\sin\beta,\cos\alpha > 0$. 
%%%----------------------------------
%%%	a new paragraph

%%%	a new paragraph
%%%----------------------------------
For notational compactness, we adopt the following 
generic notation of the MSSM couplings in 
Appendices~\ref{section:zdecay_amplitudes} and 
\ref{section:muon_decay} for the vertex and box corrections. 
Only three types of the couplings appear among the 9 types 
of the renormalizable couplings listed in ref.~\cite{mssml99}. 
The $FFV$ (fermion-fermion-vector) couplings, $g_\alpha^{F_1 F_2 V}$, 
are defined from the corresponding Lagrangian term as 
%%%--------------
\bea
{\cal L} &=& g_\alpha^{F_1 F_2 V} 
	\ov{F_1} \gamma^\mu P_\alpha F_2 V_\mu 
\label{eq:ffv}
\eea
%%%--------------
where $\alpha = L,R$. 
The $SSV$ (scalar-scalar-vector) couplings, $g^{S_1 S_2 V}$, 
are defined by 
%%%--------------
\bea
{\cal L} &=& i g^{S_1 S_2 V}  S_1^* \bdd_\mu S_2 V^\mu, 
\label{eq:ssv}
\eea
%%%--------------
where $A\bdd_\mu B = A(\partial_\mu B) - (\partial_\mu A)B$. 
Finally, the $FFS$ (fermion-fermion-scalar) couplings, 
$g_\alpha^{F_1 F_2 S}$, are defined by the following 
Lagrangian term: 
%%%--------------
\bea
{\cal L} &=& g_\alpha^{F_1 F_2 S} \ov{F_1} P_\alpha F_2 S,  
\label{eq:ffs}
\eea
%%%--------------
where $\alpha = L$ or $R$. 
These simple rules determine uniquely the magnitude and 
the phases of the MSSM couplings that appear in 
Appendices~\ref{section:zdecay_amplitudes} and 
\ref{section:muon_decay}. 
%--------------------------------------------
%
%         Appendix: Two point functions
%
%--------------------------------------------
\section{Two point functions}
\label{section:two_point}
\clean
%%%----------------------------------------
\subsection{Oblique parameters $S,T,U,R$}
%%%----------------------------------------
The transverse parts of the four gauge-boson-propagator 
functions are parametrized in ref.~\cite{hhkm94} as 
%%%-----------
\bsub
\bea
\ov{\Pi}^{\gamma \gamma}_T (q^2) &=& \ehatsq \ov{\Pi}_T^{QQ} (q^2), 
\\
\ov{\Pi}^{\gamma Z}_T (q^2) &=& \ehat \gzhat \biggl\{ 
	\ov{\Pi}_T^{3Q} (q^2) - \shatsq \ov{\Pi}_T^{QQ} (q^2)
	\biggr\}, 
\\
\ov{\Pi}^{Z Z}_T (q^2) &=& 
\gzhatsq \biggl\{ 
	\ov{\Pi}_T^{33} (q^2) - 2 \shatsq \ov{\Pi}_T^{3Q} (q^2)
	+ \shat^4 \ov{\Pi}_T^{QQ} (q^2)	\biggr\}, 
\\
\ov{\Pi}^{W W}_T (q^2) &=& \ghatsq \ov{\Pi}_T^{11} (q^2), 
\eea
\esub
%%%-----------
where $\ghatsq(\mu)=\ehatsq(\mu)/\shatsq(\mu)=
\gzhatsq(\mu)\chatsq(\mu)$ are the $\msbar$ couplings 
and all the polarization functions are renormalized in the $\msbar$ 
scheme. 
The four effective charges are defined by 
%%%-----------
\bsub
\bea
\frac{1}{\ebarsq(q^2)} &=& \frac{1}{\ehatsq(\mu)} 
	+ \re \ov{\Pi}^{QQ}_{T,\gamma}(q^2), 
\\ 
\vsk{0.2}
\frac{\sbarsq(q^2)}{\ebarsq(q^2)} &=& 
	\frac{\shatsq(\mu)}{\ehatsq(\mu)} 
	+ \re \ov{\Pi}^{3Q}_{T,\gamma}(q^2), 
\\
\vsk{0.2}
\frac{1}{\gzbarsq(q^2)} &=& \frac{1}{\gzhatsq(\mu)} 
	+ \re \ov{\Pi}^{33}_{T,Z}(q^2) 
	- 2 \shatsq \re \ov{\Pi}^{3Q}_{T,Z}(q^2) 
	+ \shat^4 \re \ov{\Pi}^{QQ}_{T,Z}(q^2),  
\\
\vsk{0.2}
\frac{1}{\gwbarsq(q^2)} &=& \frac{1}{\ghatsq(q^2)} 
	+ \re \ov{\Pi}^{11}_{T,W}(q^2), 	
\eea
\esub
%%%-----------
where 
%%%-----------
\bea
\ov{\Pi}^{AB}_{T,V} (q^2) &=& \frac{
	\ov{\Pi}^{AB}_T (q^2) - \ov{\Pi}^{AB}_T (m_V^2) }
	{q^2 -m_V^2}, 
\eea
%%%-----------
and 'overline' denotes the inclusion of the pinch 
term~\cite{pinch_term}. 
The oblique correction terms are then expressed in terms 
of these effective charges and the weak boson masses~\cite{hhkm94}; 
%%%-----------
\bsub
\bea
\frac{S}{4} &=& \frac{\sbarsq(\mzsq) \cbarsq(\mzsq)}{\abar(\mzsq)} 
	- \frac{4\pi}{\gzbarsq(0)}, 
\\
\alpha T &=&  1 - \frac{\gwbarsq(0)}{\mwsq} 
	\frac{\mzsq}{\gzbarsq(0)}, 
\\
\frac{S+U}{4} &=& \frac{\sbarsq(\mzsq)}{\abar(\mzsq)} 
	- \frac{4\pi}{\gwbarsq(0)}, 
\\
R &=& \frac{16\pi}{\gzbarsq(0)} - \frac{16\pi}{\gzbarsq(\mzsq)}, 
\eea
\esub
%%%-----------
or of the reduced functions, 
$\ov{\Pi}^{QQ}_T(q^2), \ov{\Pi}^{3Q}_T(q^2), \ov{\Pi}^{33}_T(q^2)$ 
and $\ov{\Pi}^{11}_T(q^2)$: 
%%%-----------
\bsub
\bea
S &=&
16\pi \re \biggl[ \ov{\Pi}_{T,\gamma}^{3Q}(\mzsq) 
	- \ov{\Pi}_{T,Z}^{33}(0) \biggr], 
\\
T &=& \frac{4\rttwo \gf}{\alpha}
	\biggl[ \ov{\Pi}_{T}^{33}(0) - \ov{\Pi}_{T}^{11}(0) \biggr], 
\\
U &=&
16\pi \re \biggl[ \ov{\Pi}_{T,Z}^{33}(0)
	- \ov{\Pi}_{T,W}^{11}(0) \biggr], 
\\
R &=& 16\pi \biggl[ 
	\ov{\Pi}_{T,Z}^{33}(0) - \ov{\Pi}_{T,Z}^{33}(q^2) 
	- 2\shatsq \biggl\{ 
	\ov{\Pi}_{T,Z}^{3Q}(0) - \ov{\Pi}_{T,Z}^{3Q}(q^2) 
		\biggr\}
\nonumber \\
	&& ~~~~~~~
	+ \shat^4 \biggl\{ 
	\ov{\Pi}_{T,Z}^{QQ}(0) - \ov{\Pi}_{T,Z}^{QQ}(q^2) 
	\biggr\}. 
\biggr]
\eea
\esub
%%%-----------------------------
The MSSM contribution to the $S,T,U,R$ parameters are 
then calculable from the MSSM particle contributions 
to the four $\Pi^{AB}_T$ functions. 
All the results listed below agree with those in ref.~\cite{dh90}. 
%%%-----------------------------------
\subsection{MSSM Higgs bosons}
%%%-----------------------------------
The MSSM Higgs boson contributions are summarized 
as follows: 
%%%-------------------------
\bsub
\bea
16\pi^2 \Pi_T^{QQ} &=& B_5(q^2:\mcharged, \mcharged) ,
\\ \vsk{0.1} 
16\pi^2 \Pi_T^{3Q} &=& \half B_5(q^2:\mcharged, \mcharged) , 
\\ \vsk{0.1} 
16\pi^2 \Pi_T^{33} &=& \quarter \biggl[
	\cos^2(\alpha-\beta) \bigl\{ 
	B_5(q^2:\mhlight, \mpseudo) + B_5(q^2:\mz, \mhheavy) 
				\bigr\}
\nonumber \\ \vsk{0.1} 
	&& 
	+ 
	\sin^2(\alpha-\beta) \bigl\{ 
	B_5(q^2:\mhheavy, \mpseudo) + B_5(q^2:\mz, \mhlight) 
	\bigr\}
	+ B_5(q^2:\mcharged, \mcharged) \biggl] 
\nonumber \\ \vsk{0.1} 
	&+&
	\mzsq \biggl[ 
	\sin^2(\alpha-\beta) B_0(q^2:\mz, \mhlight) 
	+ \cos^2(\alpha-\beta) B_0(q^2:\mz, \mhheavy) 
	\biggl], 
\\ \vsk{0.2} 
16\pi^2 \Pi_T^{11} &=& \quarter \biggl[
	\cos^2(\alpha-\beta) \bigl\{ 
	B_5(q^2:\mcharged, \mhlight) + B_5(q^2:\mw, \mhheavy) 
		\bigr\}
\nonumber \\ \vsk{0.1} 
	&+& 
%	~~~~~~~
	\sin^2(\alpha-\beta) \bigl\{ 
	B_5(q^2:\mcharged, \mhheavy) + B_5(q^2:\mw, \mhlight)  
		\bigr\}
%\nonumber \\ \vsk{0.1} 
%	&+& 
	+
	B_5(q^2:\mpseudo,\mcharged) \biggr]
\nonumber \\ \vsk{0.1} 
	&+& 
	\mwsq \biggl[ 
	\sin^2(\alpha-\beta) B_0(q:\mw, \mhlight)
	+
	\cos^2(\alpha-\beta) B_0(q:\mw, \mhheavy) 
	\biggr]. 
\eea
\esub
%%%-----------------------------
The $B$-functions are defined in Appendix D in ref.~\cite{hhkm94}. 
The SM Higgs boson contribution is reproduced by removing 
all terms with $H^-,H$ and $A$, and by setting 
$\sin^2(\alpha-\beta)=1$ and $\mhlight = \mh$. 
%%%---------------------------------
\subsection{Scalar fermions}
%%%---------------------------------
The scalar fermion contributions are summarized as:
%%%----------------------
\bsub
\bea
16\pi^2 \Pi_T^{QQ} &=& C_f Q_f^2 
	\sum_{\alpha=1}^2
	B_5(q^2:m_{\wt{f}_\alpha}, m_{\wt{f}_\alpha}),  
\\
\vsk{0.1}
16\pi^2 \Pi_T^{3Q} &=& C_f Q_f I_{3f} 
	\sum_{\alpha=1}^2 |(U^{\sfermi})_{\alpha1}|^2 
	B_5(q^2:m_{\wt{f}_\alpha}, m_{\wt{f}_\alpha}) , 
\\
\vsk{0.1}
16\pi^2 \Pi_T^{33} &=& C_f I_{3f}^2 
	\sum_{\alpha,\beta=1}^2
	|(U^{\sfermi})_{\alpha1}|^2|(U^{\sfermi})_{\beta1}|^2
	B_5(q^2:m_{\wt{f}_\alpha}, m_{\wt{f}_\beta}) , 
\\
\vsk{0.1}
16\pi^2 \Pi_T^{11} &=& C_f \half 
	\sum_{\alpha,\beta=1}^2
	|(U^{\sup})_{\alpha1}|^2 |(U^{\sdown})_{\beta1}|^2 
	B_5(q^2:m_{\wt{u}_\alpha}, m_{\wt{d}_\beta}) ,
\eea
\esub
%%%---------------------- 
where the color factor $C_f$ is 3 for squarks and 1 for sleptons. 
The electric charge $Q_f$ is given by $(2/3 ,-1/3, 0, -1)$ for 
$(\sup, \sdown, \sneutrino_l, \slepton)$. 
$I_{3f}$ denotes the third component of the weak isospin: 
$+1/2, -1/2$ for the up- and down-type sfermions, respectively. 
The indices $\alpha,\beta$ take $L, R$ for the first two 
generations (no left-right mixing) and $1,2$ for the third generation. 
If there is no mixing between $\sfermi_L$ and $\sfermi_R$, 
the unitary matrix $(U^{\sfermi})_{\alpha\beta}$ should be replaced 
by $\delta _{\alpha\beta}$. 
%%%------------------------
\subsection{Charginos and Neutralinos}
%%%------------------------
The chargino and neutralino contributions are as follows:
%%%------------------------
\bsub
\bea
16\pi^2 \Pi_T^{QQ} &=& 8q^2 B_3 (q^2: \mch{i}, \mch{i}), 
\\
\vsk{0.3}
16\pi^2 \Pi_T^{3Q} &=& 
	\biggl\{
	4 - \biggl| (\cunitary{L})_{2i} \biggr|^2	
	- \biggl| (\cunitary{R})_{2i} \biggr|^2
	\biggr\}
	2 q^2 B_3 (q^2 : \mch{i}, \mch{i}), 
\\
\vsk{0.3}
16\pi^2  \Pi_T^{33} &=& 
	\biggl[
	2\biggl\{ |(D_L)_{ij}|^2 + |(D_R)_{ij}|^2 \biggr\}
	(2q^2 B_3 - B_4) 
\nonumber \\ \vsk{0.1}
	&&
	+ 2 \mch{i} \mch{j} 
	\biggl\{ (D_L)_{ij} (D_R)_{ij}^*  
	+ (D_L)_{ij}^* (D_R)_{ij} \biggr\}
	B_0 \biggr](q^2: \mch{i}, \mch{j}) 
\nonumber \\ \vsk{0.1}
	&&
	+ 
	\biggl[
	\biggl\{ |(N_L)_{ij}|^2 + |(N_R)_{ij}|^2\biggr\}
	(2q^2 B_3 - B_4)
\nonumber \\ \vsk{0.1}
	&&
	+ \mn{i} \mn{j} \biggl\{ (N_L)_{ij} (N_R)_{ij}^*
		+ (N_L)_{ij}^* (N_R)_{ij} \biggr\}
	B_0 \biggr ](q^2:\mn{i}, \mn{j}), 
\\
\vsk{0.3}
16\pi^2 \Pi_T^{11} &=& 2 
	\biggl[ \biggl\{ |(C_{L})_{ij}|^2 + |(C_{R})_{ij}|^2\biggr\}
	(2q^2 B_3 - B_4)(q^2: \mn{i}, \mch{j}) 
\nonumber \\
	&& 
	+ \mn{i} \mch{j} 
 	\biggl\{ (C_{L})_{ij} (C_{R})_{ij}^* 
	+ (C_{L})_{ij}^* (C_{R})_{ij} \biggr\}
	B_0(q^2: \mn{i}, \mch{j}) 
	\biggr], 
\eea
\esub
%-----------------
where 
%----------------- 
\bsub
\bea
(C_{\alpha})_{ij} 
	&=& (\nunitary{\alpha})^*_{2i} (\cunitary{\alpha})_{1j}
 + \frac{1}{\rttwo} (\nunitary{\alpha})^*_{3i}(\cunitary{\alpha})_{2j},
\label{eq:ino_coupling_one}
\\
\vsk{0.3}
(D_{\alpha})_{ij} 
	&=& (\cunitary{\alpha})^*_{1i} (\cunitary{\alpha})_{1j} 
	+ \half (\cunitary{\alpha})^*_{2i} (\cunitary{\alpha})_{2j},  
\label{eq:ino_coupling_two}
%-----------------
\\ \vsk{0.3}
%-----------------
(N_{L})_{ij} &=& -(N_{R})^*_{ij} = 
	\half \biggl\{ (\nunitary{L})^*_{3i}(\nunitary{L})_{3j} - 
	(\nunitary{L})^*_{4i}(\nunitary{L})_{4j} \biggr\}.
\eea
\esub
%%%-----------------
Here $\alpha=L$ or $R$ in (\ref{eq:ino_coupling_one}) 
and (\ref{eq:ino_coupling_two}). 
%%%%%%%%%%%%%%%%%%%%%%%%%%%%%%%%%%%%%%%%%%%%%%%%%%%%%%%%%%%%%%%%%%%%%%%
%%%%%
%%%%%
%%%%% vertex corrections on the Z --> f_alpha f_alpha
%%%%%
%%%%%
%%%%%%%%%%%%%%%%%%%%%%%%%%%%%%%%%%%%%%%%%%%%%%%%%%%%%%%%%%%%%%%%%%%%%%%
\newpage
\section{Vertex corrections on $Z\to f_\alpha \ov{f_\alpha}$}
\label{section:zdecay_amplitudes}
\clean
%%%--------------------------------------------- 
Here we give the 1-loop corrections to the process 
$Z\to f_\alpha \ov{f_\alpha}$ in the MSSM. 
The radiative corrections to the effective coupling $g_\alpha^f$ 
is denoted by
%%%---------------
\bea
\Del g_\alpha^f &=& \frac{1}{\sqrt{4\rttwo \gf \mzsq}} 
	\biggl\{ 
	g_\alpha^{ffZ} \Sigma_{f_\alpha}'(0) 
	- \Gamma_{f_\alpha}(\mzsq) 
	\biggr\}, 
\eea
%%%---------------
where $\Sigma_{f_\alpha}'(0)$ is the derivative of the 
self energy function of the external fermion $f_\alpha$ 
with the chirality $\alpha=L$ or $R$, whose mass is neglected: 
%%%---------------
\bea
\Sigma_{f_\alpha}'(0) &=& \frac{d}{dq^2} \Sigma_{f_\alpha}(q^2)
	\biggr|_{q^2 = 0}.
\eea
%%%---------------

%%%---------------
We can express the self energy $\Sigma'_{f_\alpha}(0)$ and 
the vertex function $\Gamma_{f_\alpha}(q^2)$ mediated by a fermion 
$\psi$ and a scalar $\phi$ in a compact generic notation as 
follows: 
%%%---------------
\bsub
\bea
(4\pi)^2 \Sigma'_{f_\alpha}(0) &=& 
	C_g \biggl| g^{\psi_j f \phi_i}_\alpha \biggr|^2 
	\biggl( B_0 + B_1 \biggr) (0:m_{\phi_i}, m_{\psi_j}), 
\\
\vsk{0.3}
(4\pi)^2 \Gamma_{f_\alpha}(q^2) &=& 
	- C_g \Biggl\{ 
	\biggl( g_\alpha^{\psi_j f \phi_k} \biggr)^* 
	g_\alpha^{\psi_i f \phi_k}
	\biggl[ 
	g_\alpha^{\psi_j \psi_i Z} m_{\psi_i} m_{\psi_j} C_0 
\nonumber \\
&&
	+ g_{-\alpha}^{\psi_j \psi_i Z} 
	\biggl\{-q^2 (C_{12} + C_{23}) - 2 C_{24} + \half \biggr\}
	\biggr]
\nonumber \\
&&
	~~~~~~~~~~~~~
	\times (p_1, p_2: m_{\psi_i}, m_{\phi_k}, m_{\psi_j})
\nonumber \\
&&
	-  \biggl( g^{\psi_k f \phi_i}_\alpha \biggr)^* 
	g^{\psi_k f \phi_j}_\alpha
	g^{\phi_i \phi_j Z} 
	2 C_{24}(p_1, p_2: m_{\phi_j}, m_{\psi_k}, m_{\phi_i})
	\Biggr\}. 
\eea
\esub
%%%---------------
Here $C_g$ is $4/3$ for the gluino contribution ($\psi=\gluino$) 
and 1 for the others. 
The chirality index $-\alpha$ follows the rule: $-L=R, -R=L$. 
For each external fermion $f$, the following combination of 
$\{ \psi, \phi \}$ contribute to the vertices: 
%%%-------------
\bea
\begin{array}{l|c|c|c|c} \hline \hline 
 & \multicolumn{4}{c}{\{\psi,\phi\}} \\
	\hline 
 & f=u & d & \nu_l & l  \\
	\hline 
\psi = {\rm chargino} & \{\wt{\chi}^+,\sdown_i^*\} &\{\chargino{},\sup_i^*\} 
	& \{\wt{\chi}^+, \slepton_i^*\}
	& \{\chargino{}, \sneutrino_l^* \} \\
\psi = {\rm neutralino} &\{\neutralino{},\sup_i^*\}&\{\neutralino,\sdown_i^*\} 
	& \{\neutralino, \sneutrino_l^*\} & \{\neutralino, \slepton_i^*\} \\
\psi = {\rm gluino} & \{\gluino, \sup_i^*\} & \{\gluino, \sdown_i^*\} & 
	- & -  \\
\phi = {\rm charged~ Higgs} & \{ d, H^- \} & \{ u, H^+ \} 
	& \{ l, H^- \} & \{\nu, H^+ \} \\
\phi = {\rm neutral~ Higgs} & \{ u, (h,H,A) \} & \{ d, (h,H,A) \} & - 
	& \{l, (h,H,A)\} \\
\hline \hline 
\end{array}
\eea 
%%%-------------
The generic coupling notations of (\ref{eq:ffv}) -- (\ref{eq:ffs}) 
then suffice to calculate all the $Z f_\alpha f_\alpha$ vertex 
corrections, $\Del g_\alpha^f$. 
Summation should be taken over all non-vanishing coupling 
combinations. 
When $d=b$ or $\ell=\tau$, the summations over the sfermion 
index is taken over $i=1$ and 2. 
Otherwise, only the $i=\alpha$ sfermion contributes. 
We reproduced numerically the results reported in 
refs.~\cite{hm90,zbb_susy,yhm95}.

%%%%%%%%%%%%%%%%%%%%%%%%%%%%%%%%%%%%%%%%%%%%%%%%%%%%%%%%%%%%%%%%%%%%%%%
%%%%%
%%%%%
%%%%% vertex and box corrections on the muon decay 
%%%%%
%%%%%
%%%%%%%%%%%%%%%%%%%%%%%%%%%%%%%%%%%%%%%%%%%%%%%%%%%%%%%%%%%%%%%%%%%%%%%
%%%-------------------------
\newpage
%%%-------------------------
\section{SUSY contributions to the $\mu$-decay process}
\label{section:muon_decay}
\clean
%%%-------------------------
The $\mu$-decay constant is parametrized in ref.~\cite{hhkm94} as 
%%%-------------------------
\bea
\gf = \frac{\gwbarsq(0) + \ghatsq \delg}{4\sqrt{2}\mwsq}, 
\eea
%%%-------------------------
with
%%%-------------------------
\bea
\delg &=& (\delg)_{\rm SM} + \ddelg. 
\eea
%%%-------------------------
In the MSSM, $\ddelg$ receives contributions from the vertex and 
the box diagrams
%%%-------------------------
\bea
\ddelg &=& 2 \delta_g^{(v)} + \delta_g^{(b)}. 
\eea
%%%-------------------------
\subsection{Vertex corrections}
\clean
%%%---------------------------------------------
The $f_1 f_2 W$ vertex correction can be expressed as 
%%%---------------------------------------------
\bea
g_L^{f_1 f_2 W} \delta^{(v)} 
   &=& \Gamma^{f_1 f_2 W}(0) - \frac{1}{2} g_L^{f_1 f_2 W} 
	\biggl\{ \Sigma'_{f_1}(0) + \Sigma'_{f_2}(0)
	\biggr\}. 
\eea
%%%---------------
In the $\mu \to \nu_\mu e \ov{\nu}_e$ process, the 
$\mu \nu_\mu W^-$ and $\nu_e e W^+$ vertices give the 
some $\delta^{(v)}$, with 
$g_L^{\mu \nu_\mu W^-} = g_L^{\nu_e e W^+} = \frac{\ghat}{\rttwo}$. 
From the $\nu_e e W^+$ vertex, we find
%%%--------------- 
\bsub
\bea
(4 \pi)^2 \Sigma'_{e_L}(0) &=& 
\biggl| g_L^{\neutralino{i} e \selectron_L}\biggr|^2 
	(B_0 + B_1) (0:m_{\selectron_L}, m_{\neutralino{i}})
+
\biggl| g_L^{\chargino{j} e \sneutrino_e}\biggr|^2 
	(B_0 + B_1) (0:m_{\sneutrino_e}, m_{\chargino{j}}), 
\nonumber \\
&& \\
(4 \pi)^2 \Sigma'_{\nu_e}(0) &=& 
\biggl| g_L^{\neutralino{i} \nu_e \sneutrino_e}\biggr|^2 
	(B_0 + B_1) (0:m_{\sneutrino_e}, m_{\neutralino{i}})
+
\biggl| g_L^{\wt{\chi}^+_j \nu_e \selectron_L}\biggr|^2 
	(B_0 + B_1) (0:m_{\selectron_L}, m_{\chargino{j}}), 
\nonumber \\
\eea
\label{eq:muon_self}
\esub
%%%---------------
\bea
(4 \pi)^2 \Gamma_{e\nu_e W^+} &=& 
	-(g_L^{\wt{\chi}^+_j \nu_e \selectron_L})^* 
	g_L^{\neutralino{i} e \selectron_L}\biggl\{ 
	g_L^{\wt{\chi}^+_j \neutralino{i} W } m_{\neutralino{i}}
	m_{\chargino{j}} C_0 
	+ g_R^{\wt{\chi}^+_j \neutralino{i} W } 
	(-2 C_{24} + \half) \biggr\} 
\nonumber \\
	&&~~~~~~~~~~~~\times
	(0:m_{\neutralino{i}}, m_{\selectron_L}, m_{\chargino{j}})
\nonumber \\
\vsk{0.3}
&&
	-(g_L^{\neutralino{i} \nu_e \sneutrino_e})^* 
	g_L^{\chargino{j} e \sneutrino_e}
	\biggl\{ 
	g_L^{\neutralino{i} \chargino{j} W } 
	m_{\neutralino{i}} m_{\chargino{j}}C_0 
	+ g_R^{\neutralino{i} \chargino{j} W} 
	(-2 C_{24} + \half) \biggr\} 
\nonumber \\
	&&~~~~~~~~~~~~\times
	(0:m_{\chargino{j}},m_{\sneutrino_e}, m_{\neutralino{i}} )
\nonumber \\
\vsk{0.3}
&&
	+ (g_L^{\neutralino{i} \nu_e \sneutrino_e})^* 
	g_L^{\neutralino{i} e \selectron_L}
	g_L^{\sneutrino_e \selectron_L W} 2 C_{24}(0:m_{\selectron_L}, 
	m_{\neutralino{i}}, m_{\sneutrino_e}). 
\label{eq:muon_vertex}
\eea
%%%--------------------
In eqs.~(\ref{eq:muon_self}), (\ref{eq:muon_vertex}), 
summation over $i=1$ to 4 $(\neutralino{i})$ and 
$j=1$ to 2 $(\chargino{j})$ should be understood. 
The momentum arguments of the $C$-functions are set to 
zero $(p_i^2 = p_i p_j = 0)$. 
%%%--------------------
\subsection{Box corrections}
%%%--------------------
The box contributions to the $\mu \to \nu_\mu e \ov{\nu}_e$ 
amplitude can be expressed as 
%%%--------------------
\bea
iT &=& i \biggl\{ 
	M(1) + M(2) + M(3) + M(4) \biggr\}
	\ov{u_e} \gamma^\mu P_L v_{\nu_e} 
	\ov{u_{\nu_\mu}} \gamma^\mu P_L u_{\mu}. 
\eea
%%%------------------
By taking into account the normalization of the tree-level 
amplitude, $-\ghatsq/2\mwsq$, the box diagram contributions 
to the $\delg$ parameter is 
%%%------------------
\bea
\delta^{(b)} &=& -\frac{2 m_W^2}{\ghatsq} \sum_{i=1}^4 M(i). 
\eea
%%%------------------
Each $M(i)$ is given by 
%%%------------------
\bsub
\bea
16 \pi^2 M(1) &=& (g_L^{\neutralino{i} e \selectron_L})^* 
	    g_L^{\neutralino{i} \mu \smuon_L }
	   (g_L^{\wt{\chi}^+_j \nu_\mu \smuon_L})^* 
	    g_L^{\wt{\chi}^+_j \nu_e \selectron_L}
	D_{27}(m_{\smuon_L}, m_{\selectron_L}, m_{\wt{\chi}^+_j}, 
	m_{\neutralino{i}})
\nonumber \\
&& \\
\vsk{0.3}
16 \pi^2 M(2) &=& (g_L^{\chargino{j} e \sneutrino_e})^* 
	g_L^{\chargino{j} \mu \sneutrino_\mu } 
	(g_L^{\neutralino{i} \nu_\mu \sneutrino_\mu})^* 
	g_L^{\neutralino{i} \nu_e \sneutrino_e}
   D_{27}(m_{\sneutrino_\mu}, m_{\sneutrino_e}, 
	m_{\chargino{j}}, m_{\neutralino{i}})
\nonumber \\
&& \\
\vsk{0.3}
16 \pi^2 M(3) &=& \half m_{\neutralino{i}} m_{\chargino{j}} 
	g_L^{\wt{\chi}^+_j \nu_e \selectron_L}
	g_L^{\chargino{j} \mu \sneutrino_\mu } 
	(g_L^{\neutralino{i} \nu_\mu \sneutrino_\mu})^* 
	(g_L^{\neutralino{i} e \selectron_L})^* 
   D_0(m_{\sneutrino_\mu}, m_{\selectron_L}, m_{\chargino{j}}, 
	m_{\neutralino{i}})
\nonumber \\
&& \\
\vsk{0.3}
16 \pi^2 M(4) &=& \half m_{\neutralino{i}} m_{\chargino{j}} 
	g_L^{\neutralino{i} \nu_e \sneutrino_e}
	g_L^{\neutralino{i} \mu \smuon_L}
	(g_L^{\wt{\chi}^+_j \nu_\mu \smuon_L})^*
	(g_L^{\chargino{j} e \sneutrino_e } )^*
   D_0(m_{\smuon_L}, m_{\sneutrino_e}, m_{\chargino{j}}, 
	m_{\neutralino{i}})
\nonumber \\
\eea
\esub
All the $D$-functions are evaluated at the zero momentum transfer 
limit.  
We reproduced the results presented in ref.~\cite{hmy95}.   
%%%-------------------------
\end{appendix}
%---------------------------------------------------------
%
%
%
%REFERENCES	%REFERENCES	%REFERENCES	%REFERENCES
%
%
%
%---------------------------------------------------------

\end{document}